\documentclass[twocolumn,letterpaper,aps,prc,superscriptaddress,longbibliography,nofootinbib,floatfix]{revtex4-2}

\usepackage{graphicx}   
\usepackage{multirow}
\usepackage{amsmath}
\usepackage{subfigure}

\usepackage{xspace}     

\newcommand{\fig}[1]{Fig.~\ref{#1}}
\newcommand{\eq}[1]{Eq.~\ref{#1}}
\newcommand{\tab}[1]{Table~\ref{#1}}
\newcommand{\sect}[1]{Sec.~\ref{#1}}
\newcommand{\append}[1]{Appendix~\ref{#1}}
\newcommand{\s}[1]{\mbox{$\sqrt{s}=#1$\,GeV}}
\newcommand{\snn}[1]{\mbox{$\sqrt{s_{_{NN}}}=#1$\,GeV}}

\newcommand{\sqsn}{\mbox{$\sqrt{s_{_{NN}}}$}\xspace}
 
\newcommand{\AB}[2]{#1$+$#2}

\newcommand{\pp}{$p$$+$$p$\xspace}

\newcommand{\pbpb}{\mbox{Pb$+$Pb}\xspace}

\newcommand{\auau}{\mbox{Au$+$Au}\xspace}

\newcommand{\Npart}{\mbox{$N_{\rm part}$}\xspace}
\newcommand{\Ncoll}{\mbox{$N_{\rm coll}$}\xspace}

\newcommand{\dNch}{\mbox{$dN_{\rm ch}/d\eta$}\xspace}
\newcommand{\dNgdy}{\mbox{$dN_{\gamma}/dy$}\xspace}
\newcommand{\gev}{\mbox{GeV}\xspace}
\newcommand{\gevc}{\mbox{GeV/$c$}\xspace}
\newcommand{\gevcc}{\mbox{GeV/$c^2$}\xspace}

\newcommand{\pt}{\mbox{$p_T$}\xspace}
\newcommand{\piz}{\mbox{$\pi^0$}\xspace}
\newcommand{\mee}{\mbox{$m_{e^+e^-}$}\xspace}
\newcommand{\meeg}{\mbox{$m_{ee\gamma}$}\xspace}
\newcommand{\ee}{\mbox{$e^+e^-$}\xspace}
\newcommand{\eeg}{\mbox{$e^+e^-\gamma$}\xspace}
\newcommand{\Teff}{\mbox{$T_{\rm eff}$}\xspace}
\newcommand{\bdl}{\mbox{$\int{Bdl}$}\xspace}
\newcommand{\gammaincl}{\mbox{$\gamma^{\rm incl}$}\xspace}
\newcommand{\gammadir}{\mbox{$\gamma^{\rm dir}$}\xspace}
\newcommand{\gammahadr}{\mbox{$\gamma^{\rm hadr}$}\xspace}
\newcommand{\gammapiz}{\mbox{$\gamma^{\rm{\pi^0}}$}\xspace}
\newcommand{\gammanonp}{\mbox{$\gamma^{\rm{nonprompt}}$}\xspace}
\newcommand{\Rg}{\mbox{$R_{\gamma}$}\xspace}
\newcommand{\ef}{\mbox{$\langle\epsilon_{\gamma} f\rangle$}\xspace}
\newcommand{\Npiz}{\mbox{$N_{ee}^{\rm{\pi^0}}$}\xspace}
\newcommand{\Neetag}{\mbox{$N_{ee}^{\rm{\pi^0}, tag}$}\xspace}
\newcommand{\Nincl}{\mbox{$N_{\gamma}^{\rm{incl}}$}\xspace}
\newcommand{\Ninclsg}{\mbox{${\rm SG}^{\rm{ee}}$}\xspace}
\newcommand{\Ninclfg}{\mbox{${\rm FG}^{\rm{ee}}$}\xspace}
\newcommand{\Ninclbg}{\mbox{${\rm BG}^{\rm{ee}}$}\xspace}
\newcommand{\Minclbg}{\mbox{${\rm MBG}^{\rm{ee}}$}\xspace}
\newcommand{\Ntag}{\mbox{$N_{\gamma}^{\rm{\pi^0, tag}}$}\xspace}
\newcommand{\Ntagsg}{\mbox{${\rm SG}^{ee\gamma}$}\xspace}
\newcommand{\Ntagfg}{\mbox{${\rm FG}^{ee\gamma}$}\xspace}
\newcommand{\Ntagbg}{\mbox{${\rm BG}^{ee\gamma}$}\xspace}
\newcommand{\Ntaguncorr}{\mbox{${\rm BG}_{\rm uncorr}^{ee\gamma}$}\xspace}
\newcommand{\Ntagcorr}{\mbox{${\rm BG}_{\rm corr}^{ee\gamma}$}\xspace}

\newcommand{\Mtaguncorr}{\mbox{${\rm MBG}_{\rm uncorr}^{ee\gamma}$}\xspace}
\newcommand{\Mtagcorr}{\mbox{${\rm MBG}_{\rm cor}^{ee\gamma}$}\xspace}
\newcommand{\Mtagcomb}{\mbox{${\rm MBG}_{\rm comb}^{ee\gamma}$}\xspace}

\begin{document}
 
\title{Nonprompt direct-photon production in Au$+$Au collisions at
$\sqrt{s_{_{NN}}}=200$ GeV}
 
\newcommand{\abilene}{Abilene Christian University, Abilene, Texas 79699, USA}
\newcommand{\augie}{Department of Physics, Augustana University, Sioux Falls, South Dakota 57197, USA}
\newcommand{\banaras}{Department of Physics, Banaras Hindu University, Varanasi 221005, India}
\newcommand{\barc}{Bhabha Atomic Research Centre, Bombay 400 085, India}
\newcommand{\baruch}{Baruch College, City University of New York, New York, New York, 10010 USA}
\newcommand{\bnlcoll}{Collider-Accelerator Department, Brookhaven National Laboratory, Upton, New York 11973-5000, USA}
\newcommand{\bnlphys}{Physics Department, Brookhaven National Laboratory, Upton, New York 11973-5000, USA}
\newcommand{\caucr}{University of California-Riverside, Riverside, California 92521, USA}
\newcommand{\charlesczech}{Charles University, Faculty of Mathematics and Physics, 180 00 Troja, Prague, Czech Republic}
\newcommand{\ciae}{Science and Technology on Nuclear Data Laboratory, China Institute of Atomic Energy, Beijing 102413, People's Republic of China}
\newcommand{\cns}{Center for Nuclear Study, Graduate School of Science, University of Tokyo, 7-3-1 Hongo, Bunkyo, Tokyo 113-0033, Japan}
\newcommand{\colorado}{University of Colorado, Boulder, Colorado 80309, USA}
\newcommand{\columbia}{Columbia University, New York, New York 10027 and Nevis Laboratories, Irvington, New York 10533, USA}
\newcommand{\czechtech}{Czech Technical University, Zikova 4, 166 36 Prague 6, Czech Republic}
\newcommand{\debrecen}{Debrecen University, H-4010 Debrecen, Egyetem t{\'e}r 1, Hungary}
\newcommand{\elte}{ELTE, E{\"o}tv{\"o}s Lor{\'a}nd University, H-1117 Budapest, P{\'a}zm{\'a}ny P.~s.~1/A, Hungary}
\newcommand{\ewha}{Ewha Womans University, Seoul 120-750, Korea}
\newcommand{\famu}{Florida A\&M University, Tallahassee, FL 32307, USA}
\newcommand{\fsu}{Florida State University, Tallahassee, Florida 32306, USA}
\newcommand{\gsu}{Georgia State University, Atlanta, Georgia 30303, USA}
\newcommand{\hiroshima}{Hiroshima University, Kagamiyama, Higashi-Hiroshima 739-8526, Japan}
\newcommand{\howard}{Department of Physics and Astronomy, Howard University, Washington, DC 20059, USA}
\newcommand{\ihepprot}{IHEP Protvino, State Research Center of Russian Federation, Institute for High Energy Physics, Protvino, 142281, Russia}
\newcommand{\illuiuc}{University of Illinois at Urbana-Champaign, Urbana, Illinois 61801, USA}
\newcommand{\inrras}{Institute for Nuclear Research of the Russian Academy of Sciences, prospekt 60-letiya Oktyabrya 7a, Moscow 117312, Russia}
\newcommand{\instpasczech}{Institute of Physics, Academy of Sciences of the Czech Republic, Na Slovance 2, 182 21 Prague 8, Czech Republic}
\newcommand{\isu}{Iowa State University, Ames, Iowa 50011, USA}
\newcommand{\jaea}{Advanced Science Research Center, Japan Atomic Energy Agency, 2-4 Shirakata Shirane, Tokai-mura, Naka-gun, Ibaraki-ken 319-1195, Japan}
\newcommand{\jeonbuk}{Jeonbuk National University, Jeonju, 54896, Korea}
\newcommand{\jyvaskyla}{Helsinki Institute of Physics and University of Jyv{\"a}skyl{\"a}, P.O.Box 35, FI-40014 Jyv{\"a}skyl{\"a}, Finland}
\newcommand{\kek}{KEK, High Energy Accelerator Research Organization, Tsukuba, Ibaraki 305-0801, Japan}
\newcommand{\korea}{Korea University, Seoul 02841, Korea}
\newcommand{\kurchatov}{National Research Center ``Kurchatov Institute", Moscow, 123098 Russia}
\newcommand{\kyoto}{Kyoto University, Kyoto 606-8502, Japan}
\newcommand{\lawllnl}{Lawrence Livermore National Laboratory, Livermore, California 94550, USA}
\newcommand{\losalamos}{Los Alamos National Laboratory, Los Alamos, New Mexico 87545, USA}
\newcommand{\lund}{Department of Physics, Lund University, Box 118, SE-221 00 Lund, Sweden}
\newcommand{\lyon}{IPNL, CNRS/IN2P3, Univ Lyon, Universit{\'e} Lyon 1, F-69622, Villeurbanne, France}
\newcommand{\maryland}{University of Maryland, College Park, Maryland 20742, USA}
\newcommand{\mass}{Department of Physics, University of Massachusetts, Amherst, Massachusetts 01003-9337, USA}
\newcommand{\mate}{MATE, Laboratory of Femtoscopy, K\'aroly R\'obert Campus, Gy\"ongy\"os, Hungary}
\newcommand{\michigan}{Department of Physics, University of Michigan, Ann Arbor, Michigan 48109-1040, USA}
\newcommand{\miss}{Mississippi State University, Mississippi State, Mississippi 39762, USA}
\newcommand{\muhlenberg}{Muhlenberg College, Allentown, Pennsylvania 18104-5586, USA}
\newcommand{\nara}{Nara Women's University, Kita-uoya Nishi-machi Nara 630-8506, Japan}
\newcommand{\natmephi}{National Research Nuclear University, MEPhI, Moscow Engineering Physics Institute, Moscow, 115409, Russia}
\newcommand{\newmex}{University of New Mexico, Albuquerque, New Mexico 87131, USA}
\newcommand{\nmsu}{New Mexico State University, Las Cruces, New Mexico 88003, USA}
\newcommand{\northcg}{Physics and Astronomy Department, University of North Carolina at Greensboro, Greensboro, North Carolina 27412, USA}
\newcommand{\ohio}{Department of Physics and Astronomy, Ohio University, Athens, Ohio 45701, USA}
\newcommand{\ornl}{Oak Ridge National Laboratory, Oak Ridge, Tennessee 37831, USA}
\newcommand{\orsay}{IPN-Orsay, Univ.~Paris-Sud, CNRS/IN2P3, Universit\'e Paris-Saclay, BP1, F-91406, Orsay, France}
\newcommand{\peking}{Peking University, Beijing 100871, People's Republic of China}
\newcommand{\pnpi}{PNPI, Petersburg Nuclear Physics Institute, Gatchina, Leningrad region, 188300, Russia}
\newcommand{\pusan}{Pusan National University, Pusan 46241, Korea}
\newcommand{\riken}{RIKEN Nishina Center for Accelerator-Based Science, Wako, Saitama 351-0198, Japan}
\newcommand{\rikjrbrc}{RIKEN BNL Research Center, Brookhaven National Laboratory, Upton, New York 11973-5000, USA}
\newcommand{\rikkyo}{Physics Department, Rikkyo University, 3-34-1 Nishi-Ikebukuro, Toshima, Tokyo 171-8501, Japan}
\newcommand{\saispbstu}{Saint Petersburg State Polytechnic University, St.~Petersburg, 195251 Russia}
\newcommand{\seoulnat}{Department of Physics and Astronomy, Seoul National University, Seoul 151-742, Korea}
\newcommand{\stonybrkc}{Chemistry Department, Stony Brook University, SUNY, Stony Brook, New York 11794-3400, USA}
\newcommand{\stonycrkp}{Department of Physics and Astronomy, Stony Brook University, SUNY, Stony Brook, New York 11794-3800, USA}
\newcommand{\tenn}{University of Tennessee, Knoxville, Tennessee 37996, USA}
\newcommand{\texsu}{Texas Southern University, Houston, TX 77004, USA}
\newcommand{\titech}{Department of Physics, Tokyo Institute of Technology, Oh-okayama, Meguro, Tokyo 152-8551, Japan}
\newcommand{\tsukuba}{Tomonaga Center for the History of the Universe, University of Tsukuba, Tsukuba, Ibaraki 305, Japan}
\newcommand{\vandy}{Vanderbilt University, Nashville, Tennessee 37235, USA}
\newcommand{\weizmann}{Weizmann Institute, Rehovot 76100, Israel}
\newcommand{\wigner}{Institute for Particle and Nuclear Physics, Wigner Research Centre for Physics, Hungarian Academy of Sciences (Wigner RCP, RMKI) H-1525 Budapest 114, POBox 49, Budapest, Hungary}
\newcommand{\yonsei}{Yonsei University, IPAP, Seoul 120-749, Korea}
\newcommand{\zagreb}{Department of Physics, Faculty of Science, University of Zagreb, Bijeni\v{c}ka c.~32 HR-10002 Zagreb, Croatia}
\newcommand{\zambia}{Department of Physics, School of Natural Sciences, University of Zambia, Great East Road Campus, Box 32379, Lusaka, Zambia}
\affiliation{\abilene}
\affiliation{\augie}
\affiliation{\banaras}
\affiliation{\barc}
\affiliation{\baruch}
\affiliation{\bnlcoll}
\affiliation{\bnlphys}
\affiliation{\caucr}
\affiliation{\charlesczech}
\affiliation{\ciae}
\affiliation{\cns}
\affiliation{\colorado}
\affiliation{\columbia}
\affiliation{\czechtech}
\affiliation{\debrecen}
\affiliation{\elte}
\affiliation{\ewha}
\affiliation{\famu}
\affiliation{\fsu}
\affiliation{\gsu}
\affiliation{\hiroshima}
\affiliation{\howard}
\affiliation{\ihepprot}
\affiliation{\illuiuc}
\affiliation{\inrras}
\affiliation{\instpasczech}
\affiliation{\isu}
\affiliation{\jaea}
\affiliation{\jeonbuk}
\affiliation{\jyvaskyla}
\affiliation{\kek}
\affiliation{\korea}
\affiliation{\kurchatov}
\affiliation{\kyoto}
\affiliation{\lawllnl}
\affiliation{\losalamos}
\affiliation{\lund}
\affiliation{\lyon}
\affiliation{\maryland}
\affiliation{\mass}
\affiliation{\mate}
\affiliation{\michigan}
\affiliation{\miss}
\affiliation{\muhlenberg}
\affiliation{\nara}
\affiliation{\natmephi}
\affiliation{\newmex}
\affiliation{\nmsu}
\affiliation{\northcg}
\affiliation{\ohio}
\affiliation{\ornl}
\affiliation{\orsay}
\affiliation{\peking}
\affiliation{\pnpi}
\affiliation{\pusan}
\affiliation{\riken}
\affiliation{\rikjrbrc}
\affiliation{\rikkyo}
\affiliation{\saispbstu}
\affiliation{\seoulnat}
\affiliation{\stonybrkc}
\affiliation{\stonycrkp}
\affiliation{\tenn}
\affiliation{\texsu}
\affiliation{\titech}
\affiliation{\tsukuba}
\affiliation{\vandy}
\affiliation{\weizmann}
\affiliation{\wigner}
\affiliation{\yonsei}
\affiliation{\zagreb}
\affiliation{\zambia}
\author{N.J.~Abdulameer} \affiliation{\debrecen}
\author{U.~Acharya} \affiliation{\gsu} 
\author{A.~Adare} \affiliation{\colorado} 
\author{C.~Aidala} \affiliation{\michigan} 
\author{N.N.~Ajitanand} \altaffiliation{Deceased} \affiliation{\stonybrkc} 
\author{Y.~Akiba} \email[PHENIX Spokesperson: ]{akiba@rcf.rhic.bnl.gov} \affiliation{\riken} \affiliation{\rikjrbrc} 
\author{M.~Alfred} \affiliation{\howard} 
\author{N.~Apadula} \affiliation{\isu} \affiliation{\stonycrkp} 
\author{H.~Asano} \affiliation{\kyoto} \affiliation{\riken} 
\author{B.~Azmoun} \affiliation{\bnlphys} 
% ...
\collaboration{PHENIX Collaboration}  \noaffiliation
 
\date{\today}

 
\begin{abstract}


The measurement of the direct-photon spectrum from Au$+$Au collisions at
$\sqrt{s_{_{NN}}}=200$ GeV is presented by the PHENIX collaboration using
the external-photon-conversion technique for 0\%--93\% central collisions in
a transverse-momentum ($p_T$) range of 0.8--10 GeV/$c$. An excess of direct
photons, above prompt-photon production from hard-scattering processes, is
observed for $p_T<6$ GeV/$c$.  Nonprompt direct photons are measured by
subtracting the prompt component, which is estimated as 
$N_{\rm coll}$-scaled direct photons from $p$$+$$p$ collisions at 200 GeV, 
from the direct-photon spectrum. Results are obtained for
$0.8<p_T<6.0$ GeV/$c$ and suggest that the spectrum has an increasing 
inverse slope from ${\approx}0.2$ to 0.4 GeV/$c$ with increasing $p_T$, 
which indicates a possible sensitivity of the measurement to photons from 
earlier stages of the evolution of the collision. In addition, like the 
direct-photon production, the $p_T$-integrated nonprompt direct-photon 
yields also follow a power-law scaling behavior as a function of 
collision-system size.  The exponent, $\alpha$, for the nonprompt 
component is found to be consistent with 1.1 with no apparent $p_T$ 
dependence.

\end{abstract}

\maketitle

 
\section{\label{sec:intro}Introduction}       
 
Direct photons, defined as those not coming from hadron decays, have 
long been considered a golden probe towards our understanding of the 
evolution of relativistic heavy-ion collisions -- from the quark-gluon 
plasma (QGP) phase to the hadron-gas (HG) phase~\cite{shuryak}. Unlike 
strongly interacting probes, such as identified particles and jets, 
direct photons traverse the medium unmodified due to the small cross 
section of electromagnetic interaction. These penetrating photons encode 
information about the environment in which they were created, including 
the temperature and the collective motion of the medium. While the 
direct photons at high transverse momentum, \pt, are dominated by 
photons created from hard-scattering processes, such as quark-gluon 
Compton scattering, in the low-\pt regime, they were initially predicted 
to be of a thermal origin, being emitted from the QGP and HG phase (see 
Ref.~\cite{David:2019wpt} for a recent review).

The \pt spectrum of low-\pt direct photons from \auau collisions at 
\snn{200}, first measured by PHENIX~\cite{PHENIX:2008uif}, shows a clear 
excess above the hard-scattering contribution estimated from \pp 
measurements for \pt below 3~\gevc. Followup measurements by PHENIX have 
established that low-\pt direct-photon emission also shows a large 
anisotropy with respect to the reaction 
plane~\cite{PHENIX:2011oxq,PHENIX:2015igl}, and that the yield increases 
faster than \Npart or \dNch as a function of the centrality of the 
collision~\cite{PHENIX:2014nkk}. Low-\pt direct photons in \auau 
collisions at 200 \gev have also been measured by 
STAR~\cite{STAR:2016use} using the same basic method 
as~\cite{PHENIX:2008uif}, but different detection techniques, which leads to different systematic uncertainties between STAR and PHENIX measurements.
Quantitatively, STAR results appear to be a factor 3 smaller than those 
from PHENIX. This tension has not yet been resolved. Furthermore, low 
\pt photons have been measured in \auau at lower \sqsn of 39 GeV and 
62.9 GeV by PHENIX~\cite{PHENIX:2018for}, and in \pbpb at \snn{2760} by 
ALICE~\cite{ALICE:2015xmh}.

The excess of direct photons in \AB{A}{A} collisions, in the low-\pt 
regime, is usually interpreted as the contribution of thermal radiation 
emitted from the expanding and cooling QGP and HG phase. Due to the 
rapid anisotropic expansion of the system, the radiation is Doppler 
shifted. Over the years, several theoretical models have been developed 
and refined to describe the production rates and space-time evolution of 
thermal photons in relativistic heavy-ion 
collisions~\cite{Gale:2021emg,Linnyk:2015rco,Paquet:2015lta,McLerran:2014hza,Shen:2013vja,vanHees:2011vb,Dusling:2009ej,Dion:2011pp}. 
While most of these state-of-the-art models describe the data 
qualitatively, they fall short of simultaneously describing all the 
features of the data quantitatively. To describe the large yield, early 
emission at high temperatures is favored, while sufficient build up of 
collective motion is required to explain the large anisotropy, thereby 
favoring late-stage emission. This tension, often termed as the 
``direct-photon puzzle'', hints at an incomplete understanding of the 
different sources and mechanisms of direct-photon production. This has 
triggered more thoughts on other unconventional photon sources, such as 
emission from the pre-equilibrium stage, strong magnetic field effects, 
etc.~\cite{vanHees:2014ida,Berges:2017eom,Heffernan:2014mla,Linnyk:2015tha, 
Basar:2012bp, Basar:2014swa, Muller:2013ila, Gale:2021emg}.  For that 
very reason this paper refers to the low-\pt-excess direct photons as 
``nonprompt'' instead of ``thermal''.

To provide new insights and further understandings, the PHENIX 
collaboration presents results from the high-statistics 2014 Au$+$Au data 
at \snn{200}. With a 10-fold increase in statistics compared to previously 
published results, differential direct-photon measurements as functions of 
\pt and system size over a broad \pt range from 0.8--10~\gevc and in 10\% 
centrality classes are discussed. A new algorithm, which utilizes the 
silicon-vertex detector (VTX) as the conversion material for photons, is 
developed for this analysis.

The paper is organized as follows: Section~\ref{sec:exp} presents the 
experimental setup relevant to this measurement and the algorithm to 
reconstruct the conversion photons.  Section~\ref{sec:measurement} 
describes the double-ratio method to determine the direct-photon excess 
ratio, \Rg, and gives details of the experimental measurement. 
Section~\ref{sec:sys} investigates the systematic uncertainties. 
Section~\ref{sec:result} discusses the results.  
Section~\ref{sec:summary} presents the summary and conclusions.
Finally, there are two appendices: \append{sec:appendix-A} discusses the event
mixing procedures and their validity, while \append{sec:appendix-B}
describes the Monte-Carlo (MC)-sampling method used to derive the 
final systematic uncertainties on the direct-photon yield.

 
\section{\label{sec:exp}Experimental setup and photon measurements}

\subsection{PHENIX 2014 \auau \snn{200} data set}

In 2014, a total of 19 billion \auau collisions at \snn{200} 
were recorded by the PHENIX detector at the Relativistic Heavy Ion
Collider (RHIC) with a minimum-bias (MB) trigger, based on the 
response of two beam-beam counters (BBC)~\cite{PHENIX:BBC}. The BBCs are 
located on either side of the interaction point along the beam axis at 
$z={\pm}1.44$~m with a pseudorapidity coverage of $3.1<|\eta|<3.9$ and 
full $2\pi$ azimuthal acceptance. The MB trigger requires a coincident 
signal in both BBCs.  Each BBC, comprising 64 \v{C}erenkov counters, 
measures the total number of charged particles produced during the 
collision within its acceptance. The charged-particle multiplicity is 
used to divide the MB events into different centrality classes; 
0\%--10\% corresponds to the most central collisions which produces the 
largest number of charged particles, while 80\%--93\% corresponds to 
peripheral collisions with only a small number of charged particles. 
The BBCs also utilize the arrival time of the produced particles 
on each side to determine the collision vertex along the beam direction.

\begin{figure}[ht!]
     \includegraphics[width=1.0\linewidth]{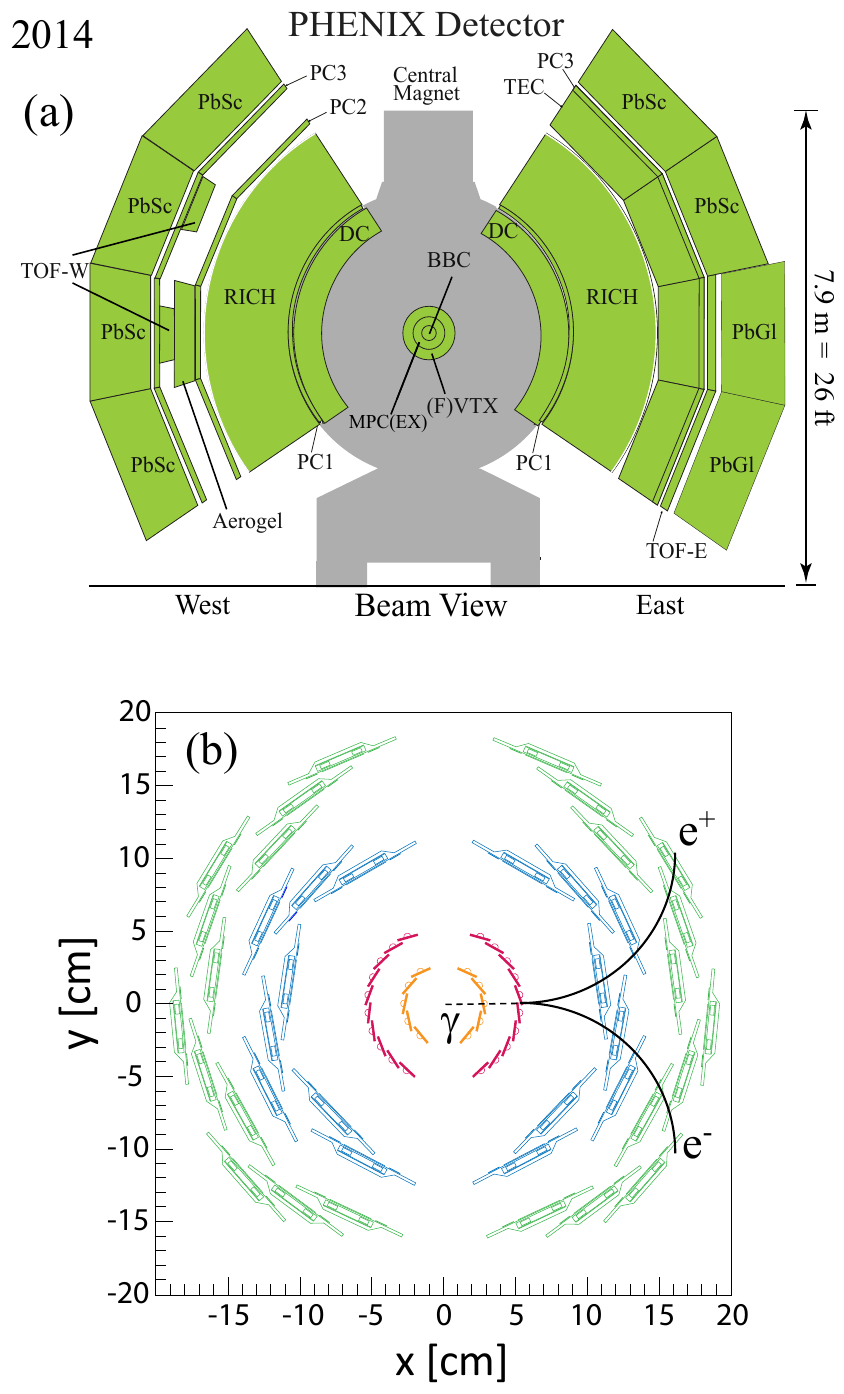} 
\caption{(a) The beam view of the PHENIX central-arm spectrometer for 
the year 2014. (b) A magnified view of the silicon-vertex detector. The 
solid curves correspond to the electron and positron tracks from photon 
conversion.}
   \label{fig:photon_reco}
\end{figure}

The direct-photon measurement, presented here, is based on the tracking 
and identification of electrons and positrons from photon conversions in 
the detector material and the direct calorimetric measurement of photons 
in the two PHENIX central arm spectrometers shown in 
\fig{fig:photon_reco}~\cite{PHENIX:overview}.
The VTX~\cite{PHENIX:VTX} comprises four 
silicon layers at nominal radii of 2.6, 5.1, 11.8, and 16.7~cm. In the 
beam direction, the active area covers approximately $\pm 11$ cm for the 
innermost layer and $\pm19$ cm for the outer layer. The VTX is not used 
as an active detector in the measurement. However, it acts as the photon 
converter, which is critical for this analysis. The total material 
thickness of the VTX in terms of radiation length, $X_0$, is 
${\approx}13\% X_0$. Events are selected with a $z$ vertex within 
$\pm10$ cm of the nominal interaction point. After applying quality 
assurance criteria, a total of $1.25\times10^{10}$ events are analyzed.

The central-arm spectrometers have three major parts: A 
charged-particle tracking system~\cite{PHENIX:tracking,PHENIX:2002wmm}, 
particle-identification detectors~\cite{PHENIX:PID}, and electromagnetic 
calorimeters (EMCal)~\cite{PHENIX:EMCal}. Each arm covers 90$^{\circ}$ 
in the azimuthal direction with $|\eta|<0.35$. The tracking system is 
located $\approx$2.2~m from the beam axis outside of an axial magnetic 
field. The main tracking detectors are drift chambers (DC) and pad 
chambers (PC1). The DC provides a precise measurement of the transverse 
momentum for charged particles with $\pt>0.2$~\gevc. The PC1 measures 
the momentum along beam direction, $p_{z}$. The effective momentum 
resolution of the central-arm tracking system, for this analysis, is 
$\sigma_p/p=0.8\%{\oplus}2\%\,p$~[GeV/$c$], where $p$ is 
the transverse momentum of the track.

Charged tracks are identified as electrons or positrons with a 
ring-imaging \v{C}erenkov detector (RICH). The RICH has a $CO_2$ gas 
radiator with a low radiation threshold for electrons (0.018~\gevc) and 
a relatively high threshold for charged pions ($>4.87$~\gevc). Requiring 
a signal in at least two phototubes in the focal plane of the RICH at 
the expected ring location effectively separates electrons below 5 
GeV/$c$ from charged hadrons. A further matching of the momentum, $p$, 
of the charged track to the energy, $E$, as measured in the EMCal within 
$-2\sigma_{E/p}<E/p<5{\sigma_{E/p}}$ removes most hadrons remaining 
in the sample. Here $\sigma_{E/p}$ is the momentum-dependent 
resolution of the energy to momentum ratio, $E/p$.

For the calorimetric identification of photons, two types of calorimeters 
are used, lead-scintillator (PbSc) and lead-glass (PbGl). The PbSc EMCal, 
which covers 3/4 of the acceptance, is a sandwich sampling detector, also 
referred to as a Shashlik type calorimeter. Based on the widths of 
reconstructed \piz mass through the $\piz \rightarrow \gamma\gamma$ decay, 
the effective photon-energy resolution in this analysis is 
$\sigma_E/E=8.1\%/\sqrt{E~{\rm [GeV]}}{\oplus}5.0\%$. The remaining 1/4 of 
the acceptance is covered by the PbGl EMCal, which is a homogeneous 
\v{C}erenkov-type detector with an effective resolution of 
$\sigma_E/E=8.7\%/\sqrt{E~{\rm\,[GeV]}}{\oplus}5.8\%$. Nominal cuts on the 
energy threshold ($E>500$~MeV) and shower shape ($\chi^2<3$) are applied 
to identify photons.

\subsection{\label{sec:external_conversions}External photon conversions 
in the VTX}

Earlier measurements of direct photons from PHENIX are based on three 
different strategies to measure photons in \AB{A}{A} collisions. The 
calorimeter method is used to measure photons with \pt of several~\gevc 
via their energy deposited in the EMCal~\cite{PHENIX:2011oxq}. To access 
lower \pt, \ee pairs from photon conversions are reconstructed with the 
tracking system. These \ee pairs are either from ``internal'' 
conversions of virtual photons emitted from the 
collision~\cite{PHENIX:2008uif} or ``external'' conversions of photons 
in the detector material~\cite{PHENIX:2014nkk}.

\begin{figure}[htb] 
        \includegraphics[width=1.0\linewidth]{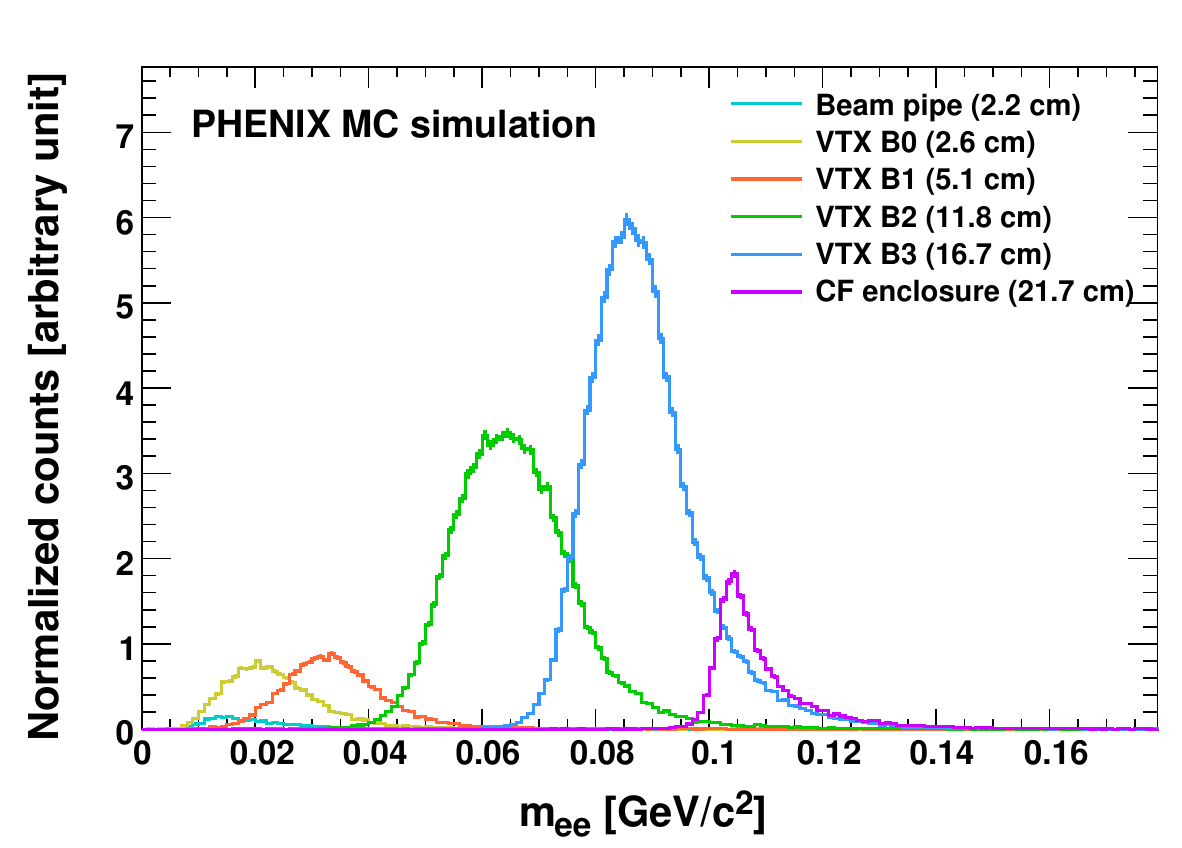} 
\caption{
Artificial \ee pair mass for external photon conversions. Each curve 
corresponds to a different radius region, which roughly maps to the 
locations of beam-pipe, layers 1 (B0) through 4 (B3) of the VTX, and the VTX 
(CF) carbon-fiber enclosure.
}
\label{fig:massPeaks}
\end{figure}

Here, external photon conversions at the VTX detector are reconstructed 
from \ee pairs. The VTX material is distributed between 2 and 25~cm 
along the radial direction. Depending on the conversion point, a 
different amount of magnetic field is traversed by the \ee pair. In the 
standard PHENIX track-reconstruction algorithm, the tracking system 
measures a part of the trajectory outside of the magnetic field at a 
radial position of $\approx$2.2~m. The momentum vector is determined by 
assuming that the particle originates at the event vertex. This 
assumption is incorrect for the \ee pairs from conversions in the VTX 
material. Both $e^+$ and $e^-$ traverse a smaller \bdl than tracks from 
the vertex and thus the azimuthal component of the momentum vector is 
mismeasured in opposing directions, leading to an artificial opening 
angle and mismeasured mass of the \ee pair.  Because the magnetic field in 
the region of the VTX detector is approximately constant at 0.9~Tesla, 
the artificial mass acquired is proportional to the radial location of 
the conversion point. \fig{fig:massPeaks} shows the mass of \ee pairs 
simulated with the {\sc geant3} PHENIX-detector 
simulation~\cite{Brun:1987ma}, different curves represent photon 
conversions in different VTX layers.  The \mee is larger 
for conversions at larger radii with most conversions occurring in the 
third and fourth layers of the VTX, where the material budget is the 
largest.

To correctly reconstruct and identify photon conversions at different 
VTX layers, a new track-reconstruction algorithm is developed. The new 
algorithm relies on the fact that the $e^+$ and $e^-$ from a conversion 
have the same origin and that their momenta were initially parallel in 
radial direction. This additional constraint eliminates the need to 
assume the origin of the track.

\begin{figure}[!htb] 
         \includegraphics[width=0.96\linewidth]{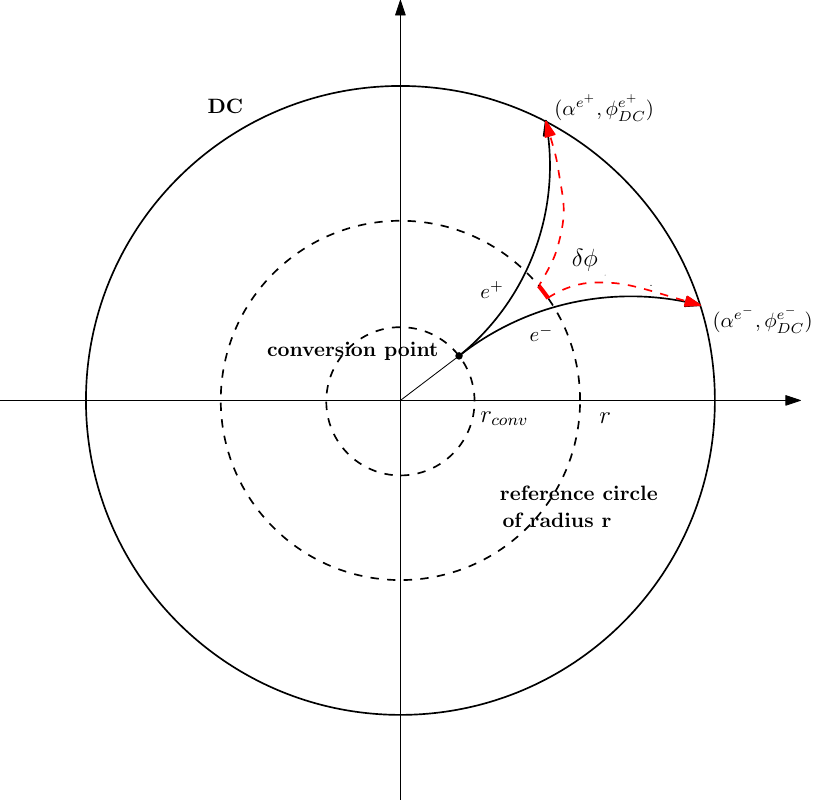} 
        \caption{
Schematic view of the conversion-reconstruction algorithm. The two 
tracks are reconstructed to the same radius r. $\delta\phi$ is the 
azimuthal-angular difference between the two tracks for a given 
reconstruction radius. $\delta\phi$ is zero at the conversion point.
        }
        \label{fig:2D}
\end{figure}

The algorithm is illustrated in~\fig{fig:2D}. For all radii between 0 
and 30~cm, all possible momenta of the $e^+$ and $e^-$ are scanned to 
identify the azimuthal location $\phi_\pm$ at which the track is 
perpendicular to the circle of the given radius, or in other words 
points back radially to the event vertex. The conversion point is 
determined by finding the radius for which the difference of the 
azimuthal angles of the \ee pair, $\delta\phi=\phi_+{ - }\phi_-$, becomes 
zero. If such radius exists, the pair is identified as a conversion 
candidate at the location $(\phi_{\rm conv},r_{\rm conv})$, where $\phi_{\rm conv}$ 
is the azimuthal angle of the conversion point, reconstructed with a 
resolution of $\approx$4~mrad, and $r_{\rm conv}$ is the radial position 
reconstructed with a resolution of $\approx$2~cm.
 
 
\section{\label{sec:measurement}Data Analysis}      
 
\subsection{Double-ratio tagging method}


The number of direct photons emitted in a \auau collision is small 
compared to the number of photons from hadron decays. To make a precise 
measurement of the direct-photon yield, a tagging method is 
employed~\cite{PHENIX:2014nkk}, which measures the ratio, \Rg, of all 
photons, referred to as inclusive photons, \gammaincl, to the photons 
from hadron decays, \gammahadr. The ratio \Rg is evaluated as double 
ratio, such that most systematic uncertainties cancel explicitly. 
The \Rg given in \eq{eq:double_ratio_tag} features three main terms:
\begin{equation}
R_{\gamma} = \frac{\gamma^{\rm incl}}{\gamma^{\rm hadr}} = \frac{\left(\frac{\gamma^{\rm incl}}{\gamma^{\pi^{0}}}\right)}{\left(\frac{\gamma^{\rm hadr}}{\gamma^{\pi^{0}}}\right)} = \frac{\langle\epsilon_{\gamma} f\rangle\left(\frac{N^{\rm incl}_{\gamma}}{N_{\gamma}^{\pi^{0},\rm tag}}\right)_{\rm Data}}{\left(\frac{\gamma^{\rm hadr}}{\gamma^{\pi^{0}}}\right)_{\rm Sim}}~,
\label{eq:double_ratio_tag}
\end{equation}

\begin{itemize}

\item [(i)] The ratio of measured photon yields \Nincl/\Ntag is the number of 
measured conversion photons in a given \pt bin, divided by the 
sub-sample of those conversion photons that are tagged by a second 
photon as resulting from a $\piz \rightarrow \gamma\gamma$ decay. This 
quantity is measured in bins of fixed conversion photon \pt.

\item [(ii)]The conditional acceptance and efficiency \ef is the conditional 
probability to detect and reconstruct the second \piz decay photon with 
the EMCal, given that the first decay photon was reconstructed as \ee 
pair from a photon conversion. The probability is averaged over all 
parent \piz \pt that can contribute to the given conversion photon \pt.

\item [(iii)]The cocktail ratio \gammahadr/\gammapiz is the ratio of all 
photons from hadron decays over 
only those photons from \piz decays.

\end{itemize} 

\noindent The following sections discuss how each term is determined.

\subsection{Ratio of the measured photon yields~{\Nincl/\Ntag}} 

Electrons and positrons in a given event are combined to \ee pairs and 
conversion candidates are selected with appropriate cuts, which results 
in a foreground sample of \ee pair \Ninclfg. All conversion candidates 
in a conversion photon \pt~bin, are combined with all photon 
showers in the EMCal above an energy threshold, $E_{cut}$. The invariant 
mass \meeg is calculated and all combinations that lie in a mass window 
around the \piz mass are considered as candidates for tagged photons 
\Ntagfg.  Due to the large particle multiplicity in \auau collisions, 
there are many false combinations where the electron, positron or photon 
are not from the same source. These background pairs must be subtracted 
statistically to obtain the signals of interest \Ninclsg and \Ntagsg.

For \ee pairs, there are two possible combinations, signal pairs of 
interest \Ninclsg and uncorrelated background \Ninclbg pairs where the 
electron and positron are from different sources. Their sum constitutes 
the foreground \Ninclfg:
\begin{equation}
\Ninclfg = \Ninclsg + \Ninclbg.
\end{equation}\label{eq:ee_prob}

\noindent When the \ee pairs are combined with photons to \eeg combinations, 
both types of \ee pairs are combined with photons that are either correlated 
or uncorrelated with the pair:
\begin{equation}
\Ntagfg = \Ntagsg + \Ntaguncorr + \Ntagcorr.
\label{eq:eeg_prob}
\end{equation}

Introducing $i,j,k$ as the source of the positron, electron, and photon, 
respectively, the terms in \eq{eq:eeg_prob} are:

\begin{itemize}

\item[(1)] The first term is the signal of interest with positron, 
electron, and photon from the same source ($i=j=k$).

\item[(2)] The second term represents the cases where the \ee pair is 
combined with uncorrelated photons.  This includes the case ($i=j\neq 
k$), where the \ee pair is correlated and randomly combined with a 
$\gamma$ as well as the case ($i\neq j\neq k$) where all three are from 
different sources.

\item[(3)] The third term represents cases ($(i\neq j=k) \vee (j\neq i=k)$), 
where the \ee pair is not from the same source but the $\gamma$ is 
correlated with either the $e^+$ or the $e^-$.  

\end{itemize}

\noindent Each of the background terms is determined with different 
event-mixing procedures, which were developed using the MC method.  The 
event-mixing procedures and their validity are discussed in detail in 
\append{sec:appendix-A}.

\subsubsection{Determination of the inclusive photon yield {\Nincl}} 
\label{sec:analysis_Nincl}

\begin{figure}[!htb] 
      \includegraphics[width=1.0\linewidth]{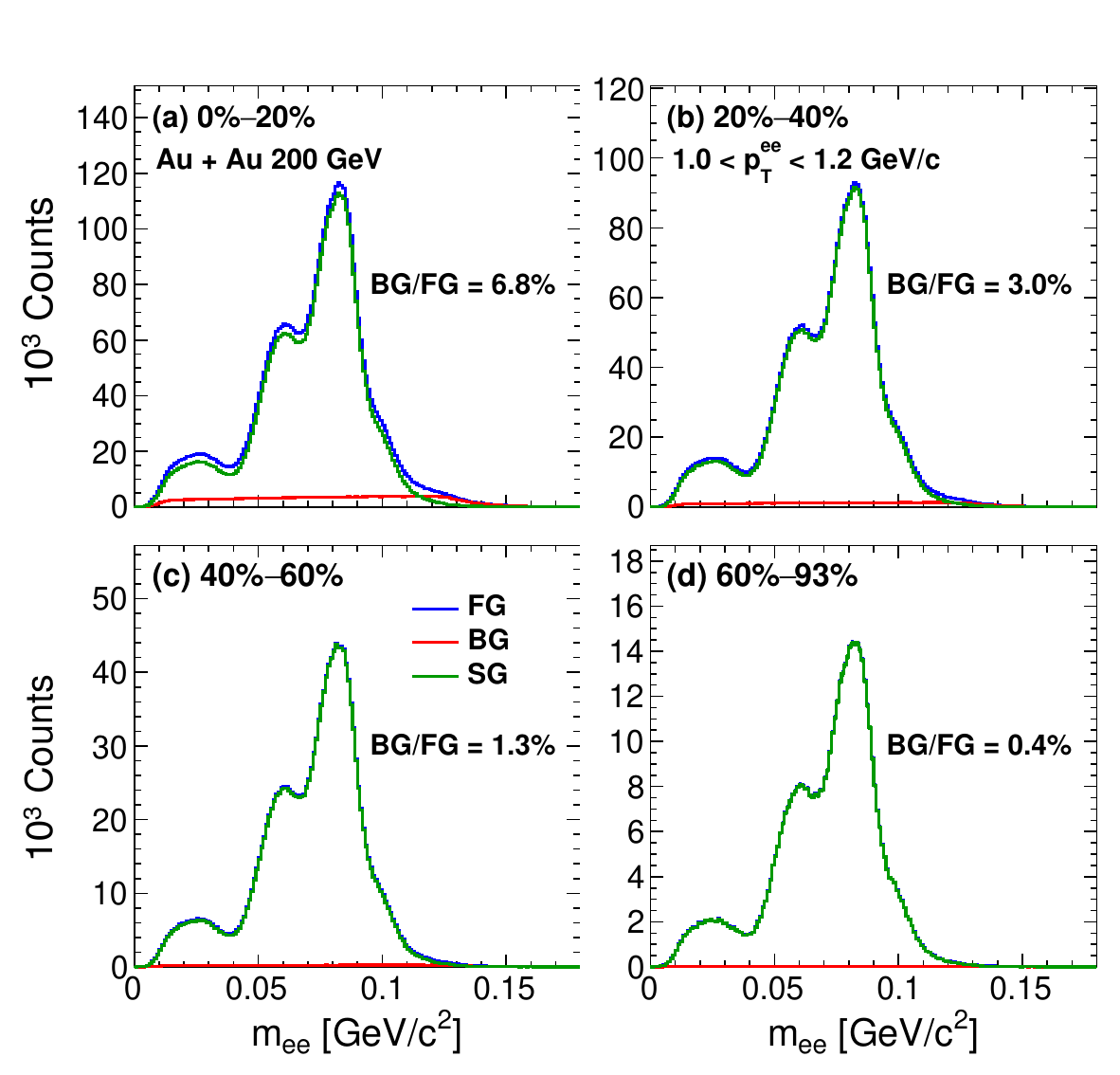} 
\caption{Mass distribution, \mee, of the \ee pairs after conversion 
selection cuts are applied. All four panels are for the same \pt range 
$1.0<\pt<1.2$~\gevc for four different centrality selections (a) 
0\%--20\%, (b) 20\%--40\%, (c) 40\%--60\% and (d) 60\%--93\%. Shown are 
the foreground \Ninclfg, background \Ninclbg and signal \Ninclsg.}

\label{fig:data_eemass}
\end{figure}

Photons that convert at the VTX detector are selected by pairing 
electron and positron tracks to \ee pairs. All pairs are required to 
have a valid conversion point at a radial location within the VTX 
detector, between 1 and 29~cm. In addition, both tracks need to match in 
the beam direction within $|\Delta z| < 4$ cm. The invariant mass 
distribution of the selected \ee conversion pairs is shown in 
\fig{fig:data_eemass} for the \pt range $1.0<\pt<1.2$~\gevc. The four 
panels correspond to four different centrality selections. Each panel 
shows the same peak structure, which is characteristic of the multilayer 
structure of the VTX detector.

The \ee pairs passing the conversion selection criteria contain 
uncorrelated \ee pairs, where the $e^+$ and $e^-$ are from different 
sources. These backgrounds are also shown in \fig{fig:data_eemass}. 
Because of its combinatorial nature, the background to foreground ratio 
increases towards more central-event selections. An event-mixing 
technique is used to estimate and subtract this background (see 
\append{sec:appendix-A} for details). In this technique, an $e^+$ from 
event A is paired with an $e^-$ from another event B to produce the 
random \ee pair sample. To assure the events A and B have similar 
topological characteristics, it is required that both events:

\begin{itemize}

\item[(a)] are from the same 10$\%$ centrality selection, 

\item[(b)] have their interaction vertex within $\Delta z = 2.5$ cm, 

\item[(c)] have their reaction planes aligned within $\Delta\phi = \pi/6$. 

\end{itemize}

After the subtraction of the uncorrelated background, more than 99\% of the 
pairs are from photon conversion in the VTX materials. The remaining pairs are 
from internal virtual photon conversions that passed the conversion selection 
criteria. The sources of these pairs are similar to those of the photon 
conversion pairs, with the majority resulting from \piz Dalitz decays. 
An additional lower mass cut at 0.04 \gevcc removes about 90\% of theses 
internal conversions, rendering the remainder negligible. 
Finally, \Nincl is calculated by integrating 
  the counts in the mass range from 0.04 to 0.12~\gevcc, corresponding to 
layers 3 and 4 of the VTX. The analysis is repeated for bins in \pt and 
in centrality.

\subsubsection{Tagged photon raw yield {\Ntag}} 
\label{sec:analysis_Ntag}

\begin{figure*}
\begin{minipage}{0.99\linewidth}
  \includegraphics[width=0.99\linewidth]{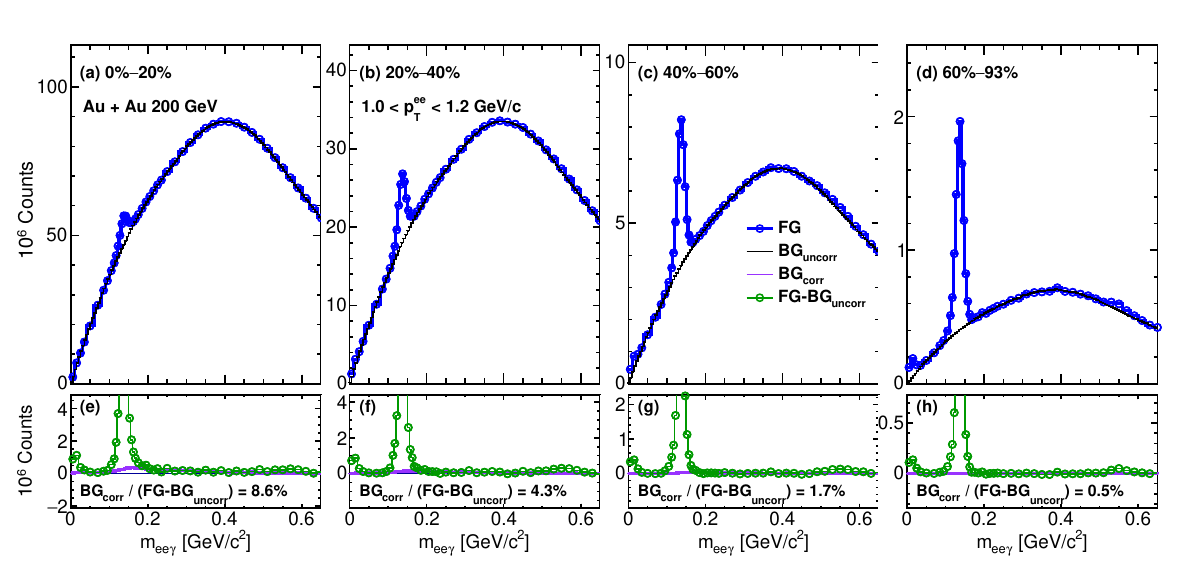} 
\caption{Mass distribution, \meeg, for \ee pairs with \pt from 1.0 to 
1.2~\gevc, for four centrality selection (a,e) 0\%--20\%, (b,f) 
20\%--40\% (c,g) 40\%--60\%, and (d,h) 60\%--93\%.  Panels (a) through 
(d) show the foreground \Ntagfg and the uncorrelated background 
\Ntaguncorr. Panels (e) through (h) show the difference 
$\Ntagfg-\Ntaguncorr$, together with the correlated background 
\Ntagcorr.}
  \label{fig:eeg_data_cent}
\end{minipage}
\begin{minipage}{0.99\linewidth}
  \includegraphics[width=0.99\linewidth]{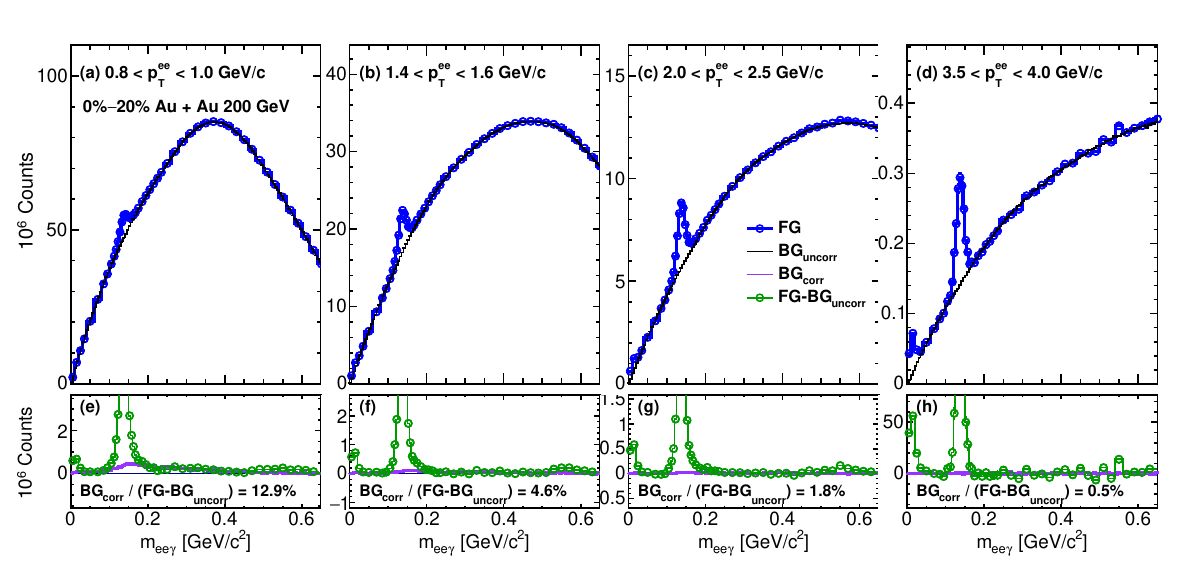} 
\caption{Mass distribution, \meeg, for \ee pairs from the 0\%--20\% 
centrality selection for four different \ee pair \pt regions, (a,e) 0.8 
to 1.0, (b,f) 1.4 to 1.6, (c,g) 2.0 to 2.5 and (d,h) 3.5 to 4~\gevc.  
Panels (a) through (d) show the foreground \Ntagfg and the uncorrelated 
background \Ntaguncorr.  Panels (e) through (h) show the difference 
$\Ntagfg-\Ntaguncorr$, together with the correlated background 
\Ntagcorr. }
  \label{fig:eeg_data_pt}
\end{minipage}
\end{figure*}

Next, the subset of \ee pairs in the \Nincl sample that can be tagged as 
photons from a \piz decay, \Ntag, is determined. For a given event, each 
\ee conversion candidate, in the mass window in which \Nincl is counted, 
is paired with all reconstructed showers in the EMCal with shower shape 
$\chi^2<3$ and energy larger than $E_{cut}=0.5$~GeV, excluding those 
matched to the \ee pair itself. The energy cut, together with the \pt cut 
of 0.2~\gevc on the $e^+$ and $e^-$, constitutes an implicit asymmetry cut 
on the \piz decay photons that depends on the \pt of the \piz. For all 
\eeg combinations, the invariant mass \meeg is calculated. This 
constitutes the foreground \Ntagfg, for which an example is given in 
\fig{fig:eeg_data_cent} for the \ee pair in the \pt range 
$1.0<\pt<1.2$~\gevc. The four panels (a) to (d) correspond to four 
centrality selections 0\%--20\%, 20\%--40\%, 40\%--60\%, and 60\%--93\%, 
respectively.

Despite the large background, the signal, \Ntagsg, is clearly visible as a 
peak around the \piz mass, even in panel (a), which is the most central 
event selection.  As discussed above, the background \Ntagbg has two 
components:
\begin{equation}
    \Ntagbg = \Ntaguncorr + \Ntagcorr,
\end{equation}
\noindent for which the shape and normalization are obtained from the 
event-mixing procedures described in \append{sec:appendix-A}. The 
results are also shown in \fig{fig:eeg_data_cent}. The uncorrelated 
background, \Ntaguncorr, is given in panels (a) to (d). The much smaller 
correlated background, \Ntagcorr, is only revealed after \Ntaguncorr is 
subtracted from the foreground, \Ntagfg. The differences are given in 
panels (e) to (h) for central to peripheral events, respectively. 
Figure~\ref{fig:eeg_data_cent} indicates that the correlated background 
decreases with centrality from $\Ntagcorr/(\Ntagfg-\Ntaguncorr)=8.6\%$ 
in central collisions to 0.5\% in the most-peripheral collisions.

For the 0\%--20\% centrality selection, \fig{fig:eeg_data_pt} shows the 
mass distributions \meeg for four different \ee pair \pt ranges. The 
representation is the same as for \fig{fig:eeg_data_cent}. Panels (a) 
through (d) all show a clear peak around the \piz mass. The backgrounds 
are the largest for low \pt and the most central events. As \pt 
increases and the event multiplicity decreases, the backgrounds are 
significantly reduced.

Because of the complexity of the particle correlations present in the 
real Au$+$Au collision events, including effects of collective 
expansion, jet production, hadron decays, etc., there is a small 
residual background that is not captured by the event-mixing procedure. 
To remove this background, a low-order polynomial, $f_{ee\gamma}$, is 
fitted to the ratio $(\Ntagfg-\Ntagbg)/\Ntaguncorr$ in the mass range 
0.05--0.08 and 0.23--0.45~\gevcc. This function is used to correct 
\Ntaguncorr before it is finally subtracted. Thus, the final 
distribution for \Ntag is:
\begin{equation}
\Ntag = \Ntagfg - \Ntagcorr - (1+f_{ee\gamma})\times\Ntaguncorr.
\end{equation}\label{eq:Ntag}

\begin{figure}[htb] 
        \includegraphics[width=1.0\linewidth]{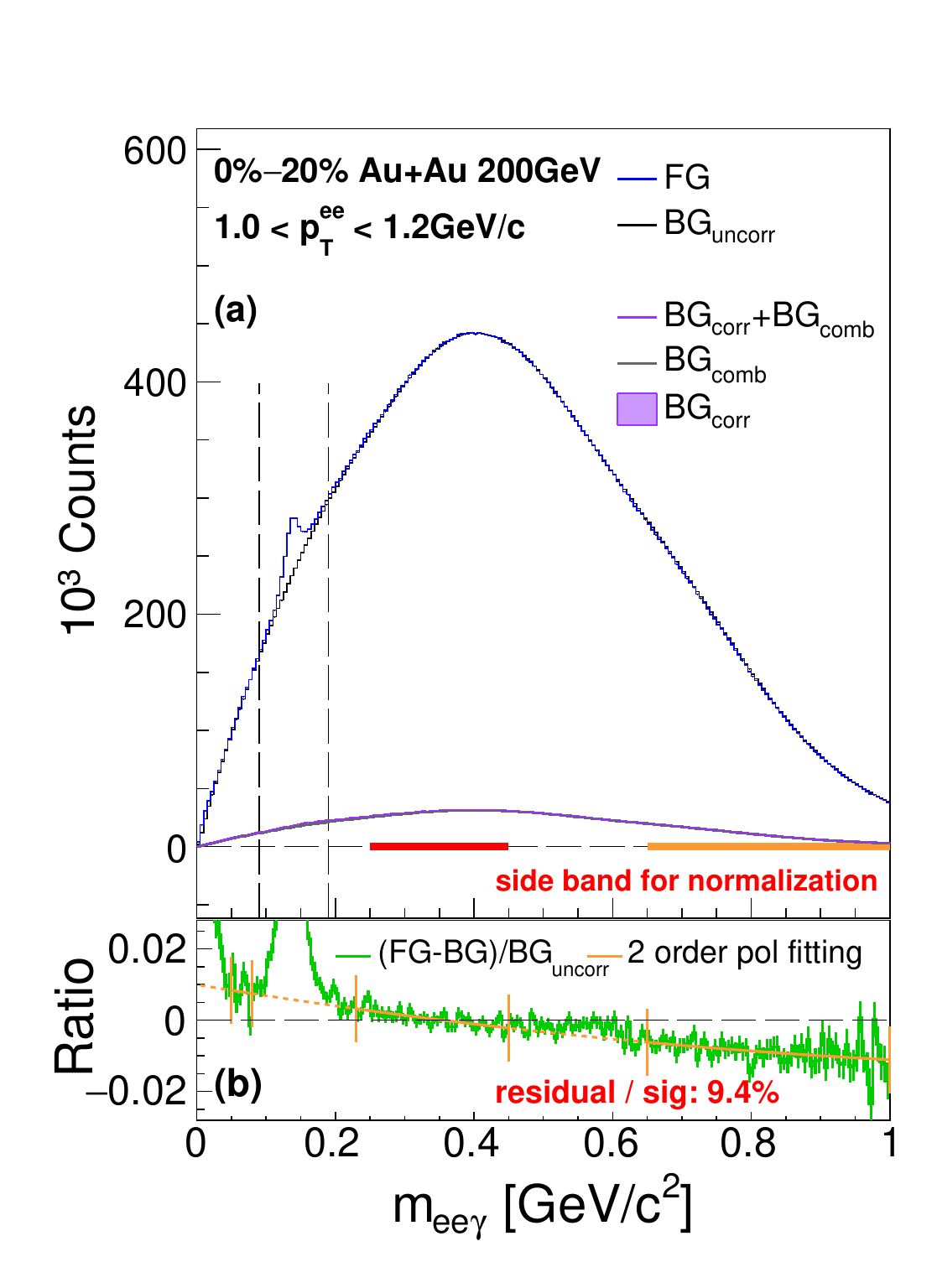} 
\caption{(a) An example for \Ntagfg and the various 
background components after normalization in the indicated regions. 
(b) The ratio $(\Ntagfg-\Ntagbg)/\Ntaguncorr$ and the polynomial 
fit to determine the residual-background correction $f_{ee\gamma}$.
}
        \label{fig:data_eegmass_dzed_lowpt}
\end{figure}

An example of the residual background is given in 
\fig{fig:data_eegmass_dzed_lowpt} for the \ee pair \pt range of 1 to 1.2 
\gevc and 0\%--20\% centrality selection. In panel (a), \Ntagfg with all 
the background components are shown. Panel (b) gives a second-order 
polynomial fit to the ratio $(\Ntagfg-\Ntagbg)/\Ntaguncorr$ ratio, 
$f_{ee\gamma}$, which is used to determine the residual background. Due 
to the unfavorably small signal-to-background ratio in this case, the 
residual background in the \piz mass region is $\approx$9.4\%. The 
residual background quickly drops with \pt and centrality bins, for 
example as \pt increases to 3~\gevc, the residual background reduces to 
2.7\%. For each \pt-centrality bin combination, \Ntag is extracted by 
integrating the number of entries in a window around the \piz peak 
($0.09 < \meeg < 0.19$)~\gevcc after all background subtractions are 
applied.

Note that the extracted \Ntag described in the this section can also be 
used to measure the $\pi^{0}$ invariant yield once corrected with 
detector acceptance and efficiency, which can potentially extend the 
previous PHENIX $\pi^{0}$ measurements ~\cite{ PHENIX:2003qdj} to lower 
$p_T$ regions. However, to establish such a measurement, in particular 
the evaluation of systematic uncertainties requires significant 
additional work that is beyond the scope of this manuscript.

\subsection{Conditional probability {\ef} } 
\label{sec:analysis_ef}

The probability, \ef, that the second photon is in the acceptance and is 
reconstructed, given a conversion \ee pair from a \piz decay, is 
extracted from the single \piz simulation. In this simulation, 
individual \piz are tracked through the PHENIX MC-simulation framework. 
The \piz are generated with $d^2N/d\pt dy$ spectra that were fitted to 
$\pi^\pm$ and $\pi^0$ data measured by PHENIX (see 
Sec.~\ref{sec:analysis_gh}), uniform in the rapidity range $|y|<0.5$, 
and uniform over $2\pi$ in azimuthal angle, $\phi$.

The energy scale and resolution of the EMCal in the MC simulation is 
tuned as closely as possible to resemble the one observed in data by 
comparing the mean and width of the measured and simulated \piz mass 
distribution. The \piz are reconstructed through the $\piz\rightarrow 
\gamma\gamma$ decay channel. For this purpose an asymmetry of less than 
20\% between the energies of the two decay photons was applied to keep 
the two-photon energies similar.

In the single \piz MC simulation, \ee pairs in the mass window $0.04 
<\mee< 0.12$~\gevcc are counted to determine \Npiz, the number of 
reconstructed \ee pairs in a given \ee pair \pt bin. The sub-sample for 
which the second photon of the \piz decay is reconstructed as a shower 
in the EMCal is counted as \Neetag. The value of \ef is then determined 
as:
\begin{equation}
 \ef = \frac{\Neetag}{{N_{ee}^{\rm \pi^0}}}.
\label{eq:ef}
\end{equation}

For the extraction of \Neetag the presence of other showers in the EMCal 
needs to be taken into account. This is done by embedding the showers 
from the simulated single \piz into the EMCal response from \auau 
collisions at the tower level. The combined EMCal information is then 
reclustered to form new showers.  All of the showers that contain energy 
deposited by the embedded singe \piz (identified by the MC ancestry 
information) are combined with the \ee pair.

\begin{figure}[!htb] 
        \includegraphics[width=1.0\linewidth]{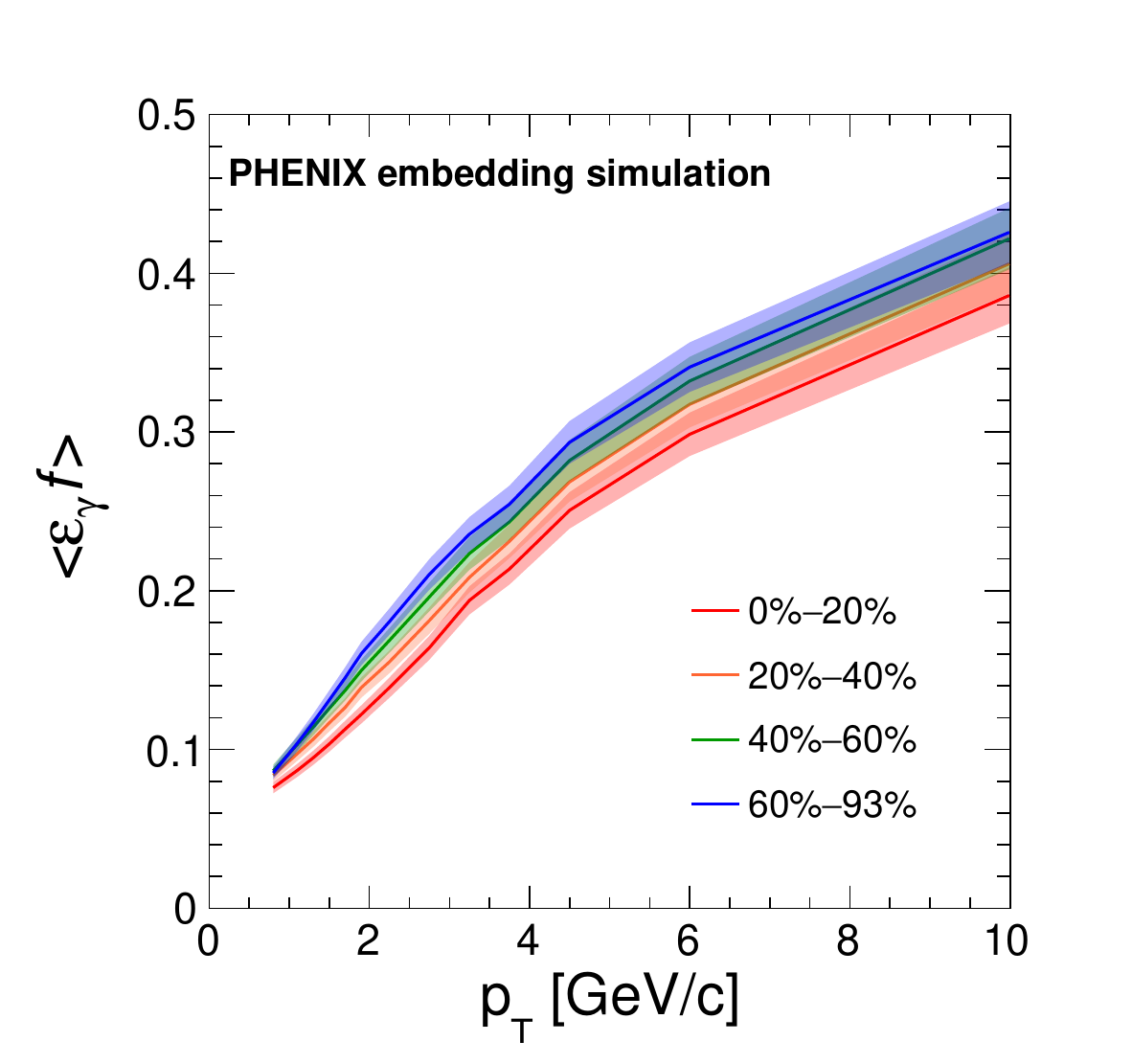} 
\caption{Conditional probability \ef as a function of $p_T$ in 
0\%--20\%, 20\%--40\%, 40\%--60\% and 60\%--93\% centrality classes.
}
        \label{fig:ef_vs_pt}
\end{figure} 

Similar to the \Ntag extraction from data, a residual background 
subtraction is applied. This eliminates any remaining background inside 
the \piz counting window. The residual background is estimated by a 
second order polynomial function fit in the mass range 0.05--0.08 and 
0.23--0.45~\gevcc. This residual background mainly comes from events 
where both decay photon convert to $e^{+}e^{-}$ pairs, and the 
reconstructed conversion photon gets paired with the EMCal cluster of 
the $e^+$ or $e^-$ from the other conversion. The extracted \ef is shown 
in \fig{fig:ef_vs_pt} as a function of the \ee pair \pt for the four 
centrality selections.

The increasing trend of \ef with increasing conversion photon \pt is 
partly due to the decrease in the opening angle between the conversion 
photon and the second photon so that the second photon is more likely to 
fall into the acceptance of the EMCal. Another important factor is that 
the average energy of the second photon increases with increasing 
conversion photon \pt, and hence, the efficiency of the energy threshold 
cut increases towards higher \pt. The difference in \ef between 
different centrality classes is mainly related to the shower shape 
($\chi^2$) selection, because the showers are more distorted in central 
Au$+$Au collisions due to the larger detector occupancy, resulting 
in more accidental overlaps from the underlying event, and the 
centrality dependent parent \piz \pt distributions.

\begin{table}[htb]  
\caption{Parameters for the modified Hagedorn function 
Eq.~\ref{eqn:hagedorn} to PHENIX data~\cite{PHENIX:2003iij, 
PHENIX:2003qdj, PHENIX:2008saf} from \auau collisions at \snn{200}. } 
{\label{tab:hagedorn}}
\begin{ruledtabular} \begin{tabular}{cccccc} 
 centrality   & $A$                & $a$                & $b$                & $p_0$   & $n$  \\
              & c(GeV/$c$)$^{-2}$ & (GeV/$c$)$^{-1}$ & (GeV/$c$)$^{-2}$ & GeV/$c$ &    \\
\hline
    min.bias  & 504.5            & 0.5169           & 0.1626           & 0.7366  & 8.274  \\
    0\%--10\%    & 1331.0           & 0.5654           & 0.1945           & 0.7429  & 8.361  \\
    10\%--20\%   & 1001.0           & 0.5260           & 0.1628           & 0.7511  & 8.348  \\
    20\%--30\%   & 750.7            & 0.4900           & 0.1506           & 0.7478  & 8.229  \\
    30\%--40\%   & 535.3            & 0.4534           & 0.1325           & 0.7525  & 8.333  \\
    40\%--50\%   & 364.5            & 0.4333           & 0.1221           & 0.7385  & 8.261  \\
    50\%--60\%   & 231.2            & 0.4220           & 0.1027           & 0.7258  & 8.220  \\
    60\%--70\%   & 118.1            & 0.4416           & 0.0559           & 0.7230  & 8.163  \\
    70\%--80\%   & 69.2             & 0.2850           & 0.0347           & 0.7787  & 8.532  \\
    80\%--93\%   & 51.1             & 0.2470           & 0.0619           & 0.7101  & 8.453  \\
\end{tabular} \end{ruledtabular}
\end{table} 

\subsection{Cocktail ratio {\gammahadr/\gammapiz} } 
\label{sec:analysis_gh}


The last ingredient to calculate \Rg~ is the cocktail ratio 
\gammahadr/\gammapiz of photons from \piz, $\eta$, $\omega$, and $\eta'$ 
decays over those from \piz decays. The cocktail ratio is obtained using 
the PHENIX meson decay generator EXODUS, which simulates mesons 
according to given input \pt spectra, decays them based on the known 
decay kinematics and branching ratios, and aggregates the decay photons 
in the PHENIX detector acceptance.

The photons from \piz decays are generated from distributions obtained 
by fitting a modified Hagedorn function (\eq{eqn:hagedorn}) to charged 
pion~\cite{PHENIX:2003iij} and neutral pion~\cite{PHENIX:2003qdj, 
PHENIX:2008saf} data measured by PHENIX 
for the rapidity range $|y|<0.5$.

\begin{equation}    
 E\frac{d^3N_\pi^0}{dp^3}  
 = A \ \Big( e^{-(a p_T +b p_T^2)} + \frac{\pt}{p_0}\Big)^{-n}.  \label{eqn:hagedorn} 
\end{equation}

\noindent The fit parameters are summarized in Table~\ref{tab:hagedorn} for MB 
collisions, as well as for nine centrality bins. The $\eta$ meson \pt 
spectra are obtained by multiplying the \piz spectrum with the 
$\eta/\piz$ ratio, following the procedure suggested 
in~\cite{Ren:2021xbh}. 

\begin{equation}    
 E\frac{d^3N_\eta} {dp^3} 
 = E\frac{d^3N_\pi^0}{dp^3}  \times \eta/\pi^0 \times R_{\rm flow},
\label{eqn:etapi} 
\end{equation}

\noindent where $R_{\rm flow}$ is the ratio of $K^{\pm}/\pi^{\pm}$ for a given 
centrality over $K^{\pm}/\pi^{\pm}$ in \pp collisions. This procedure 
makes use of the world data for $\eta/\piz$ from \pp and small system 
collisions (see \cite{Ren:2021xbh} for references), and it avoids the 
assumption of $m_T$ scaling used in earlier work~\cite{PHENIX:2014nkk}, 
which has been shown to overestimate the number of $\eta$ mesons 
produced below 2 \gevc in \pt in \pp and small system collisions. It 
also includes the centrality dependent modification, $R_{\rm flow}$, of 
the $\eta$ \pt spectra in \auau collision due to radial flow, which was 
not taken into account in earlier work~\cite{PHENIX:2014nkk}. The 
modification $R_{\rm flow}$ is estimated using measured kaon spectra 
\cite{Adare:2013esx}. For peripheral \auau collisions, the new approach 
to determine the $\eta$ yield results in a few percent reduction of the 
number of predicted decay photons in the range 1-2 \gevc, compared to 
the $m_T$ scaling approach based on Eq.~\ref{eqn:hagedorn} that was 
taken in earlier work \cite{PHENIX:2014nkk}. The difference is within 
the systematic uncertainties quoted in that work. For central and 
semicentral collisions the new and old approach agree better in the 
sense that they predict very similar decay photon yields above 1 \gevc, 
with any differences being much smaller than the quoted systematic 
uncertainties. This agreement arises when accounting for the 
modification of the $\eta$ meson spectrum due to radial flow, which 
shifts $\eta$ mesons from low to mid \pt. This shift results in more 
decay photons above 1 \gevc in the presence of radial flow, and 
moving the predicted yield closer to the one derived 
from $m_T$ scaling. At high \pt, the $\eta/\piz$ ratio demonstrates a 
universal value at high \pt, consistent with 0.487$\pm$0.024, 
independent of collision energy, system size, or centrality. The values 
for $dN/dy$ for $\eta/$\piz, $K^{\pm}/\pi^{\pm}$ and $R_{\rm flow}$ are 
summarized in Table~\ref{tab:rflow} for $1.0 < p_T < 2.0$ \gevc, where 
the effects of flow are expected to be the largest for different 
centralities for \auau collisions at 200 GeV.

\begin{figure}[!htb] 
        \includegraphics[width=1.0\linewidth]{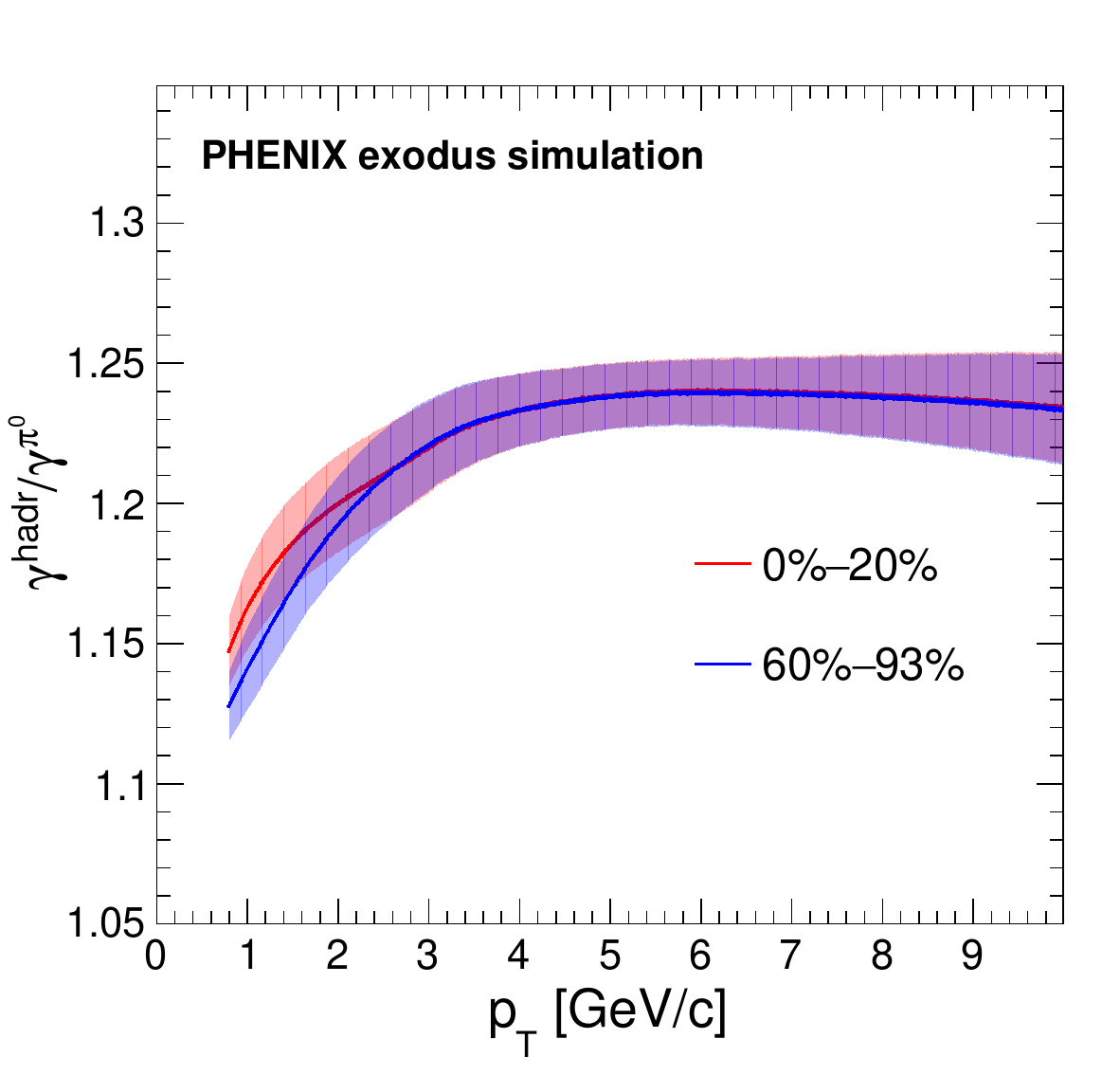} 
\caption{Cocktail ratio as a function of $p_T$ in the most central 
(0\%--20\%) and the most peripheral (60\%--93\%) centrality classes.}
        \label{fig:cocktail_vs_pt}
\end{figure}

The contribution from $\omega$ and $\eta'$ decays are based on \pt 
distributions using the \piz spectrum and replacing by 
$f$($\sqrt{p_{T}^{2}+m^2_{\rm meson}-m^2_{\pi^0}}$). The normalization 
of $\omega$ and $\eta'$ are fixed at \pt= 5~\gevc to $0.9\pm0.06$ and 
$0.25\pm0.075$, respectively~\cite{PHENIX:2014nkk}. The cocktail ratio 
\gammahadr/\gammapiz is shown in Fig.~\ref{fig:cocktail_vs_pt}.

\begin{table}[htb]  
\caption{$dN/dy$ for $\eta/$\piz, $K^{\pm}/\pi^{\pm}$ and $R_{\rm flow}$ 
for $1 < p_T < 2$ \gevc for \auau collisions at \snn{200}. There is an 
overall scale uncertainty of 0.03 on $R_{\rm flow} \times 
(\eta/\pi^{0})_{{\rm universal}}$.} {\label{tab:rflow}}
\begin{ruledtabular} \begin{tabular}{cccc} 
 centrality  & $K^{\pm}/\pi^{\pm}$  & $R_{\rm flow}$ & $R_{\rm flow} \times (\eta/\pi^{0})_{{\rm universal}}$  \\
\hline
    0\%--20\%    & 0.411 $\pm$ 0.003   & 1.20 $\pm$ 0.02  & 0.250 $\pm$ 0.004 \\
    20\%--40\%   & 0.396 $\pm$ 0.002   & 1.15 $\pm$ 0.02  & 0.237 $\pm$ 0.004 \\
    40\%--60\%  & 0.371 $\pm$ 0.002   & 1.08 $\pm$ 0.02   & 0.220 $\pm$ 0.004 \\
    60\%--93\%  & 0.337 $\pm$ 0.002   & 0.98 $\pm$ 0.02   & 0.199 $\pm$ 0.004 \\
    
\end{tabular} \end{ruledtabular}
\end{table}

\section{\label{sec:sys}Systematic uncertainties}      
 
This section describes the sources of systematic uncertainties for each 
of the three components for the calculation of \Rg. The systematic 
uncertainties are categorized into three types according to the 
correlation between the measured data points:
\begin{itemize}

\item Type A: No (or unknown) correlation between data points -- 
uncertainties on the individual data points can fluctuate independently, 
in the same way as the statistical uncertainties. 

\item Type B: The uncertainties are correlated between data points -- 
the fluctuation of each data point can be determined by the fluctuation 
of the neighboring points.

\item Type C: A special form of type B uncertainty -- every data point 
fluctuates with the exact same fraction.

\end{itemize}

\noindent In the final results, type A systematic uncertainties are 
combined with the statistical uncertainties and type B and C are 
combined to obtain the total systematic uncertainty.

\begin{figure}[!htb] 
\begin{minipage}{0.99\linewidth}
         \includegraphics[width=1.0\linewidth]{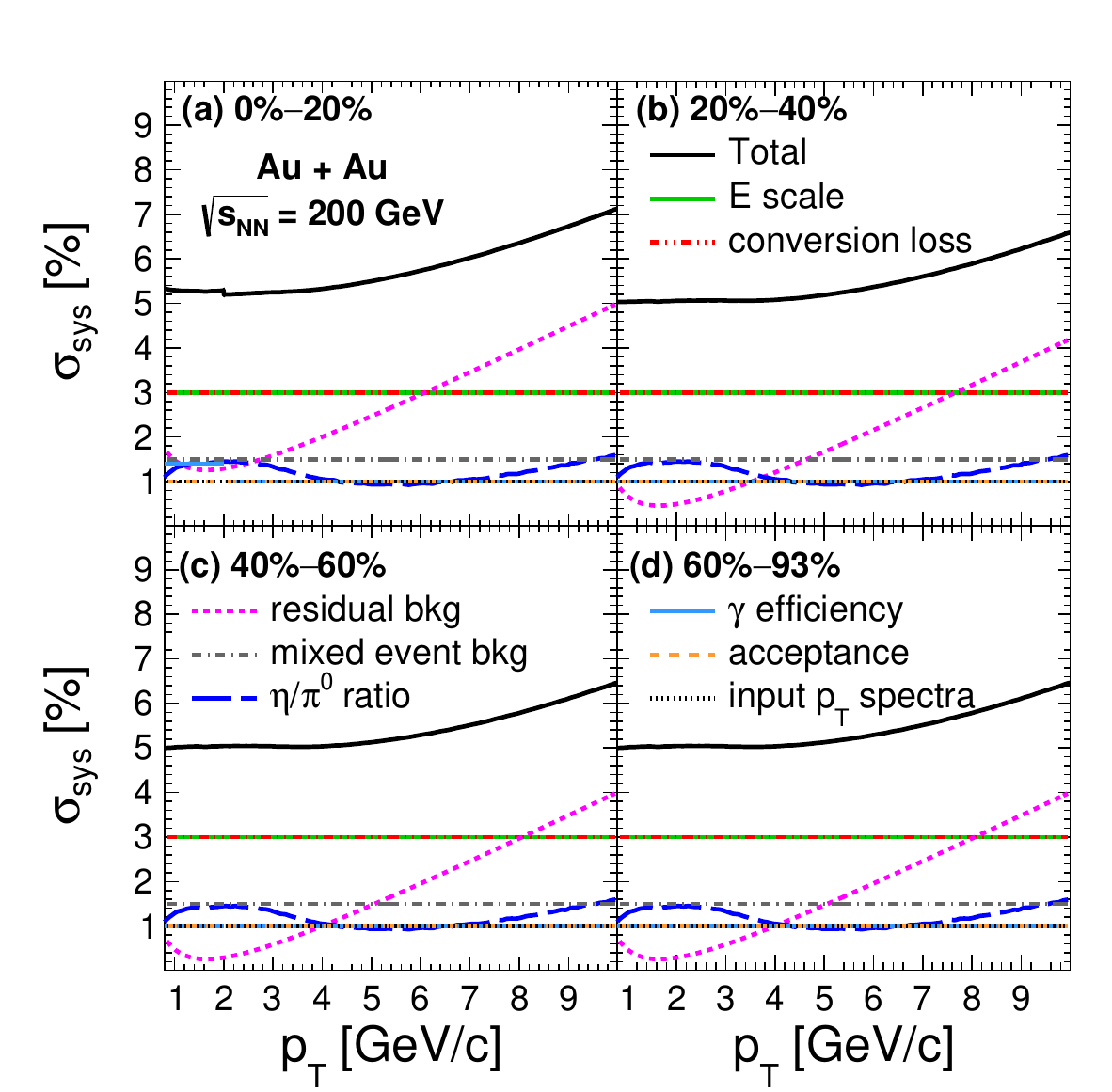} 
\caption{Systematic uncertainties of \Rg as a function of conversion 
photon \pt in 0\%--20\%, 20\%--40\%, 40\%--60\% and 60\%--93\% 
centrality bins. The black curve corresponds to total uncertainty, and 
colored curves correspond to individual sources. The lines representing 
uncertainties from the energy scale and the conversion loss overlap at 
3\%, so do the lines representing uncertainties from the $\gamma$ 
reconstruction efficiency, acceptance and input \pt spectra.}
        \label{fig:Rg_sys}
\end{minipage}
\begin{minipage}{0.99\linewidth}
        \includegraphics[width=1.0\linewidth]{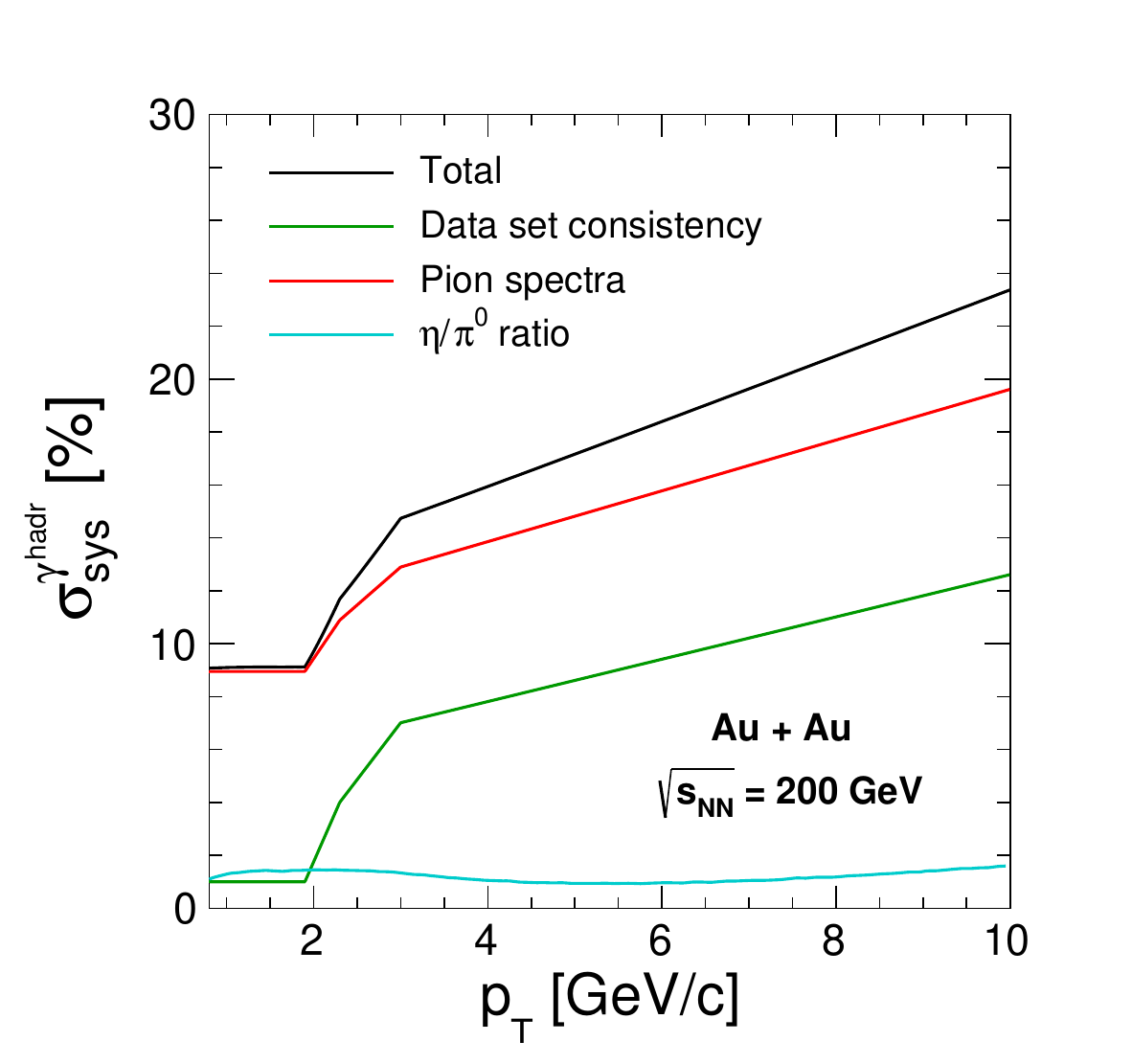} 
\caption{Systematic uncertainties of \gammahadr as a function of photon \pt.}
        \label{fig:decay_sys}
\end{minipage}
\end{figure}

The following subsections discuss the major individual sources 
contributing to the systematic uncertainties on \Rg and on the 
direct-photon yield. All contributions are summarized in 
\tab{tab:syserr} and depicted in \fig{fig:Rg_sys} and 
\fig{fig:decay_sys} as functions of \pt for \Rg and \gammadir. The 
final systematic uncertainties on \gammadir and on all quantities 
derived from \gammadir are determined using the error-sampling method 
discussed in \append{sec:appendix-B}

\begin{table*}[t]
\caption{Summary of systematic uncertainties for \Rg and \gammadir. 
Uncertainties for which ranges are given vary with \pt. 
For details see Figs.~\protect\ref{fig:Rg_sys} 
and \protect\ref{fig:decay_sys}.}
\begin{ruledtabular} \begin{tabular}{ccccccccc}
Observable & Factor & Source & correlation & correlation   & 0\%--20\% & 20\%--40\% & 40\%--60\% & 60\%--93\% \\
           &        &        & in \pt      & in centrality  \\
\hline
\Rg & \Nincl/\Ntag & \Nincl purity      & Type B  & Type B & $<1\%$  & $<1\%$    & $<1\%$    & $<1\%$ \\
    &       & \Ntag residual background & Type A  & Type A &  1.5\%--4.5\% & 0.5\%--4\%   & 0.5\%--4\%    & 0.5\%--4\% \\
    &       & \Ntag event mixing        & Type B  & Type B  & 1.5\%   & 1.5\%     & 1.5\%     & 1.5\% \\
    & \ef   & energy scale              & Type B  & Type B & $3\%$   & $3\%$     & $3\%$     & $3\%$ \\
    &       & conversion loss           & Type C  & Type C & $3\%$   & $3\%$     & $3\%$     & $3\%$ \\
    &       & $\gamma$ efficiency       & Type B  & Type A & $<1.4\%$ & $<1\%$   & $<1\%$    & $<1\%$ \\
    &       & active area \& acceptance & Type C  & Type C &  $1\%$   & $1\%$     & $1\%$     & $1\%$ \\
    &       & input \piz $p_T$ spectra  & Type B  & Type A  & $1\%$   & $1\%$     & $1\%$     & $1\%$ \\
    & $\gamma^{hadr}/\gamma^{\pi^0}$
            & $\eta/\pi^0$              & Type B  & Type C & 1--1.5\% & 1--1.5\%   & 1--1.5\%   & 1--1.5\% \\
   &        & $\omega, \eta'$           & Type B  & Type C & $<1\%$  & $<1\%$    & $<1\%$    & $<1\%$ \\ 
   & $\gamma^{hadr}$ &  input \piz \pt spectrum & Type B & Type A  & 10\%--24\% & 10\%--24\%  & 10\%--25\%   & 10\%--24\% \\
\end{tabular} \end{ruledtabular}
\label{tab:syserr}
\end{table*}

\subsection{Systematic uncertainties on {\Nincl/\Ntag}}

\subsubsection{Purity of the conversion photon sample}

Due to the high multiplicity of photons produced in \auau collisions, 
the background in the conversion sample from uncorrelated \ee pairs can 
be as large as 10\% for the most central collisions and the lowest 
\pt from 0.8 to 1.0~\gevc. This background is subtracted statistically 
with a certain accuracy. To estimate the effect on the final results, 
significantly more and less stringent conversion selection cuts were 
applied, hence, increasing or reducing the purity. The value of \ef 
\Nincl/\Ntag, obtained from the different cuts, varies by less than 1\%. 
This range is quoted as systematic uncertainty due to the limited purity 
of the conversion sample.

\subsubsection{{\piz} yield extraction}

One of the main sources of systematic uncertainty on the \Rg measurement 
is the tagged photon or \piz yield extraction. The uncertainty of \piz 
yield extraction arises from two sources: (i) from the residual 
background subtraction, which is highly correlated with the statistical 
accuracy of the mixed event background normalization, and (ii) imperfect 
description of the large backgrounds using event-mixing techniques.

To evaluate the size of the uncertainty from the residual background 
subtraction, different estimates are compared. These include using 
different functional forms for the fit and different fit ranges to 
anchor the residual background fit. In addition, the counting window for 
\piz signal extraction is varied. This gives a spread of \ef 
\Nincl/\Ntag values in each \pt and centrality bin. The standard 
deviation of the spread is quoted as the uncertainty. Due to the 
correlation with the statistical accuracy of the foreground in the 
background region, this uncertainty depends on \pt and centrality.

To test the validity of the event-mixing techniques, an MC simulation 
with high multiplicity \piz events is performed. Details are discussed 
in \append{sec:appendix-A}. The simulation shows that \Ntag/\ef can be 
determined with the event-mixing technique to better than 1.5\%.

\subsection{Systematic uncertainty on {\ef}}
\subsubsection{Energy scale}\label{sec:Escale}

The accuracy of the energy scale of the EMCal is the main source of 
systematic uncertainties in the \ef evaluation. Because of the energy 
threshold cut, the second photon is reconstructed only for $\approx$25\% 
of the \ee pairs with the lowest \pt, even though the photon was in the 
EMCal acceptance. Any potential mismatch of the energy scale between the 
simulation and real data will cause \ef to be off; a higher (lower) 
energy scale in simulation will lead to an underestimate (overestimate) 
of \ef. As mentioned earlier, to improve the accuracy, the EMCal 
response in the simulation is carefully tuned to the data using the \piz 
mass measurement in the $\piz\rightarrow\gamma\gamma$ decay channel. The 
tuning includes scaling the MC energy scale by 0.3\% and 2.2\% for the 
PbSc and PbGl calorimeters, respectively. In addition, the nonlinearity 
of the energy response is adjusted by up to 5\% at the lowest measured 
energies. After the tuning, the \piz peak positions in data and MC are 
consistent to better than 1\%. Considering the additional uncertainty 
due to the adjustment of the nonlinearity, the energy scale is known to 
better than 2\%. Changing the energy scale by $\pm2$\% introduces a 
systematic uncertainty on \ef of 3\% at low \pt and decrease towards 
high \pt. The uncertainty on the energy resolution has a negligible 
effect.

\subsubsection{Conversion photon loss} 

Another important source of systematic uncertainty on \ef is related to 
the probability that the second photon converts to an \ee pair before 
reaching the EMCal. Depending on the location of the conversion point, 
the second photon may not be properly reconstructed, thereby reducing 
\ef. To account for the ``conversion loss'', the material budget, i.e. 
thickness and location of material, implemented in the simulation 
framework must accurately reflect reality. If there is a mismatch, the 
probability for conversions to occur will be different and, hence, \ef 
will by systematically off.  As there is essentially no magnetic field 
after the DC exit window, the \ee pair from conversions between the DC 
and the EMCal will likely merge into one shower in the EMCal. Therefore, 
the value of \ef is most sensitive to differences in the material budget 
of the VTX.  Comparison of the available information about the materials 
and their thickness for all detector subsystems, reveals that the 
conversion probability in material within the magnetic field is known to 
better than 3\%, which directly translates into and uncertainty of $3\%$ 
on \Rg.

\subsubsection{Photon efficiency} 

An EMCal shower shape, $\chi^{2}$, cut is used to identify photon 
candidates among the EMCal energy clusters and to reduce the number of 
hadrons in the sample. Similar to the energy scale uncertainty, a 
difference between the shower shape in simulation and the data will 
translate directly into a systematic shift of \ef. To evaluate this 
uncertainty, the $\chi^{2}$ is varied simultaneously in both data and 
simulation and \ef \Nincl/\Ntag is recalculated. It changes by $1.4\%$ 
for 0.8--2~\gevc in the 0\%--20\% centrality bin and by less than 1$\%$ 
for all the other cases.

\subsubsection{Active area and geometric acceptance}

Due to the limited geometrical acceptance of EMCal and some inactive 
areas, the second photon is registered only for $\approx$35\% of the \ee 
pairs at the lowest \pt. Therefore, the accuracy with which the 
acceptance and dead areas are known will contribute to the systematic 
uncertainties on \ef. The uncertainty of the acceptance is the result of 
the accuracy with which the radial location of the EMCal sectors can be 
determined. The possible remaining offset leads to $<0.3\%$ difference 
in acceptance along $\phi$ direction and $<0.9\%$ in $z$ direction. The 
dead areas in the real EMCal are carefully matched to the MC simulation 
and the accuracy of the dead area determination is found to be better 
than 0.6$\%$. It is due to the cases when a tower malfunctioned only in 
a very small number of events, and not masked out in the simulation. 
Combining all these effects, the systematic uncertainty on \Rg from the 
acceptance is set to 1\%.

\subsubsection{Input {\piz} distribution}

Because \ef is averaged over all parent \piz \pt that contribute to a 
given \ee pair \pt bin, the \pt dependence of \ef is sensitive to the 
shape of the \piz distribution. The \piz parent distribution was 
determined for each centrality selection by a fit to the best available 
data from \auau collisions at \snn{200} measured by the same 
experiment~\cite{PHENIX:2003iij, PHENIX:2003qdj, PHENIX:2008saf}. The 
remaining uncertainty on \ef is smaller than 1\%.

\subsubsection{Weak decays and secondary interactions}

The tagged photon samples include decay photons from \piz from weak 
decays and \piz produced in secondary interactions. Because these \piz 
do not originate from the event vertex, \ef may be modified. Secondary 
interactions contribute less than 0.1\% of the \piz yield above \pt of 1 
\gevc and thus any distortions of \ef are negligible. Decays of $K^0_s$ 
are the predominate source of \piz from weak decays. They contribute 
between 5.8\% to 3\% of the inclusive \piz yield above 1 \gevc. With 
$c\tau = 2.68$cm, a fraction of 20\% to 25\% of those decays occur after 
the $3^{rd}$ but before the 4$^{th}$ layer of the VTX, which corresponds 
to the conversion photon sample used in this measurement. Therefore, in 
the data there are 1-2\% more conversions in the 4$^{th}$ relative to 
the 3$^{rd}$ layer compared to the MC simulation of primary \piz. The 
potential difference of \ef was estimated to be smaller than 1\%.

\subsection{Systematic uncertainty on {\gammahadr/\gammapiz}} 

The ratio \gammahadr/\gammapiz accounts for photons from hadron decays 
that occur after the kinematic freeze, other than those from \piz. The 
three largest contributors are decays of $\eta$, $\omega$, and $\eta'$ 
mesons, which contribute $\approx$23\% of the decay photons at high \pt. 
All other contributions to $\gammahadr$ are negligible.  Of the 
additional decay photons more than 80\% are from the 
$\eta\rightarrow\gamma\gamma$ decay, hence the accuracy with which 
$\eta/\pi^{0}$ is known will determine the systematic uncertainties on 
\Rg from this source. The \pt and centrality dependent upper and lower 
bounds on $\eta/\pi^{0}$ for \auau collisions at \snn{200} are taken 
from~\cite{Ren:2021xbh}. Together with the much smaller uncertainty on 
the contribution from $\omega$ and $\eta'$ decays, the systematic 
uncertainty on \Rg is below 2\% for the entire \pt range.

\subsection{Systematic Uncertainties on {\gammadir}}
\label{sec:decay_sys}

Once \Rg is determined, the direct-photon yield \gammadir is calculated 
as:
\begin{equation}
\gammadir= (\Rg-1) \ \gammahadr.
\label{eq:gammadir}
\end{equation}

\noindent In addition to the uncertainties on \Rg, the uncertainty on 
\gammahadr needs to be determined. These systematic uncertainties have 
been studied in detail in~\cite{PHENIX:2014nkk}. The main sources of 
uncertainty come from the accuracy with which the \piz \pt spectrum can 
be determined. These largely cancel in \Rg, but propagate directly to 
\gammahadr. The input \piz spectrum is based on measurements of charged 
pions, and \piz from different data taking periods (see 
\sect{sec:analysis_gh}). Each data set comes with its own systematic 
uncertainties, and in addition, the differences between different 
measurements are of the order of 10\%~\cite{PHENIX:2010xji}. The latter 
is the dominant uncertainty. The uncertainty on the spectra of other 
mesons ($\eta$, $\eta'$, $\omega$) also contributes to the uncertainty 
on \gammahadr, but to a much smaller extent.

 
\section{\label{sec:result}Results}       

\subsection{Direct photon {\Rg}}

\begin{figure}[htb]
         \includegraphics[width=1.0\linewidth]{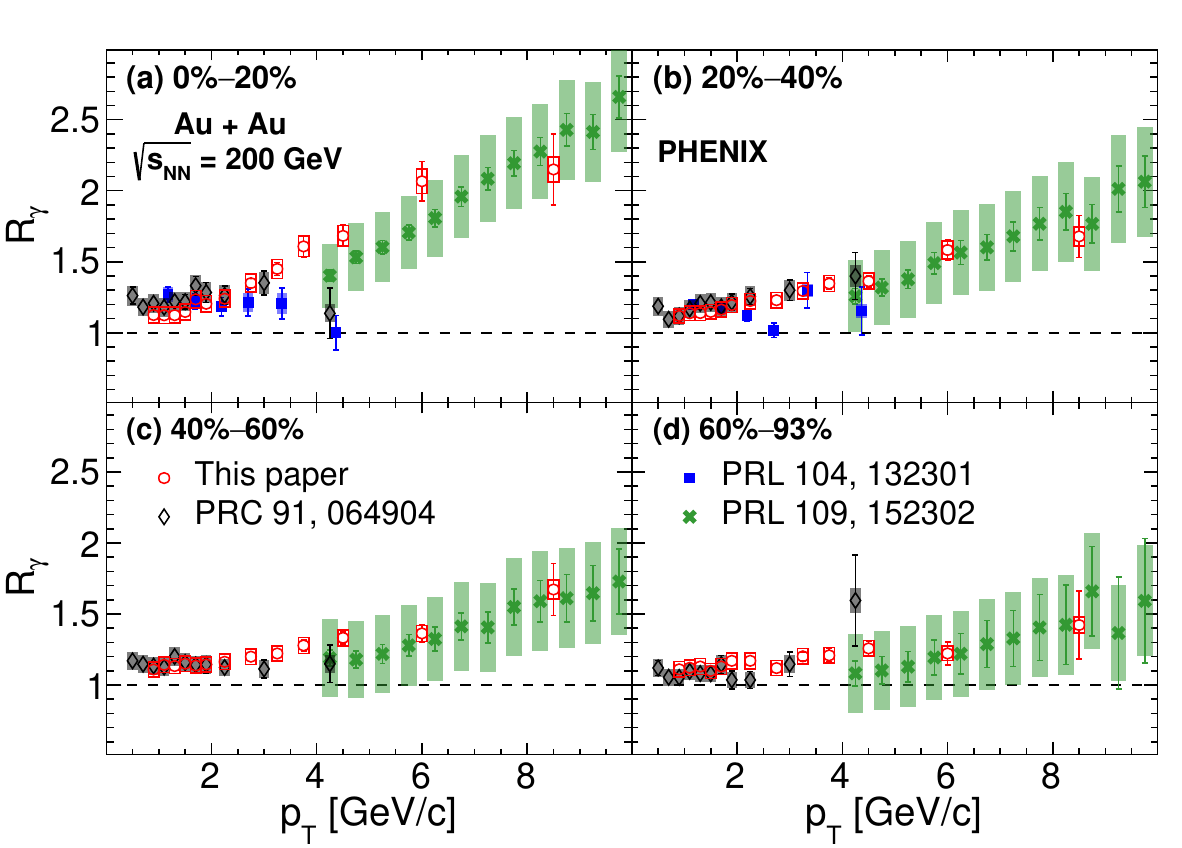} 
\caption{The ratio, $\Rg=\gammaincl/\gammahadr$, as a function of 
conversion photon \pt in 0\%--20\%, 20\%--40\%, 40\%--60\% and 
60\%--93\% centrality bins. The 2014 \auau data at \snn{200} are 
compared to results from previous PHENIX publications 
(see Refs.~\cite{PHENIX:2008uif,PHENIX:2014nkk,PHENIX:2012jbv})
}
        \label{fig:Rg_20}
\end{figure}

Figure~\ref{fig:Rg_20} shows \Rg as function of photon \pt for every 
20\% centrality class. The vertical error bar on each point corresponds 
to the statistical uncertainty, while the box gives the systematic 
uncertainty. The new results are compared with all other published 
PHENIX results for \auau at \snn{200}; these were obtained with 
different methods and have largely independent systematic uncertainties. 
The open circles were determined using the external conversion method 
deploying the HBD detector as converter~\cite{PHENIX:2014nkk}, the full 
squares are from a virtual photon internal conversion 
measurement~\cite{PHENIX:2011oxq}, and the open squares were measured 
with the EMCal alone~\cite{PHENIX:2012jbv}. All measurements agree well 
within their independent systematic uncertainties.

The 2014 data presented here have smaller statistical uncertainties than 
in previous publications at RHIC due to the increased luminosity and 
significantly larger amount of conversion material. The new results 
provide a continuous measurement across a wide range of \pt from 0.8 to 
10~\gevc. This range has previously been covered by measurements done 
with different techniques with different systematics.  Up to 3 to 4~\gevc 
internal or external photon conversions to \ee pairs have been used, 
while above 4~\gevc photons were measured in the EMCal. For all 
centrality selections, \Rg shows a significant excess that is rather 
constant below $\approx$3~\gevc. Beyond that, \Rg increases with \pt, 
the increase being most pronounced for central collisions, and 
\Rg continuously decreases towards more peripheral collisions. This is 
expected as phenomena such as jet quenching reduce the number of decay 
photons from hadron decays in more central collisions, which in turn
increases \Rg~\cite{PHENIX:2003qdj,PHENIX:2008saf}.

\begin{figure}[htb]
         \includegraphics[width=1.0\linewidth]{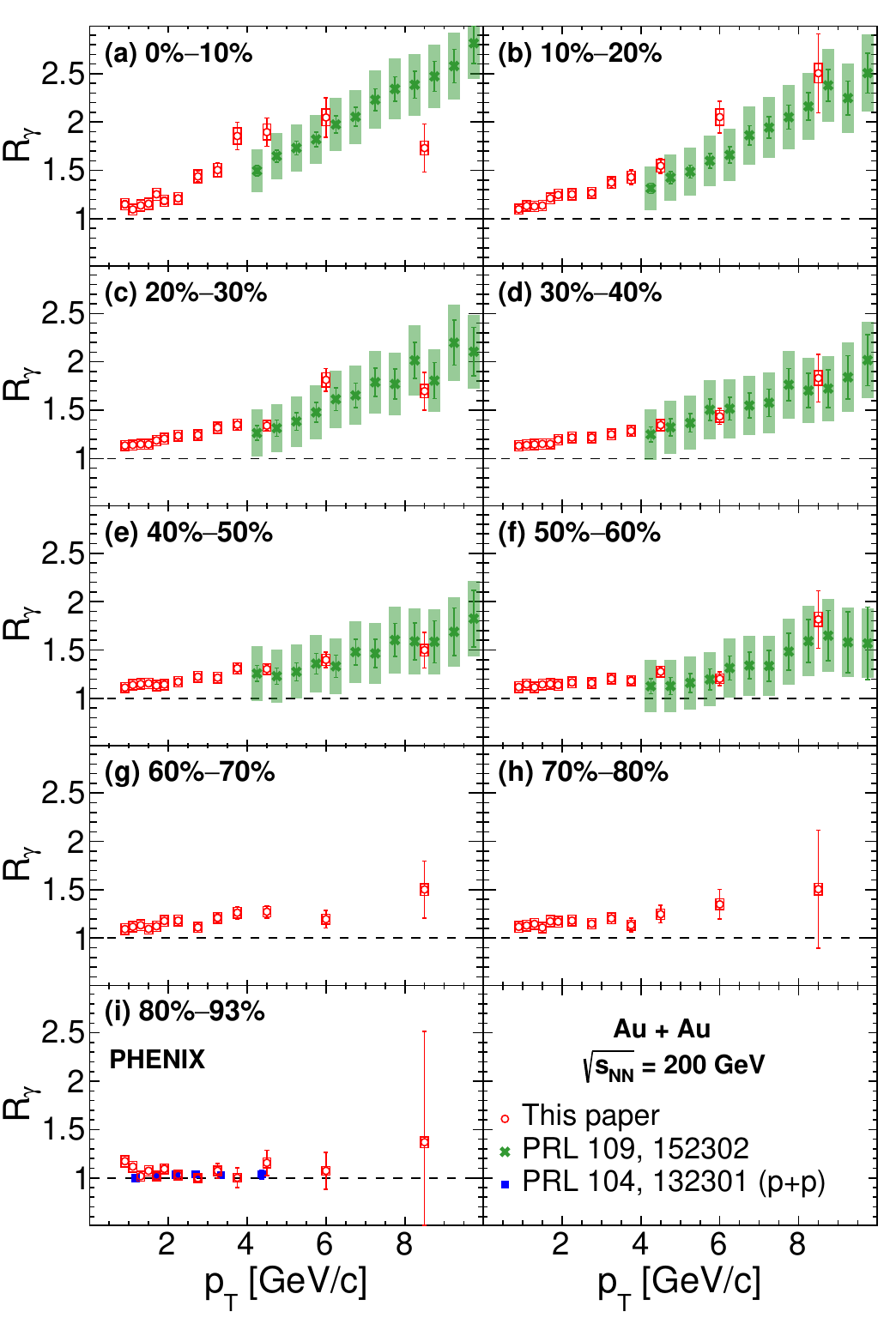} 
\caption{\Rg of direct photons as a function of conversion photon \pt in 
0\%--10\% to 80\%--93\% centrality bins
The 2014 \auau data at \snn{200} are 
compared to results from previous PHENIX publications 
(see Refs.~\cite{PHENIX:2008uif,PHENIX:2012jbv}).
}
        \label{fig:Rg_10}
\end{figure}

The high statistics of the 2014 data set allows to divide the data 
sample into nine centrality bins, from 0\%--10\% to 80\%--93\%, 10\% 
bins each, except for the last one which is slightly larger. The 
resulting \Rg are shown in \fig{fig:Rg_10}. Up to 50\%--60\% centrality, 
data from the earlier calorimeter measurement~\cite{PHENIX:2012jbv} are 
also shown.

For most bins the overall shape of $R_{\gamma}$ as a function of $p_T$ 
is similar to what is observed in \fig{fig:Rg_20}, with a notable 
difference for panel (i), which is the most-peripheral centrality 
80\%--93\%. 
Below 5~\gevc, the most-peripheral \auau data show no significant excess 
above unity and are very consistent with the direct-photon result from 
\pp collisions, which is also shown in panel (i).


The MC sampling method is used to calculate both the statistical 
and systematic uncertainties on \gammadir and all quantities derived 
from the direct photon \pt spectra. This method propagates the error 
correctly in the presence of unphysical values of $\Rg<1$ and \pt and 
centrality dependent correlations of uncertainties; it is discussed in 
detail in \append{sec:appendix-B}.

\begin{figure}[hpt]
         \includegraphics[width=1.0\linewidth]{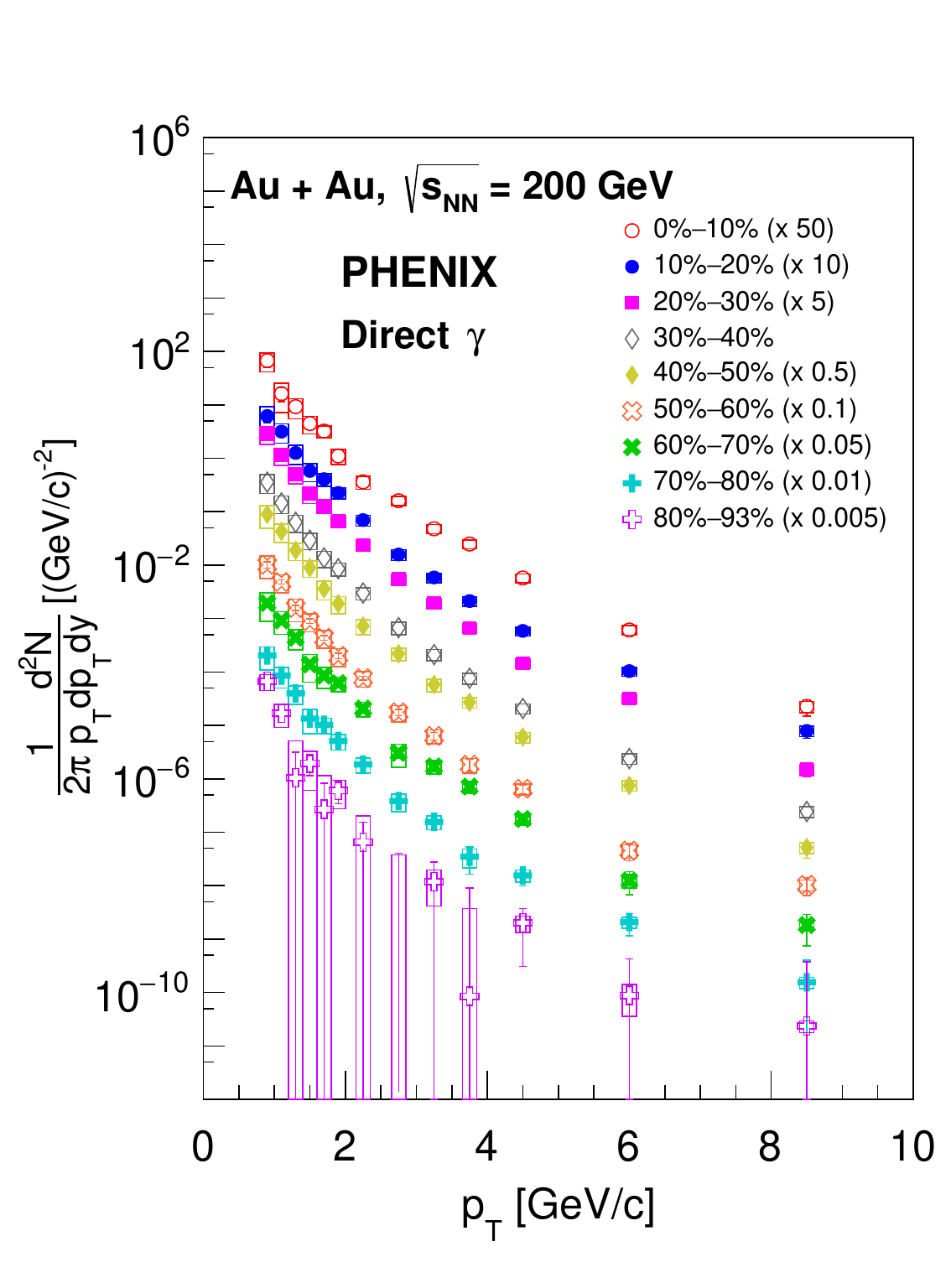} 
\caption{Invariant yield of direct photons as a function of conversion 
photon $p_{T}$ in 0\%--10\% to 80\%--93\% centrality bins.}
        \label{fig:Yield_10_allcent}
\end{figure}

\begin{figure}[hpt]
\includegraphics[width=1.0\linewidth]{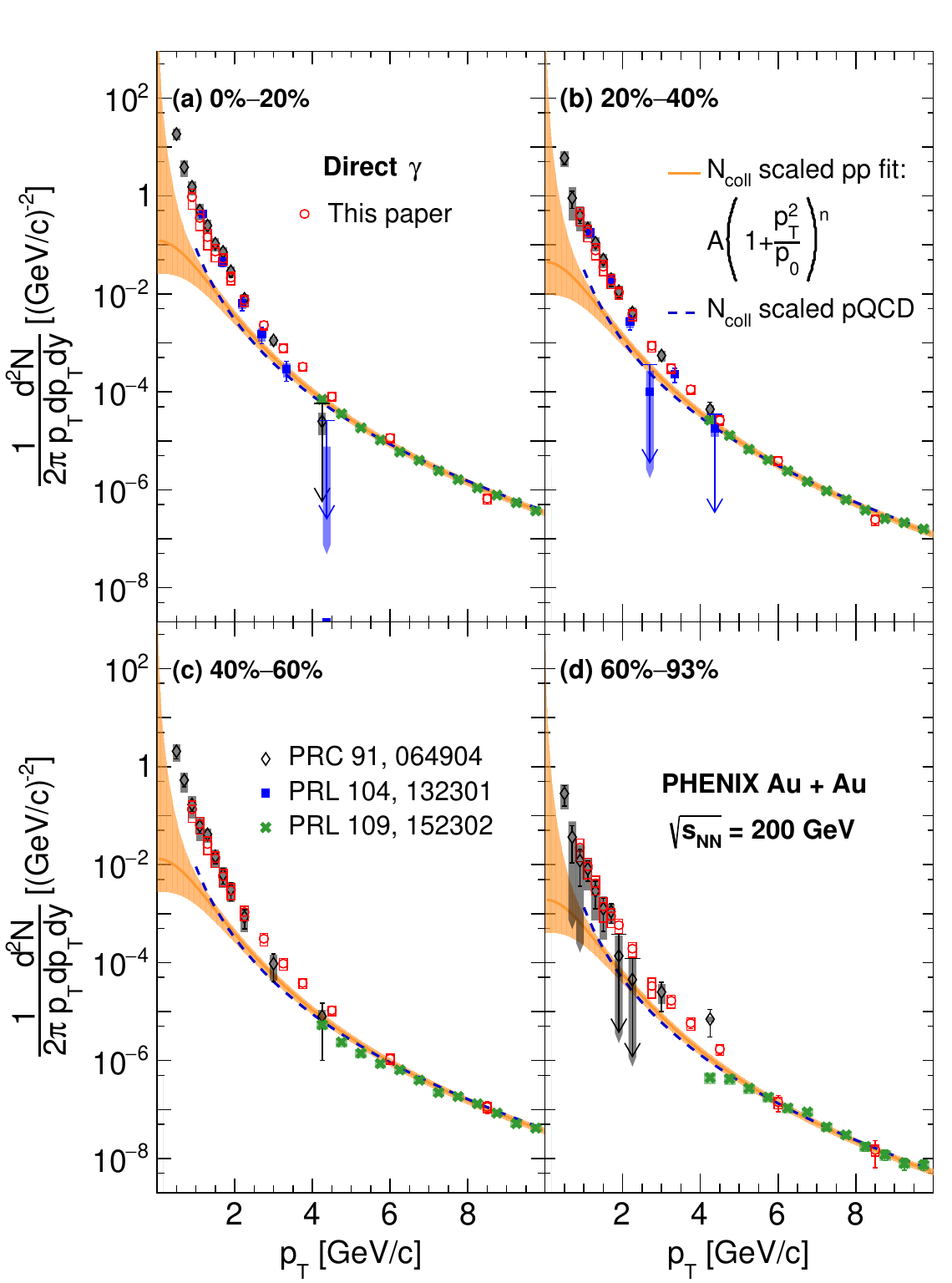} 
\caption{Invariant yield of direct photons as a function of conversion 
photon $p_{T}$ in (a) 0\%--20\%, (b) 20\%--40\%, (c) 40\%--60\% and 
(d) 60\%--93\% centrality bins.
The 2014 \auau data at \snn{200} are
compared to results from previous PHENIX publications 
(see Refs.~\cite{PHENIX:2008uif,PHENIX:2014nkk,PHENIX:2012jbv}).
}
        \label{fig:Yield_20}
\end{figure}

\subsection{Direct-photon invariant yield}

The direct-photon spectra are calculated from \Rg and \gammahadr 
using \eq{eq:gammadir}. The results for all 10\% centrality 
selections are given in~\fig{fig:Yield_10_allcent}.~\footnote{As the 
yields in the most-peripheral bin, 80\%--93\%, are mostly upper 
limits on the measurement, this bin will not be included for 
estimation of any further derived quantities in every 10\% centrality 
selection.} Figure~\ref{fig:Yield_20} compares the direct-photon 
spectra with previous measurements, as shown in broader centrality 
bins (a) 0--20\%, (b) 20--40\%, (c) 40--60\%, and (d) 60--93\%. Each 
panel also presents the \Ncoll-scaled perturbative quantum 
chromodynamics (pQCD) calculation~\cite{Paquet:2015lta} and a fit to 
direct-photon data from \pp collisions at 
\s{200}~\cite{PHENIX:2006duh,PHENIX:2012jgx,PHENIX:2012krx}. The \pp 
fit is performed with a pQCD-inspired functional 
form~\cite{PHENIX:2018che}:

\begin{equation}
\frac{d^3N}{d^2\pt dy} =  \frac{A_{pp}}{(1 + (\frac{p_{T}}{p_{0}})^2 )^{n}},
\label{eqn:ppfit}
\end{equation}
where the parameters are $A_{pp}=1.60\!\cdot\!10^{-4}$~(GeV/$c$)$^{-2}$, 
$p_{0}=1.45$~GeV/$c$ and $n=3.3$.  The error band around the central fit 
function represents the uncertainty propagated from both the data and 
the unknown true functional form of the spectrum down to very low \pt. 
The \pp fit and the pQCD calculation agree well above 2~\gevc, and can 
be used as an estimate for the prompt-photon contribution.

Figure~\ref{fig:Yield_20} also shows that the direct-photon yield for \pt 
larger than 5~\gevc is well described by the \Ncoll-scaled \pp result 
and pQCD calculations, which confirms that the high-\pt direct photons 
are predominately from initial hard-scattering processes. Below 
4--5~\gevc a clear direct-photon excess develops above the prompt 
component, gradually becoming larger towards lower \pt.

\begin{figure}[htb]
\includegraphics[width=1.0\linewidth]{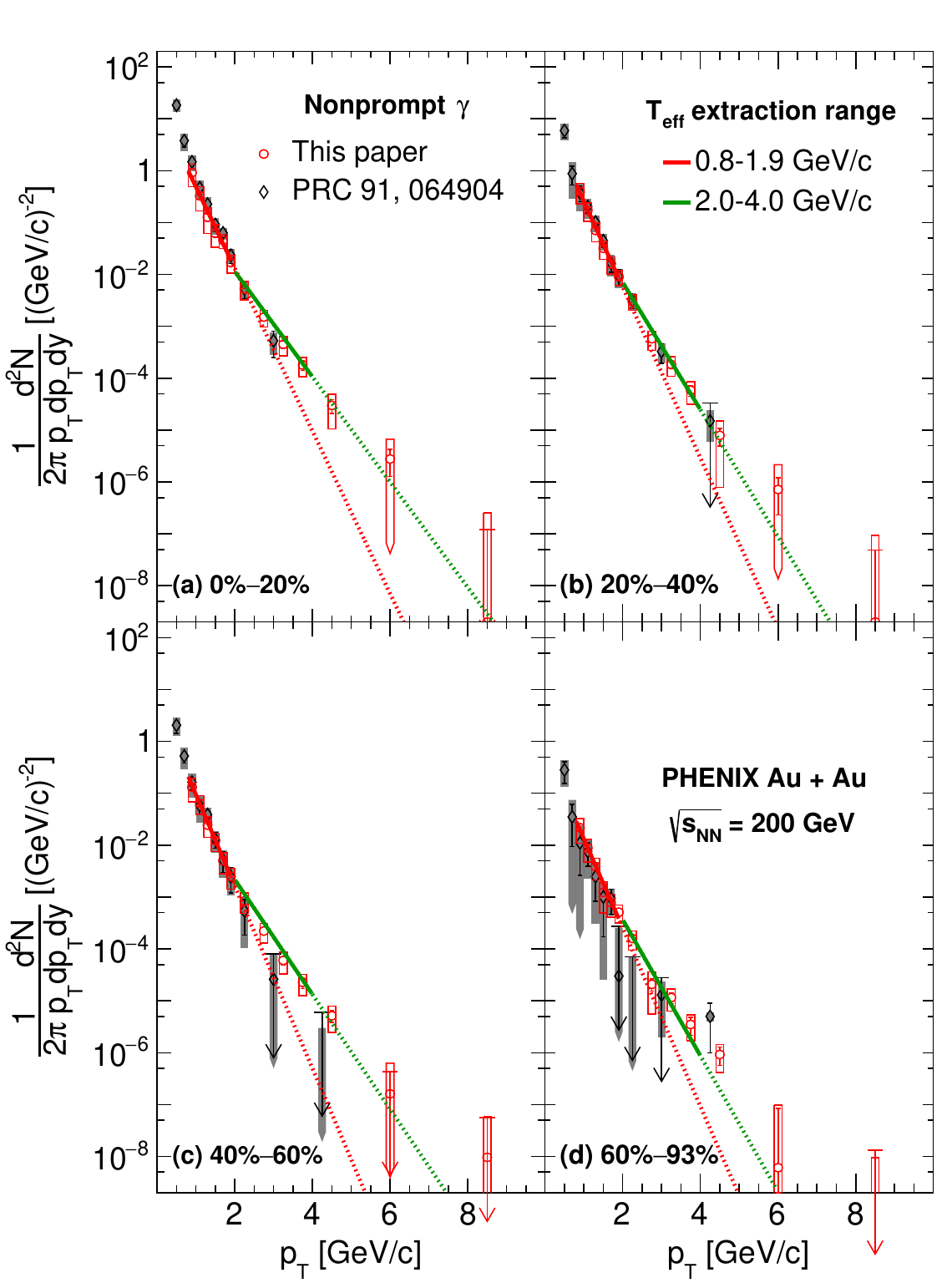} 
\caption{nonprompt direct-photon yield as a function of conversion 
photon $p_{T}$ in (a) 0\%--20\%, (b) 20\%--40\%, (c) 40\%--60\%, 
and (d) 60\%--93\% centrality bins.}
        \label{fig:Excess_20}
\end{figure}

\subsection{Nonprompt direct-photon excess}

To extract the direct-photon excess above the prompt-photon 
contribution, the \Ncoll scaled \pp fit is subtracted from the \auau 
data. This excess is thought to be mostly the radiation that is emitted 
during the collision from the hot-expanding fireball, and will 
be referred to here as nonprompt direct-photon spectra. 
Figure~\ref{fig:Excess_20} compares the nonprompt direct-photon spectra 
to previously published results from \auau collisions at 
\snn{200}~\cite{PHENIX:2014nkk}, which had significantly lower 
statistics.  The new 2014 data extend the coverage, both in \pt and 
centrality.

\begin{table*}[htb]
\caption{Inverse slopes fitted to the direct-photon spectra in different 
{\pt} ranges, and for different centrality selections. For each 
centrality range, \Ncoll and \dNch values are quoted, which are taken 
from previous work~\cite{PHENIX:2004vdg,PHENIX:2015tbb}, except for the 
\dNch values for the two most peripheral bins. Those were extrapolated 
using a fit of the form $\dNch=B(\Ncoll)^\beta$.
    } 
\label{Tab:Teff}
   \begin{ruledtabular} 
   \begin{tabular}{ccccc}
    centrality &  \dNch & \Ncoll &  \Teff (\gevc) & \Teff (\gevc)   \\   
   &&& $0.8< \pt < 1.9$~\gevc & $2< \pt < 4$ \\
   \hline  \rule{0pt}{2.6ex} 
     0\%--20\%   &  $519.2\pm26.3$  &   $770.6 \pm 79.9$ &  $0.277 \pm 0.017 \ ^{+0.036}_{-0.014}$  &  $0.428 \pm 0.031 \ ^{+0.031}_{-0.030}$ \rule{0pt}{2.6ex} \\   
    20\%--40\%  &  $225.4 \pm 13.2$  &   $282.4 \pm 28.4$ &  $0.264 \pm 0.010 \ ^{+0.014}_{-0.007}$  &  $0.354 \pm 0.019 \ ^{+0.020}_{-0.030}$ \rule{0pt}{2.6ex} \\   
    40\%--60\%  &  $85.5  \pm 8.0$   &   $82.6  \pm 9.3$  &  $0.247 \pm 0.007 \ ^{+0.005}_{-0.004}$  &  $0.392 \pm 0.023 \ ^{+0.022}_{-0.022}$ \rule{0pt}{2.6ex} \\   
    60\%--93\%  &  $16.4  \pm 2.8$   &   $12.1  \pm 3.1$  &  $0.253 \pm 0.011 \ ^{+0.012}_{-0.006}$  &  $0.331 \pm 0.036 \ ^{+0.031}_{-0.041}$ \rule{0pt}{2.6ex} \\ 
\\
     0\%--10\%   &  $623.9 \pm 32.2$    & $951 \pm 98.5$ &   $ 0.268 \pm 0.024 \ ^{+0.026}_{-0.012}$ &   $ 0.514 \pm 0.061 \ ^{+0.066 }_{-0.039}$  \rule{0pt}{3.6ex} \\ 
    10\%--20\%  &  $414.2 \pm 20.2$  & $590.1 \pm 61.1$ &   $ 0.303 \pm 0.024 \ ^{+0.062}_{-0.021}$ &   $ 0.358 \pm 0.033 \ ^{+0.024 }_{-0.035}$  \rule{0pt}{2.6ex}  \\
    20\%--30\%  &  $274   \pm 14.8$  & $357.2 \pm 35.5$ &   $ 0.263 \pm 0.011 \ ^{+0.014}_{-0.007}$ &   $ 0.351 \pm 0.024 \ ^{+0.020 }_{-0.030}$  \rule{0pt}{2.6ex}  \\
    30\%--40\%  &  $176.8 \pm 11.6$   & $207.5 \pm 21.2$ &   $ 0.256 \pm 0.011 \ ^{+0.009}_{-0.005}$ &   $ 0.333 \pm 0.024 \ ^{+0.020 }_{-0.032}$  \rule{0pt}{2.6ex}  \\
    40\%--50\%  &  $109.4 \pm 9.1$   & $111.1 \pm 10.8$ &   $ 0.244 \pm 0.009 \ ^{+0.003}_{-0.005}$ &   $ 0.389 \pm 0.029 \ ^{+0.020 }_{-0.021}$  \rule{0pt}{2.6ex}  \\
    50\%--60\%  &  $61.6  \pm 7.1$     & $54.1 \pm 7.9$ &   $ 0.246 \pm 0.010 \ ^{+0.005}_{-0.005}$ &   $ 0.345 \pm 0.031 \ ^{+0.019 }_{-0.032}$  \rule{0pt}{2.6ex}  \\
    60\%--70\%  &  32 $\pm$ 5  & $24 \pm 6$ &   $ 0.261 \pm 0.015 \ ^{+0.020}_{-0.008}$ &   $ 0.319 \pm 0.049 \ ^{+0.037 }_{-0.042}$  \rule{0pt}{2.6ex}  \\
    70\%--80\%  &  16 $\pm$ 4  & $10 \pm 3$ &   $ 0.263 \pm 0.016 \ ^{+0.016}_{-0.007}$ &   $ 0.335 \pm 0.044 \ ^{+0.020 }_{-0.035}$  \rule{0pt}{2.6ex}  \\
    80\%--93\%  &  7  $\pm$ 2  & $4 \pm 1$ &   $-$ &   $-$  \rule{0pt}{2.6ex}  \\
   \end{tabular}
   \end{ruledtabular} 
\end{table*}

\begin{figure}[htb] 
        \includegraphics[width=1.0\linewidth]{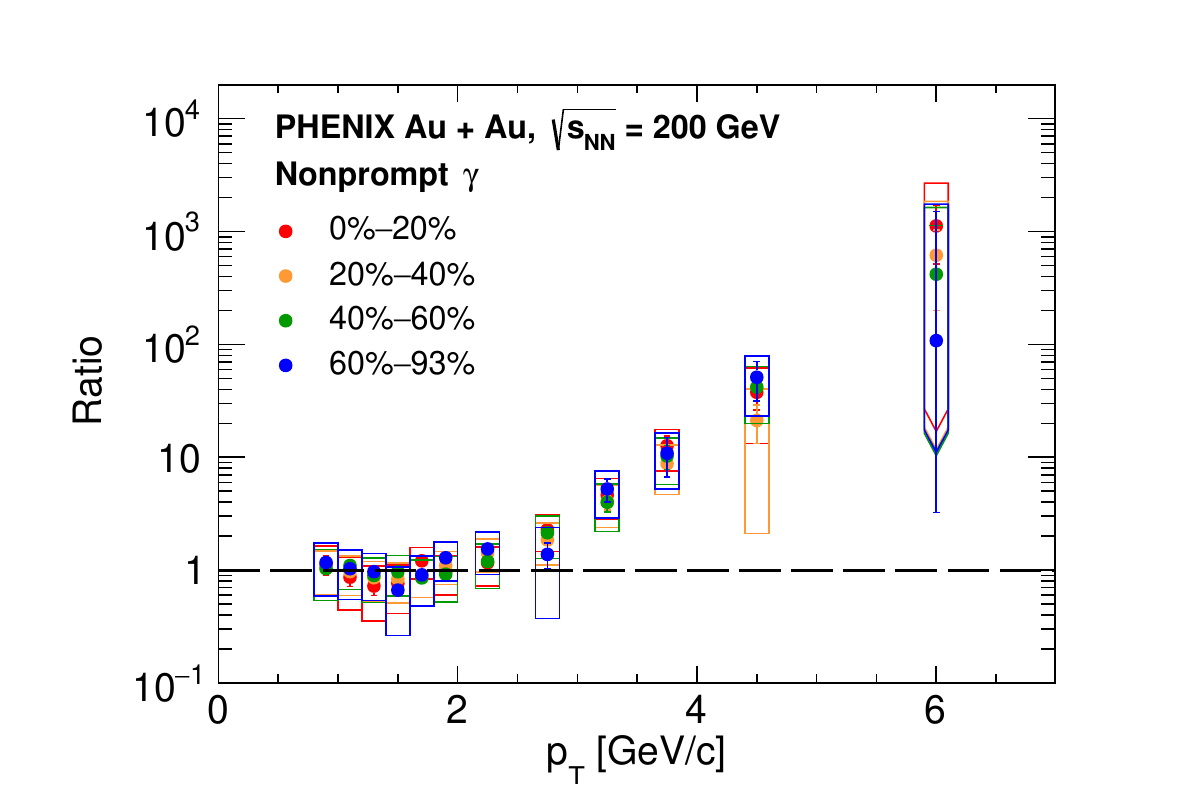} 
\caption{Ratio of the yield of nonprompt direct photons over the exponential 
fit result ($T_{\rm eff}$ fixed to 0.26~\gevc) as a function of photon 
$p_{T}$.}
        \label{fig:ratio_sub_allcent}
\end{figure}

The data are very consistent in the region of overlap. In the range 0.8 to 
1.9~\gevc, the data are fitted with an exponential function and the 
results are also shown on the panels of \fig{fig:Excess_20}. The slope 
values are given in \tab{Tab:Teff}. All centrality selections are 
consistent with an average inverse slope, $\Teff$, of 
${\approx}0.260{\pm}0.011$~\gevc. However, it is evident from 
\fig{fig:Excess_20} that the nonprompt direct-photon spectra are not 
described by a single exponential but rather have a continually increasing 
with \pt inverse slope, \Teff.  Figure~\ref{fig:ratio_sub_allcent} brings 
this out more clearly where each nonprompt direct-photon spectrum is 
divided by a fit with a fixed slope, \Teff = 0.260~\gevc. All centrality 
selections follow the same trend. Over the \pt range of up to 2~\gevc the 
ratios are consistent with unity, but above 2~\gevc, they start to rise 
monotonically.

\begin{figure}[htb]
        \includegraphics[width=1.0\linewidth]{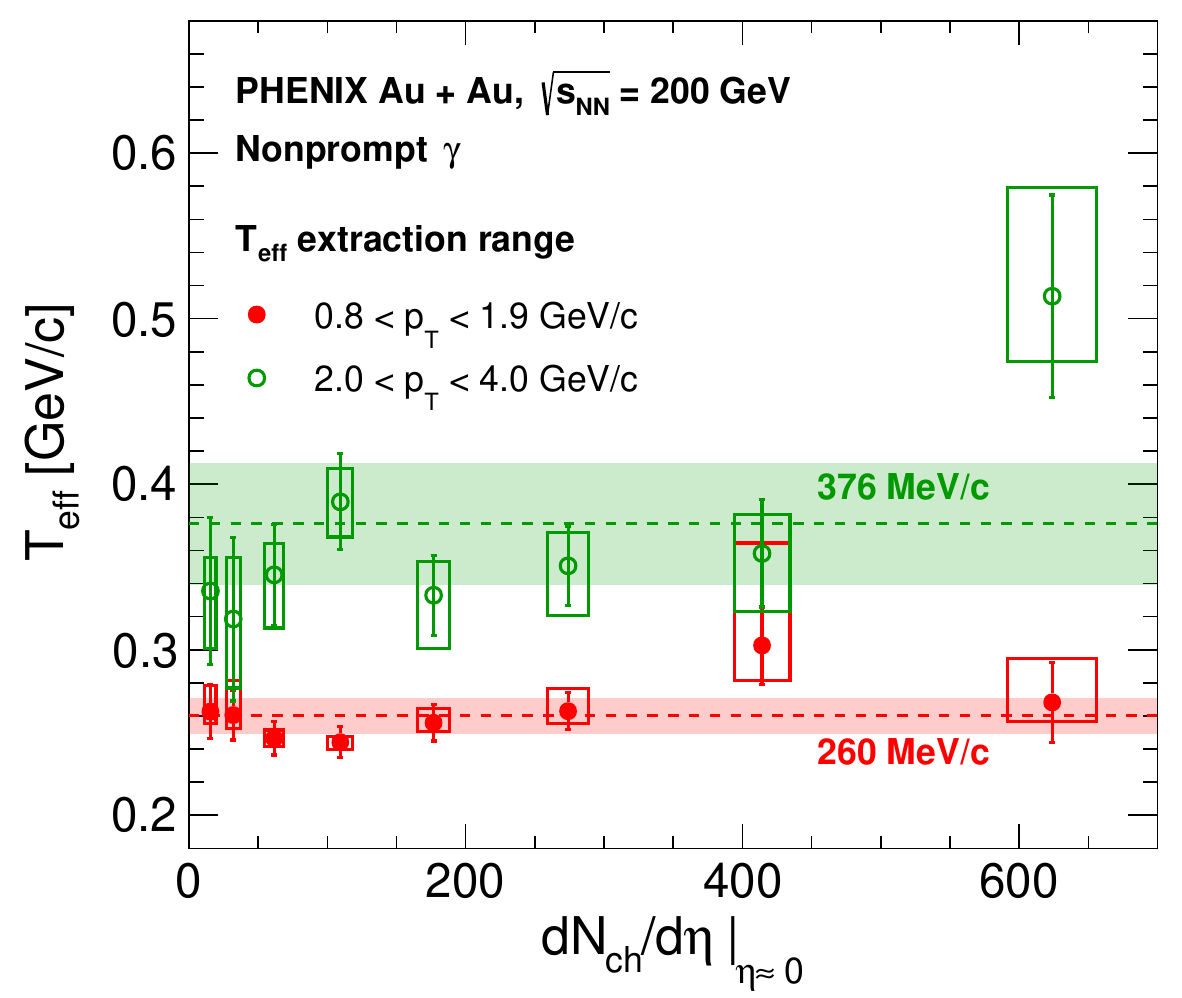} 
\caption{$T_{\rm eff}$ as a function of charged-particle multiplicity at 
midrapidity.}
        \label{fig:Teff_vs_Nch}
\end{figure}

To quantify this changing slope, the nonprompt direct-photon spectra are 
fitted with a second exponential function in the \pt range from 2 to 
4~\gevc; the results are also included in \fig{fig:Excess_20}. All data 
are consistent with an average inverse slope of $0.376\pm0.037$~\gevc, 
which is significantly larger than the slope observed below \pt= 
2~\gevc. Above 4~\gevc, the statistical and systematic uncertainties 
from the prompt-photon subtraction become too large for a detailed 
analysis.

To establish any dependence on the system size, the nonprompt direct 
photon spectra are determined for each 10\% centrality bin, and 
subsequently fitted by two exponential functions in the \pt ranges $0.8 
<\pt<1.9$~\gevc and $2<\pt<4$~\gevc.  The resulting \Teff values are 
tabulated in \tab{Tab:Teff} and depicted in \fig{fig:Teff_vs_Nch} as a 
function of \dNch. The figure also shows the average of the inverse 
slope values from fitting \fig{fig:Excess_20}. The \Teff values are 
consistent with a constant value, independent of \dNch. However, given 
the uncertainties on the data, a possible increase of \Teff with \dNch 
can not be excluded.

\begin{figure}[htb]
         \includegraphics[width=1.0\linewidth]{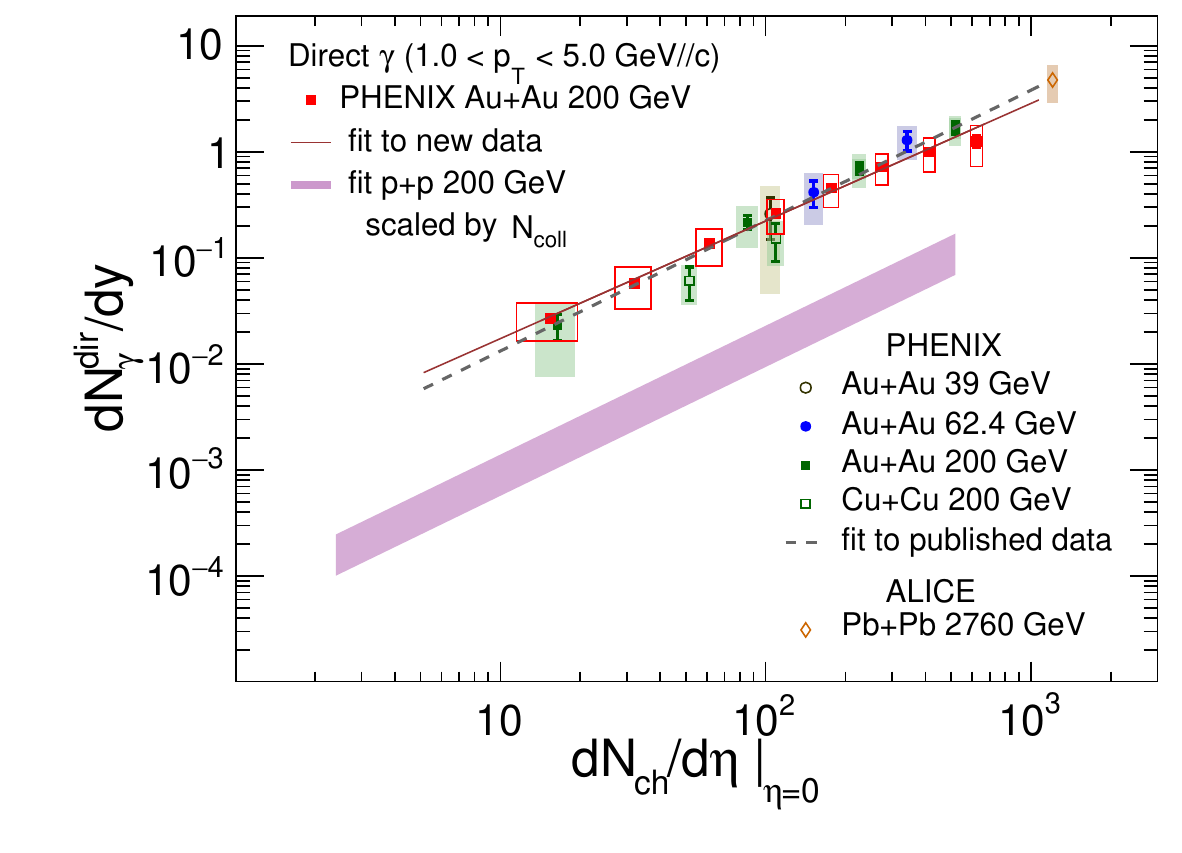} 
\caption{Integrated direct-photon yield (1--5 GeV$/c$) versus 
charged-particle multiplicity at midrapidity. The present data is 
compared to a previous compilation of data 
from~\cite{PHENIX:2018for,PHENIX:ppg225} and the \Ncoll scaled fit to 
\pp data. Also given are fits with \eq{eq:scaling} to different data; 
the solid line is a fit to the present data resulting in $\alpha = 1.11 
\pm 0.02 ({\rm stat}) \ _{-0.08}^{+0.09} ({\rm sys}) $, and the dashed 
line is from fitting previously published PHENIX 
data~\cite{PHENIX:ppg225} that gave $\alpha=1.23$ $\pm$ 0.06 $\pm$ 0.18.
The ALICE data is from Ref.~\cite{ALICE:2015xmh}.
       }
        \label{fig:IntYield_cent9}
\end{figure}

In addition to investigating the \pt and system-size dependence of the 
shape of the nonprompt direct-photon spectra, one can also look at the 
dependence of the yield on system size and \pt. As reported previously, 
the integrated direct-photon yield scales with \dNch to a power 
$\alpha$~\cite{PHENIX:2018for}:
\begin{equation}
\frac{dN_{\gamma}}{dy}
= \int_{p_{T, {\rm min}}}^{p_{T, {\rm max}}}
\frac{dN_{\gamma}^{\rm dir}}{d\pt dy} d\pt
= A\times\left(\frac{dN_{\rm{ch}}}{d\eta}\right)^{\alpha},
\label{eq:scaling}
\end{equation}
where all rapidity densities are densities at midrapidity. The 
direct-photon spectra shown in \fig{fig:Yield_10_allcent} are integrated 
from $\pt_{,{\rm min}}=1$~\gevc to $\pt_{,{\rm max}}=5$~\gevc and 
plotted as a function of \dNch in \fig{fig:IntYield_cent9}. They are in 
reasonable agreement with a compilation of other direct-photon 
results~\cite{PHENIX:2018for,PHENIX:ppg225}, also shown in the figure. 
All data follow a trend similar to the \Ncoll scaled \pp fit, shown as 
band, but at a roughly 10 times larger yield. Scaling with \Ncoll 
corresponds to $\alpha=1.25$ $\pm$ 0.02~\cite{PHENIX:2018for}. The 
current high statistics data allow for finer centrality binning and 
changes this picture somewhat at the lowest and highest \dNch. Fitting 
only the new results in \fig{fig:IntYield_cent9} gives a value of 
$\alpha = 1.11 \pm 0.02 ({\rm stat}) \ _{-0.08}^{+0.09} ({\rm sys})$. 
This value is lower, but consistent within systematic uncertainties, 
with $\alpha=1.23$ $\pm$ 0.06 $\pm$ 0.18, found by fitting all 
previously published PHENIX \AB{A}{A} data~\cite{PHENIX:ppg225}.

Note that the previous PHENIX measurements obtained the $\eta$ spectrum 
by $m_T$-scaling the \piz spectrum, while in the current measurement the 
$\eta$ spectrum is obtained from the $\eta/$\piz ratio using the world 
data.  There are significant differences between the two approaches in 
the low-\pt region~\cite{Ren:2021xbh}. Because the integration range 
starts at low \pt and is wide (1--5~\gevc), the power $\alpha$ is 
smaller than previously published values, but is consistent within 
stated systematic uncertainties.  However, it is also consistent with 
unity within uncertainties.

\begin{table}[htb]  
\caption{
Scaling power, $\alpha$, of the \dNch dependence of nonprompt and 
direct-photon yields in various integration ranges.
\label{tab:non_prompt_alpha}} 
\begin{ruledtabular}  \begin{tabular}{ccc} 
$p_{T}$ (\gevc)  & $\alpha (\gammanonp)$ & $\alpha (\gammadir)$ \\ 
\hline
0.8--1.2 & $1.119 \pm 0.038 \ _{-0.094}^{+0.116}$ & $1.124 \pm 0.036 \ _{-0.089}^{+0.121}$ \rule{0pt}{2.6ex} \\ 
1.2--1.6 & $1.107 \pm 0.029 \ _{-0.082}^{+0.108}$ & $1.118 \pm 0.027 \ _{-0.073}^{+0.097}$ \rule{0pt}{2.6ex} \\ 
1.6--2.0 & $1.136 \pm 0.034 \ _{-0.091}^{+0.129}$ & $1.152 \pm 0.029 \ _{-0.077}^{+0.113}$ \rule{0pt}{2.6ex} \\ 
2.0--3.0 & $1.087 \pm 0.032 \ _{-0.092}^{+0.108}$ & $1.120 \pm 0.025 \ _{-0.065}^{+0.095}$ \rule{0pt}{2.6ex} \\ 
3.0--4.0 & $1.119 \pm 0.078 \ _{-0.134}^{+0.206}$ & $1.171 \pm 0.048 \ _{-0.076}^{+0.114}$ \rule{0pt}{2.6ex} \\ 
4.0--5.0 & $0.950 \pm 0.176 \ _{-0.205}^{+0.315}$ & $1.137 \pm 0.077 \ _{-0.082}^{+0.108}$ \rule{0pt}{2.6ex} \\ 
5.0--10.0 & & $1.296 \pm 0.078 \ _{-0.091}^{+0.129}$ \rule{0pt}{2.6ex} \\ 
\end{tabular}  \end{ruledtabular} 
\end{table} 

\begin{figure}[htb]
        \includegraphics[width=1.0\linewidth]{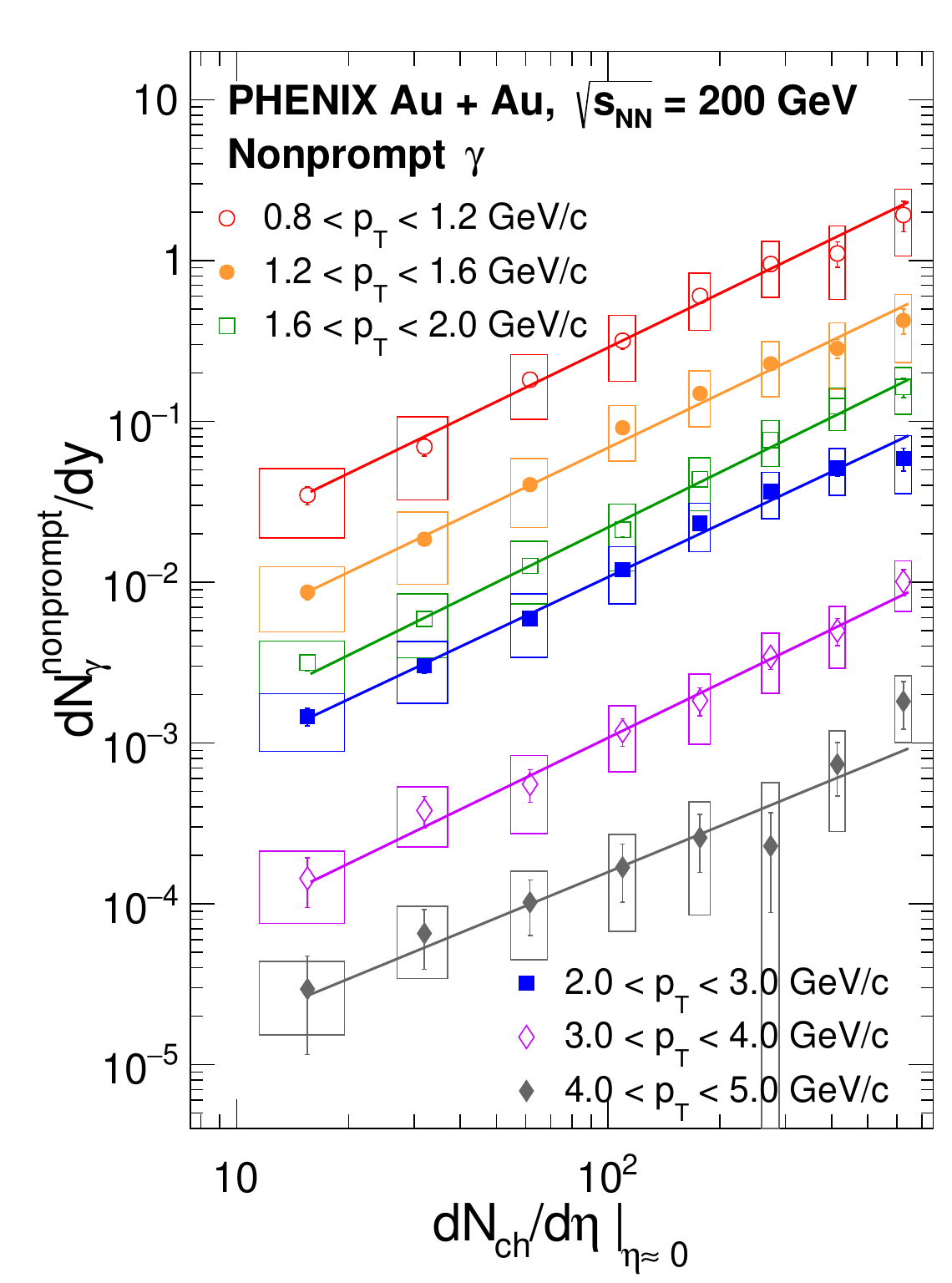} 
\caption{Integrated nonprompt direct-photon yield versus charged 
particle multiplicity at midrapidity for different $p_{T}$ integration 
ranges. 
}
        \label{fig:IntYield_cent9_finept}
\end{figure}

\begin{figure}[htb]     
         \includegraphics[width=1.0\linewidth]{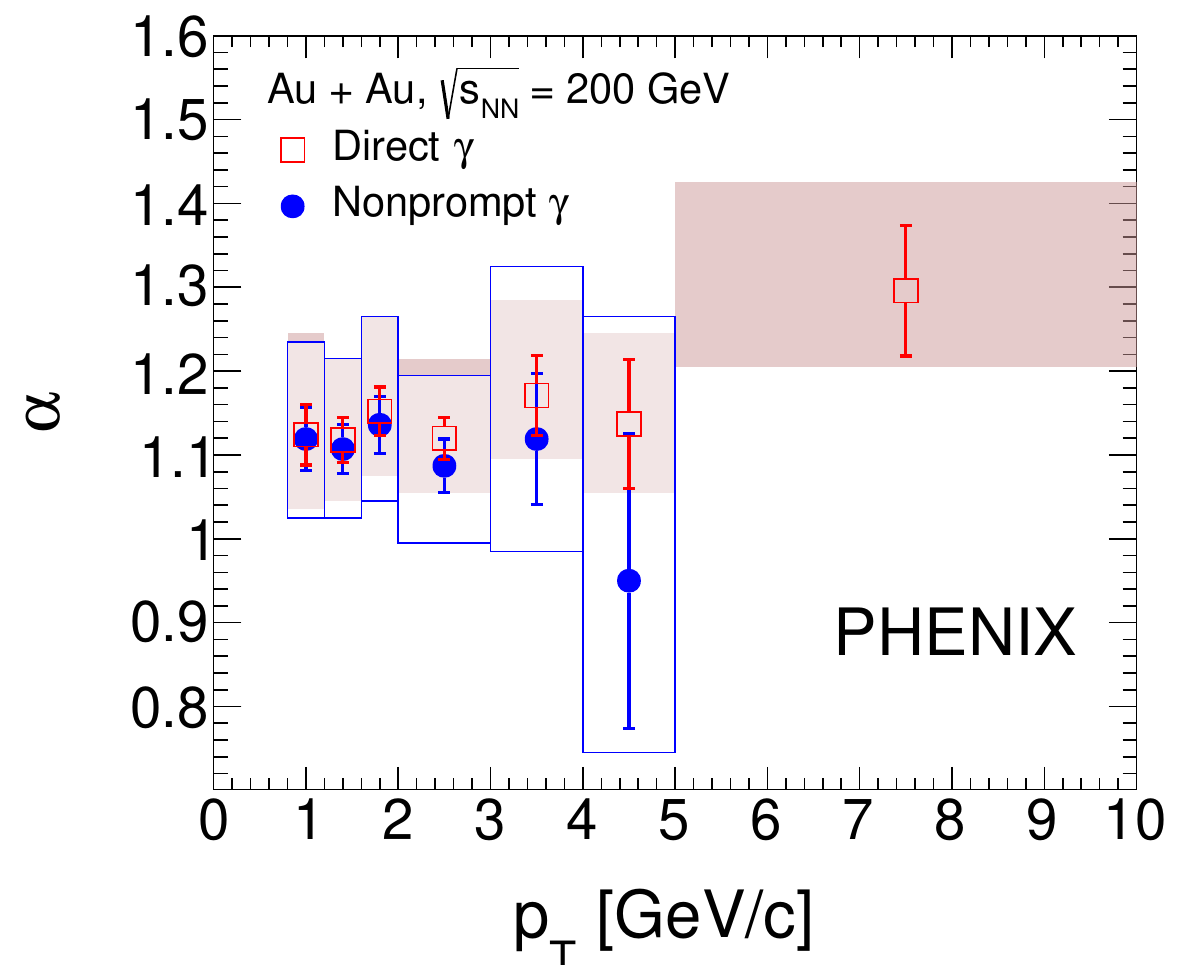} 
\caption{Scaling factors, $\alpha$, extracted from fitting 
\eq{eq:scaling} to integrated direct and nonprompt-photon yields as 
a function of \dNch. Values were obtained for different 
$p_{T}$ integration ranges tabulated in \tab{tab:non_prompt_alpha}. }
        \label{fig:alpha_vs_pt}
\end{figure}

To better understand the behavior of the scaling power, $\alpha$, in 
more detail, the direct-photon yield and its nonprompt component are 
integrated for six different nonoverlapping finer \pt regions and for 
10\% centrality classes.  The integrated nonprompt yields are shown in 
\fig{fig:IntYield_cent9_finept}. The $\alpha$ values are determined for 
each \pt selection by fitting the data with \eq{eq:scaling}. The fits 
are also shown in the figure. All $\alpha$ values, both for the direct 
photon yield and the nonprompt component, are tabulated in 
\tab{tab:non_prompt_alpha} and shown in \fig{fig:alpha_vs_pt}. It is 
evident that the values for the direct component, for higher \pt ranges, 
are consistent with the prompt component, $\alpha = 1.25\pm 0.02$, 
corresponding to \Ncoll scaling. However, they tend to be smaller, but 
still consistent within systematic uncertainties, with previous 
measurements~\cite{PHENIX:2018for} for the lower \pt ranges.

With increasing \pt, the $\alpha$ values for the nonprompt component are 
slightly lower than those from direct photons. The systematic 
uncertainties are larger due to the subtraction. The values of $\alpha$ 
for the nonprompt component, as shown in \fig{fig:alpha_vs_pt}, are 
remarkably constant with no evident \pt dependence.

\begin{figure}[htb]
         \includegraphics[width=1.0\linewidth]{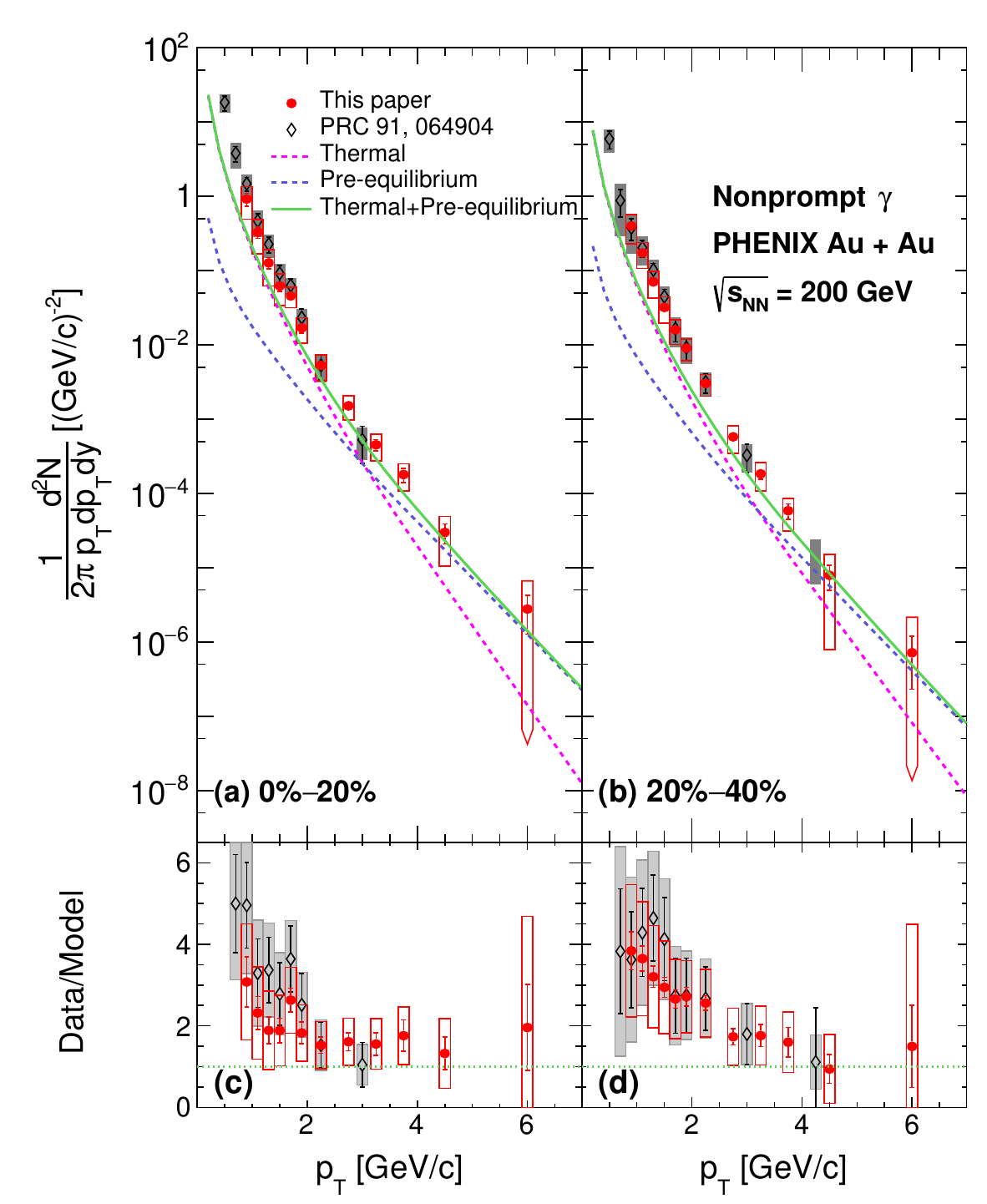} 
\caption{Nonprompt direct-photon yields for (a) 0\%--20\%
and (b) 20\%--40\% compared with model predictions 
from Refs.~\protect\cite{Gale:2021emg,paquet:2022}. 
(c,d) ratios of the yields from data to the sum of yields 
from thermal and pre-equilibrium contributions.
The 2014 \auau data at \snn{200} are
compared to results from a previous PHENIX publications 
(see Ref.~\protect\cite{PHENIX:2014nkk}).
}
        \label{fig:thermal}
\end{figure}

 
\section{\label{sec:summary}Concluding discussion of the results}   
 
The PHENIX collaboration has measured direct-photon production in \auau 
collisions at \snn{200} using photon conversions to \ee pairs. A large 
yield of direct photons below a \pt of 3~\gevc is observed for all 
centrality bins except for the most peripheral bin of 80\%--93\% with 
\dNch~=~7.4, where it seems to be consistent with the prompt-photon 
production. The next centrality bin from 70\%--80\% with \dNch = 15.5 
already shows a significant yield with properties very similar to that 
of the radiation from the more central bins.

The nonprompt direct-photon spectra are isolated by subtracting the 
prompt-photon contribution, which is estimated through a fit to the 
direct-photon data from \pp collisions at \s{200}, measured by PHENIX, 
and scaled by \Ncoll. Results are obtained for the \pt range from 0.8 to 
5~\gevc and for 0\%--93\% central collisions, covering a system size 
spanning two orders of magnitude in \dNch from $\approx$7 to 620. The 
wealth of data enabled PHENIX to carry out double-differential analyses 
of the shape of the momentum spectra and the rapidity density \dNgdy in 
\pt and \dNch.
 
For the centrality selections from 0\%--10\% to 70\%--80\%, all 
nonprompt direct-photon spectra are very similar in shape, exhibiting
increasing \Teff from 0.2 to 0.4~\gevc over the \pt range from 0.8 to 
4~\gevc. The changing \Teff is not surprising, because the spectra are 
time integrated over the full evolution of the expanding fireball, from 
its earliest pre-equilibrium state, through the QGP phase, crossing over 
to a HG, and further expanding and cooling until hadrons eventually stop 
interacting. Throughout the evolution the system cools, and thus earlier 
phases are characterized by higher temperatures. In turn, the 
contributions from the earliest times of the evolution are likely to 
dominate the emission at higher \pt, consistent with the observation of 
an increasing \Teff with \pt.

In the lower \pt range from 0.8 to 1.9~\gevc, the spectra are well 
described by a \Teff = 0.26~\gevc. This is consistent with what is 
expected for radiation from the late QGP stage until 
freeze-out~\cite{Shen:2013vja}. During this period of the evolution, the 
temperature drops from $\approx$170 MeV near the transition to 
$\approx$110~MeV when the system freezes out. At the same time the 
system is rapidly expanding and thus, the radiation is blue shifted. 
This compensates the temperature drop and results in an average \Teff 
$\approx$ 0.26~\gevc, with only minor variations with centrality of the 
collision. In Ref.~\cite{Shen:2013vja}, a moderate increase of \Teff 
with centrality was predicted. While the data favors a \Teff independent 
of centrality, they are not precise enough to exclude a moderate change.

Above a \pt of 2~\gevc, the inverse slope of the spectra continues to 
increase with \pt. Between \pt = 2 and 4~\gevc the average inverse slope 
is \Teff $\approx$ 0.376~\gevc.  This \Teff is larger than what model 
calculations for a rapidly expanding HG can accommodate, thus 
suggesting that emissions from the QGP phase and earlier times in the 
evolution starts to dominate the spectra. Expected initial temperatures 
at RHIC are $\approx$ 375 MeV with maximum \Teff in the range of 0.35 to 
0.4~\gevc, depending on viscosity~\cite{Shen:2013vja}. Thus, it is 
likely that in addition to photons from the QGP phase, photons from the 
pre-equilibrium stage are also needed to account for the measured \Teff.

In \fig{fig:thermal}, the measured nonprompt direct-photon spectra are 
compared to a recent calculation including contributions from the 
pre-equilibrium phase~\cite{Gale:2021emg,paquet:2022}. These 
calculations predicted that the pre-equilibrium radiation becomes the 
dominant source above a \pt of 3~\gevc. In the range $2<\pt<4$~\gevc, a 
fit of the thermal contribution with an exponential function results in 
an inverse slope of $\approx$0.36~\gevc, while for the pre-equilibrium 
contribution a larger inverse slope of $\approx$0.52~\gevc is found, for 
the more central collisions. Fitting the same \pt range for the combined 
thermal and pre-equilibrium spectra from the model gives an inverse 
slope of $\approx$0.425~\gevc. While the shape is reproduced well, the 
overall yield predicted by the calculations falls short compared to the 
data, in particular, below 2~\gevc where the nonprompt-photon yield 
appears to be a factor of two to three larger.

The integrated nonprompt direct-photon yield exhibits a power-law 
relation with $(\dNch)^\alpha$~\cite{PHENIX:2018for}. Fitting the power 
$\alpha$ for multiple nonoverlapping \pt ranges results in values 
consistent with
$\alpha=1.12\pm0.06 {\rm stat}\pm 0.14{\rm sys}$ with no apparent 
dependence on \pt. The model calculations in~\cite{Shen:2013vja} predict 
that the radiation from the HG phase scale with $\alpha$ close to 1.2, 
while those from the hot and dense QGP phase exhibit closer to a 
$(\dNch)^2$ dependence. Because the QGP phase has a larger relative 
contribution to the \pt spectrum with increasing \pt, it is expected 
that $\alpha$ increases with \pt. However, the \pt dependence of 
$\alpha$ from the pre-equilibrium phase needs further theoretical 
understanding.

In conclusion, the 10-fold increase in statistics compared to previous 
samples of \auau collisions recorded by PHENIX enabled detailed 
measurements of the radiation from the hot and expanding fireball. The 
experimentally observed inverse slopes of the \pt spectra are 
qualitatively consistent with predictions for thermal and 
pre-equilibrium radiation. However, there seems to be more photons 
emitted from \auau collisions than can be accounted for in model 
calculations. Furthermore, although this work presents no new data on 
the azimuthal anisotropy, maximum anisotropy is observed for photons 
$\approx$2--3~\gevc. In this \pt range, the yield is larger than what 
would be expected from a rapidly but anisotropically expanding hadronic 
fireball. Finally, the centrality dependence of the nonprompt 
direct-photon yield, expressed in terms of the scaling power 
$\alpha(\pt)$, shows no indication of changing with \pt.

\begin{figure*}[htb]
\begin{minipage}{0.48\linewidth}
\includegraphics[width=0.99\linewidth]{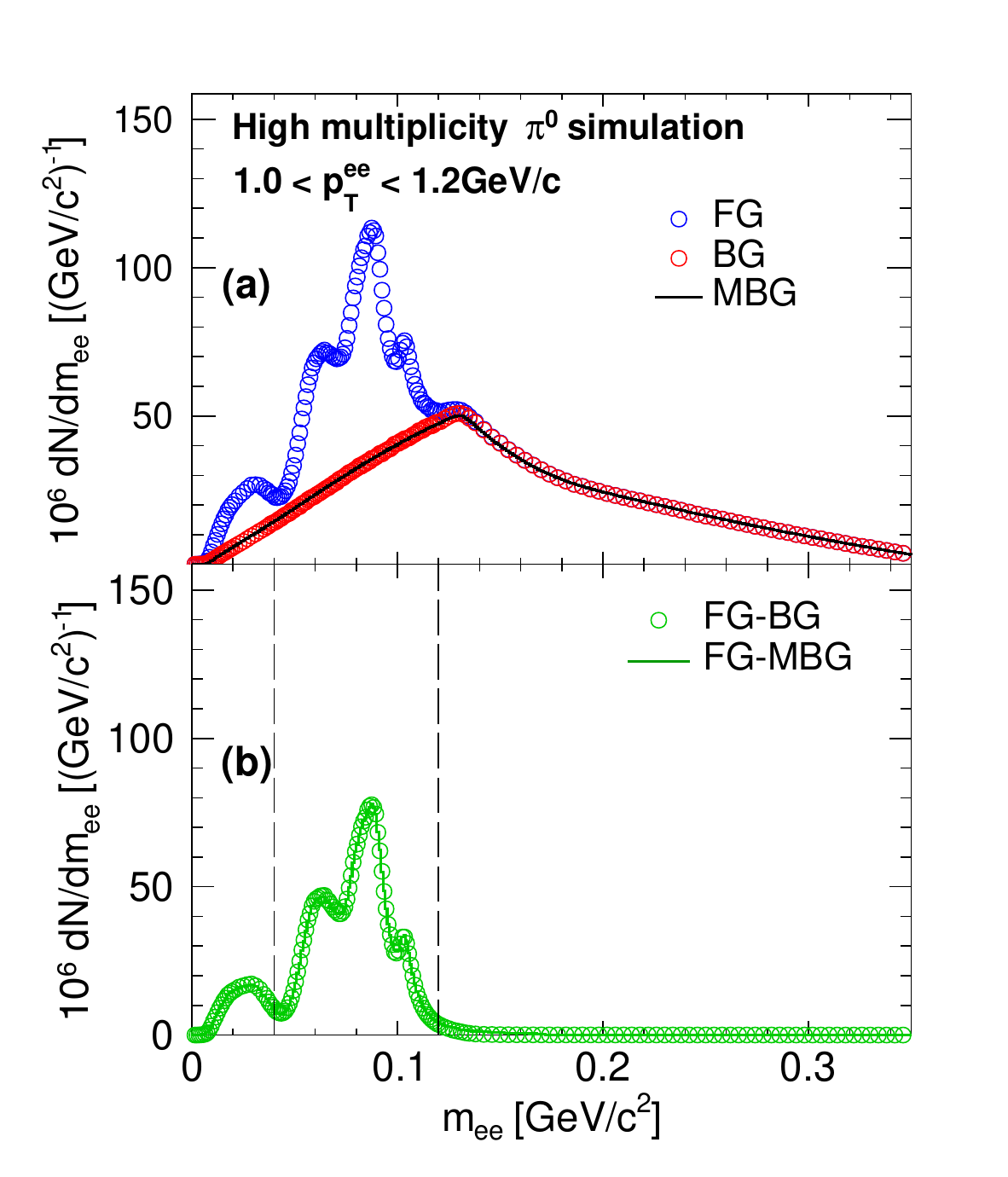} 
\caption{Invariant mass distributions of $e^+e^-$ pairs reconstructed 
from the high-multiplicity \piz pseudodata in the \pt range $0.8 < \pt 
< 1.0$~\gevc. The least-restrictive conversion selection cuts are 
applied, which only require that the reconstruction algorithm has identified 
the \ee pair as a conversion candidate. Panel (a) compares foreground, 
\Ninclfg, the true background, \Ninclbg, and the background determined 
from the mixed event technique, \Minclbg. Panel (b) gives the extracted 
conversion photon signal. }
        \label{fig:mee1}
\end{minipage}
\hspace{0.2cm}
\begin{minipage}{0.48\linewidth}
\vspace{-0.7cm}
         \includegraphics[width=0.99\linewidth]{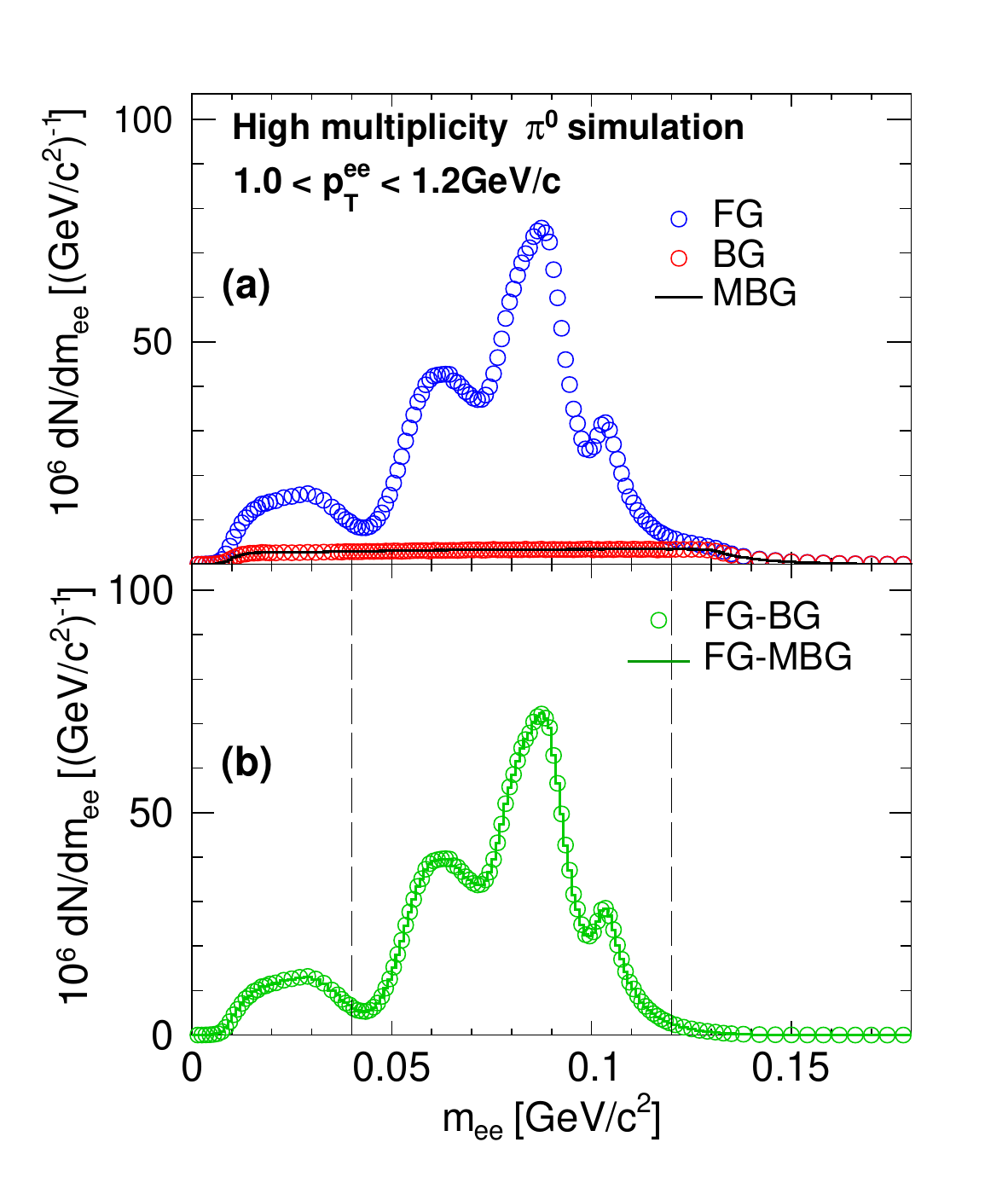} 
\caption{Invariant mass distributions of $e^+e^-$ pairs reconstructed 
from the high-multiplicity \piz pseudodata. Same as \fig{fig:mee1}, but 
with an additional constraint that the $e^+$ and $e^-$ match in beam 
direction. Panel (a) compares foreground, \Ninclfg, the true background, 
\Ninclbg, and the background determined from the mixed event technique, 
\Minclbg. Panel (b) gives the extracted conversion photon signal.}
        \label{fig:mee2}
\end{minipage}
\end{figure*}

\begin{figure}[htb]
        \includegraphics[width=1.0\linewidth]{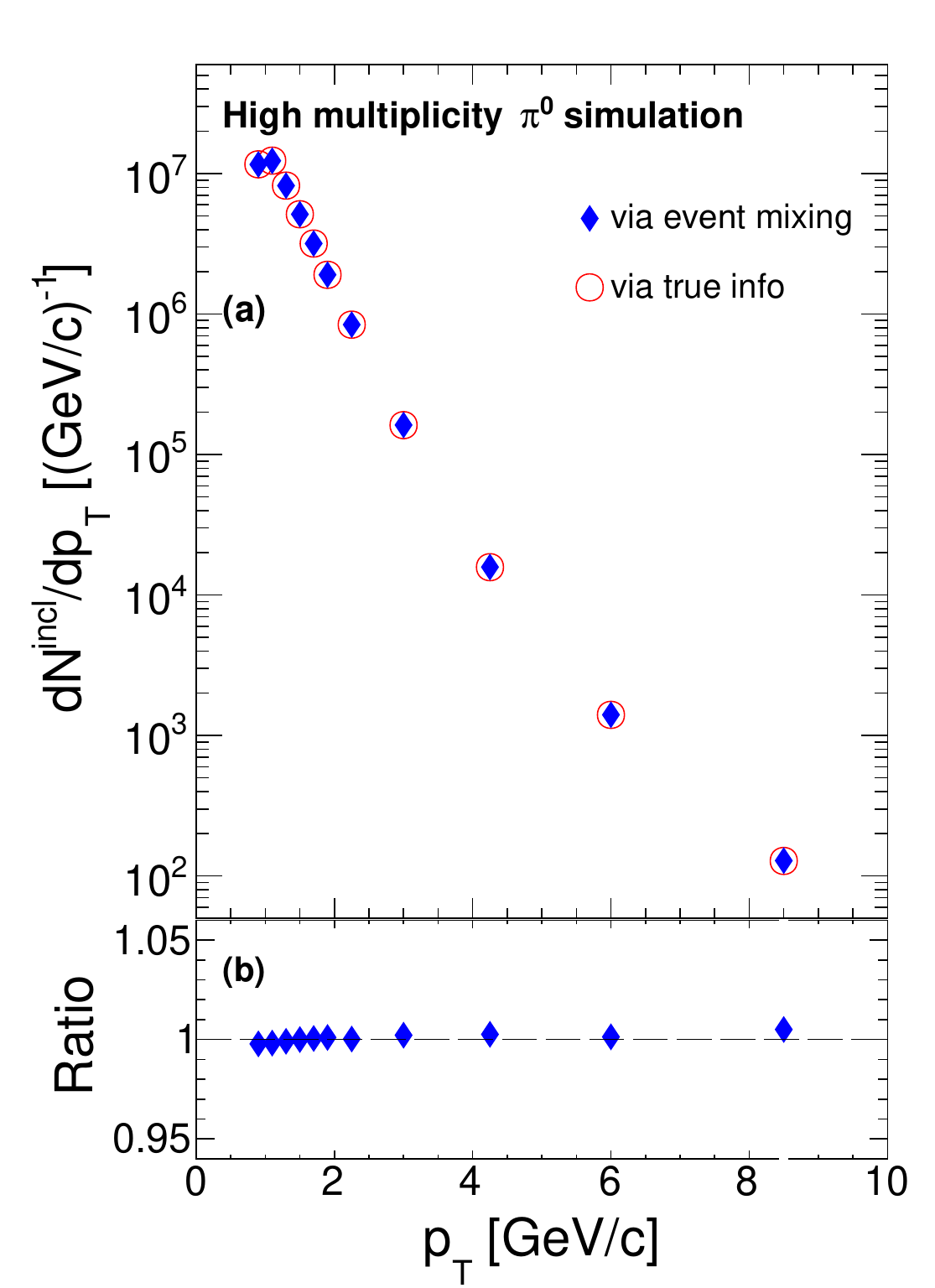} 
         \label{fig:N_incl_compare_dzed}
\caption{Extracted $N_{\gamma}^{\rm incl}$ after the background 
subtraction, as a function of conversion photon \pt. The diamonds are 
obtained by subtracting the background from the mixed event technique; 
they are compared to the open symbols for which the true background was 
subtracted. Panel (b) shows the ratio of the event-mixing result over 
the true information result.}
        \label{fig:incl_compare}
\end{figure}

\begin{figure}[hpt]
         \includegraphics[width=0.95\linewidth]{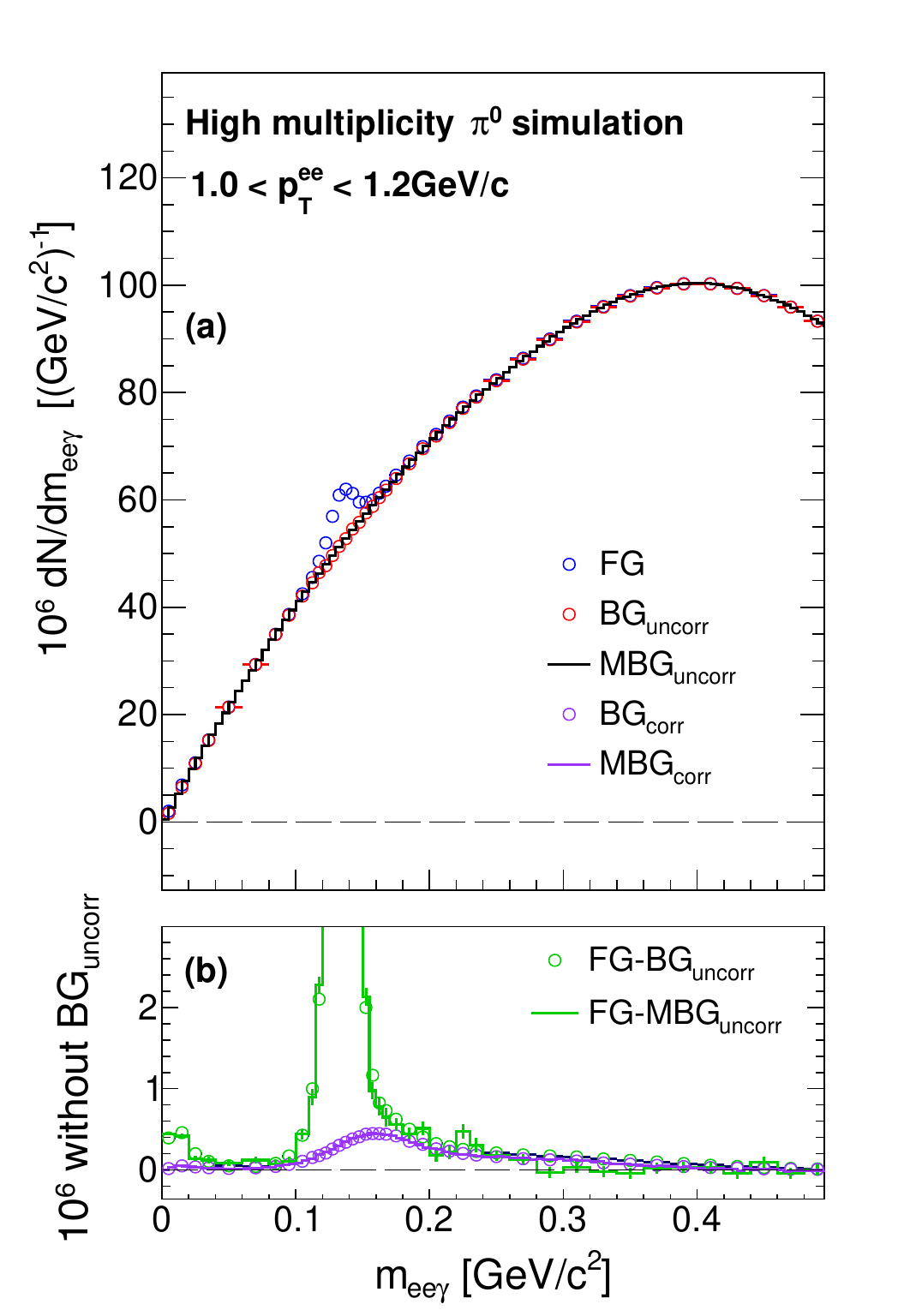} 
\caption{Invariant mass distributions of $e^+e^-\gamma$ pairs.}
        \label{fig:meeg}
\end{figure}

\begin{figure}[hpt]
        \includegraphics[width=1.0\linewidth]{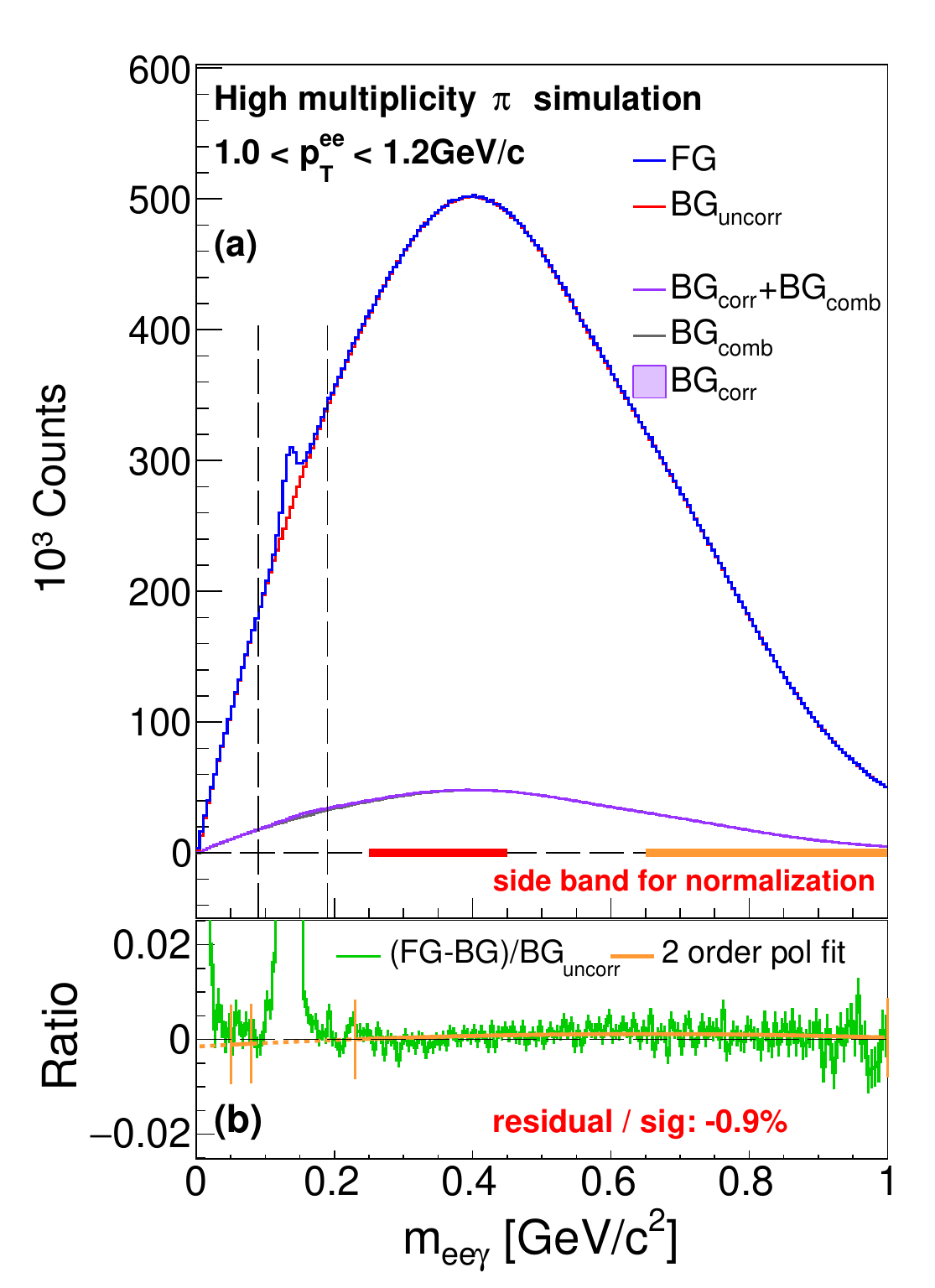} 
\caption{Invariant mass distributions of $e^+e^-\gamma$ pairs from the 
same event (FG) and different event-mixing setups. }
        \label{fig:meeg_norm}
\end{figure}

\begin{figure}[hpt]
        \includegraphics[width=1.0\linewidth]{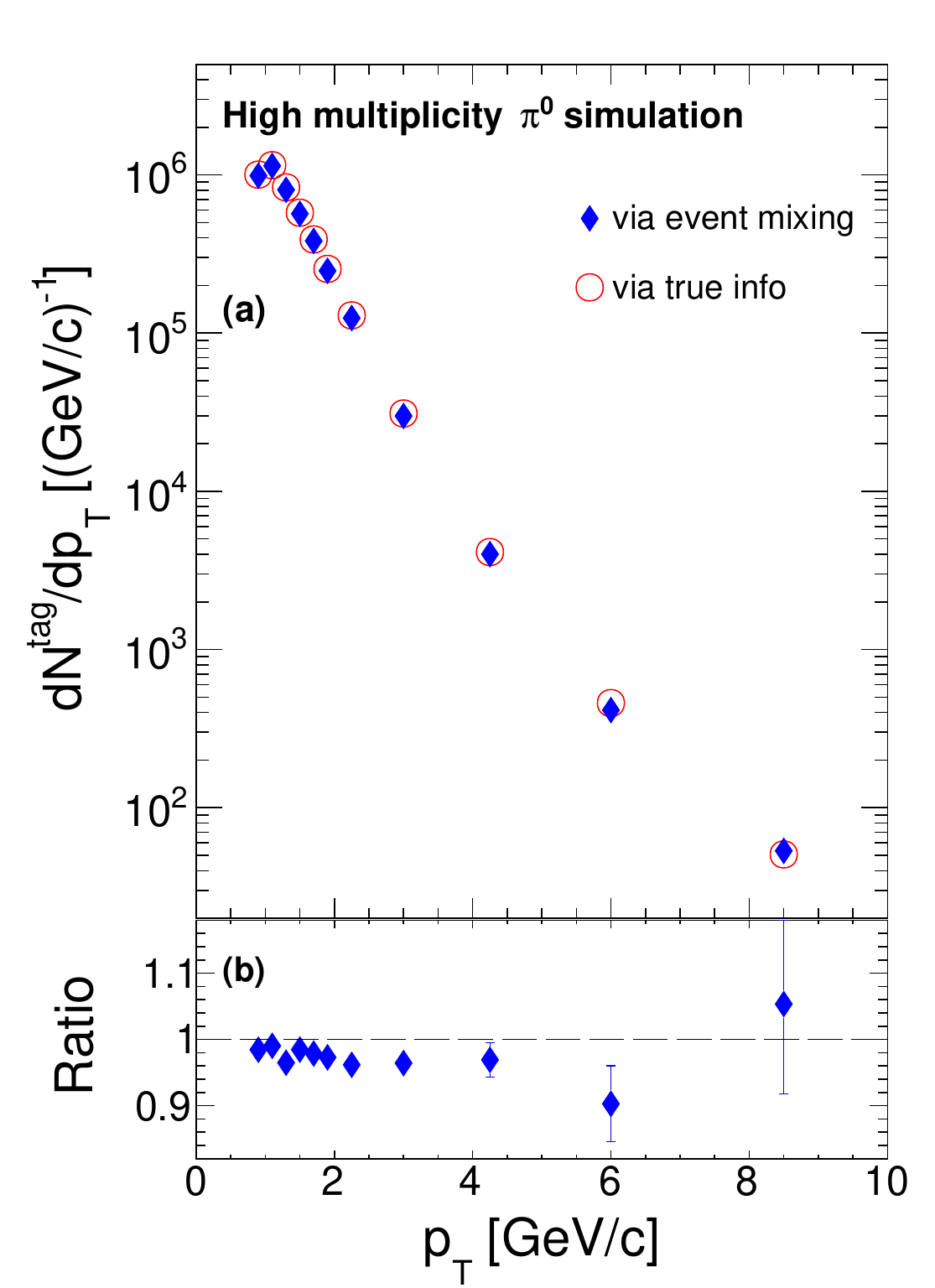} 
\caption{Extracted \Ntag as a function of conversion-photon $p_{T}$ 
using the (red) true information and (blue) event-mixing technique. The 
bottom panel shows the ratio of the event-mixing result over the true 
information result.}
        \label{fig:tag_compare}
\end{figure}

\begin{figure}[htb] 
\begin{minipage}{0.99\linewidth}
         \includegraphics[width=1.0\linewidth]{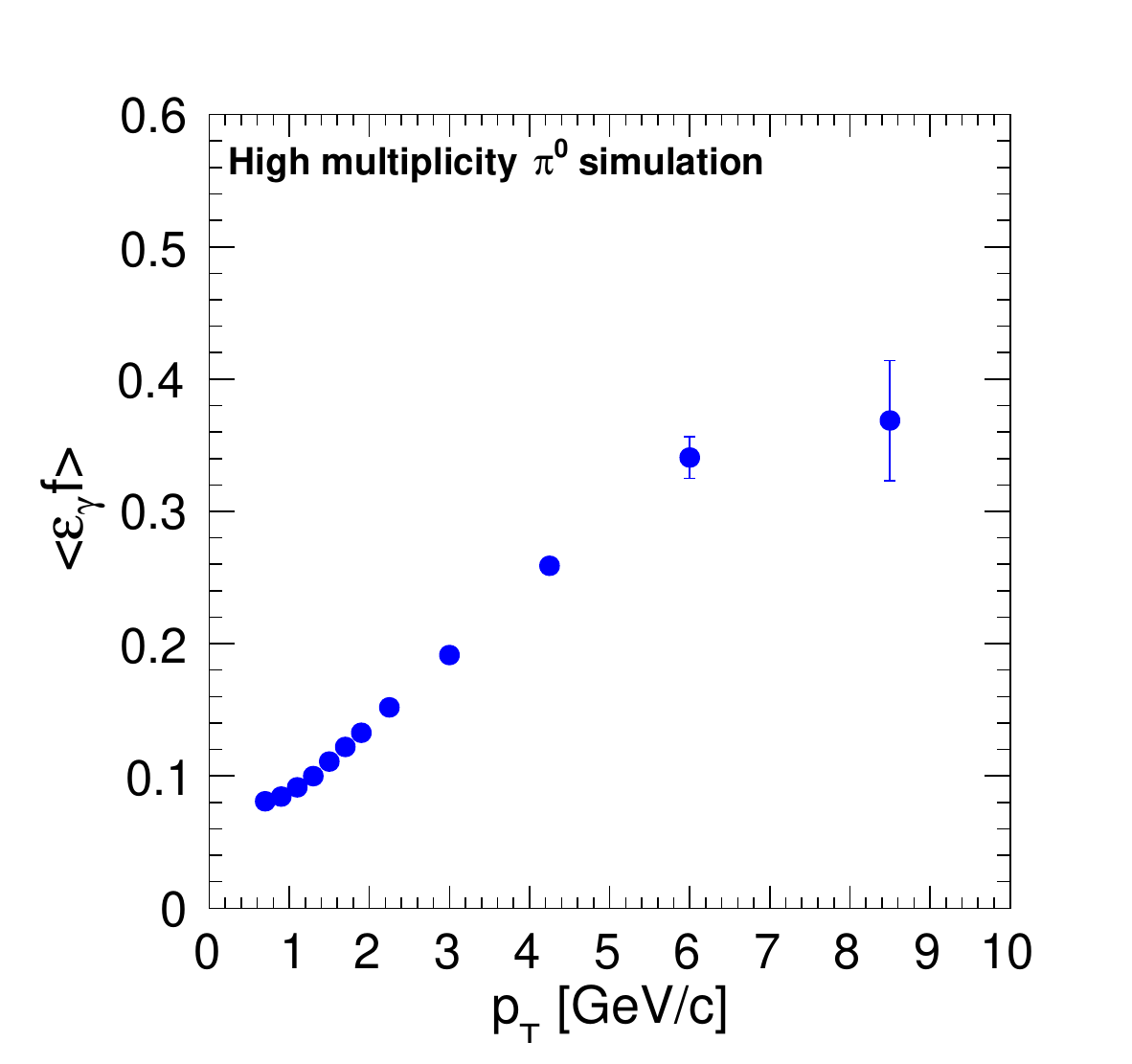} 
\caption{Average conditional probability \ef as a function of conversion 
photon $p_{T}$. }
        \label{fig:280pi0ef}
\end{minipage}
\begin{minipage}{0.99\linewidth}
        \includegraphics[width=1.0\linewidth]{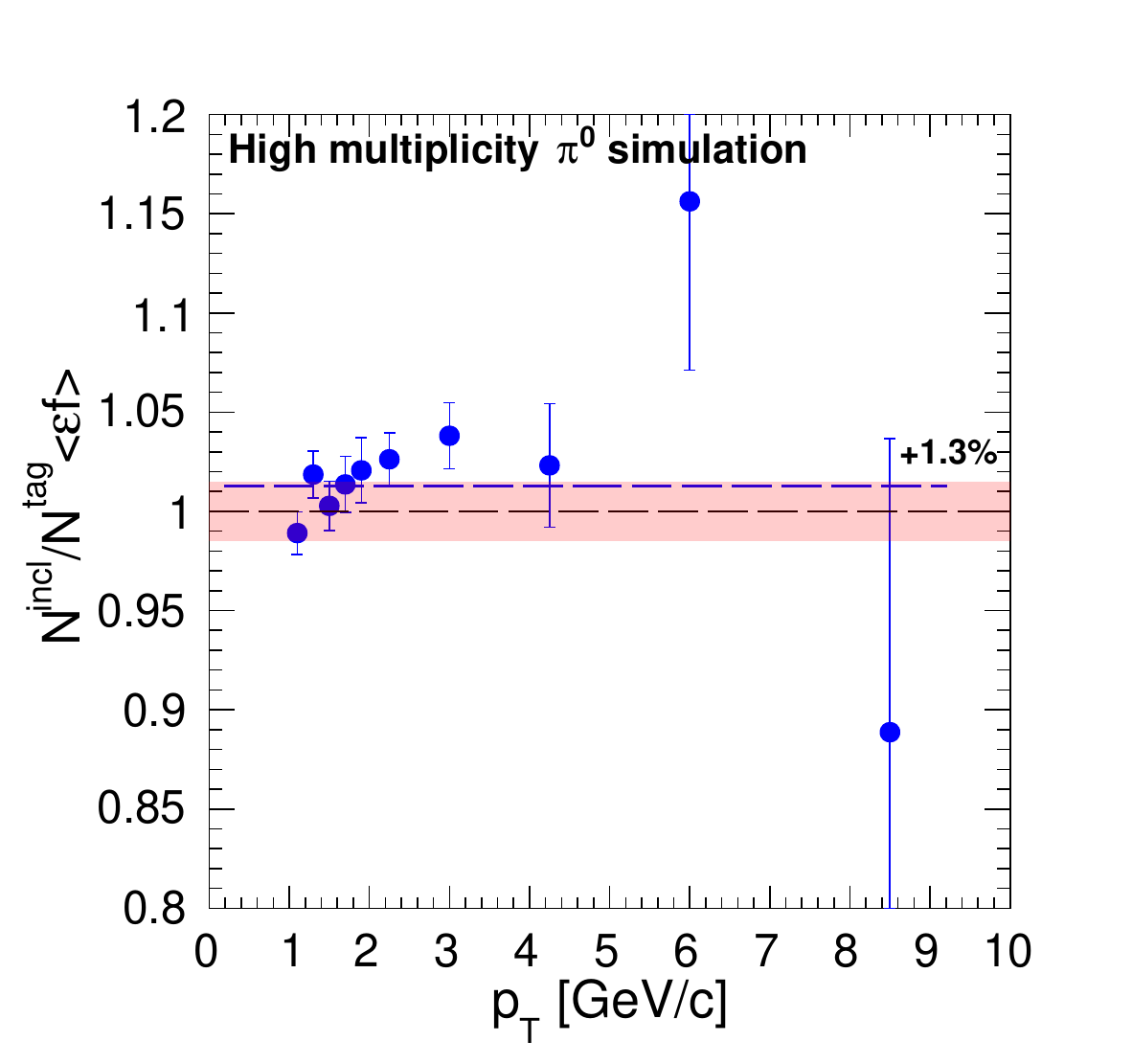} 
\caption{Ratio $R_{\gamma}^{\rm pseudo}$ as a function of conversion 
photon $p_{T}$. The dashed line 
gives a constant offset of 1.3\% fit to the points, and the dashed band 
represents a $\pm1.5\%$ range around unity. }
        \label{fig:pseudo_Rg}
\end{minipage}
\end{figure}



\begin{acknowledgments}

We thank the staff of the Collider-Accelerator and Physics
Departments at Brookhaven National Laboratory and the staff of
the other PHENIX participating institutions for their vital
contributions.  
We also thank J.F. Paquet for many fruitful discussions and
sharing additional information.
We acknowledge support from the Office of Nuclear Physics in the
Office of Science of the Department of Energy,
the National Science Foundation,
Abilene Christian University Research Council,
Research Foundation of SUNY, and
Dean of the College of Arts and Sciences, Vanderbilt University
(USA),
Ministry of Education, Culture, Sports, Science, and Technology
and the Japan Society for the Promotion of Science (Japan),
Natural Science Foundation of China (People's Republic of China),
Croatian Science Foundation and
Ministry of Science and Education (Croatia),
Ministry of Education, Youth and Sports (Czech Republic),
Centre National de la Recherche Scientifique, Commissariat
{\`a} l'{\'E}nergie Atomique, and Institut National de Physique
Nucl{\'e}aire et de Physique des Particules (France),
J. Bolyai Research Scholarship, EFOP, the New National Excellence
Program ({\'U}NKP), NKFIH, and OTKA (Hungary),
Department of Atomic Energy and Department of Science and Technology
(India),
Israel Science Foundation (Israel),
Basic Science Research and SRC(CENuM) Programs through NRF
funded by the Ministry of Education and the Ministry of
Science and ICT (Korea).
Ministry of Education and Science, Russian Academy of Sciences,
Federal Agency of Atomic Energy (Russia),
VR and Wallenberg Foundation (Sweden),
University of Zambia, the Government of the Republic of Zambia (Zambia),
the U.S. Civilian Research and Development Foundation for the
Independent States of the Former Soviet Union,
the Hungarian American Enterprise Scholarship Fund,
the US-Hungarian Fulbright Foundation,
and the US-Israel Binational Science Foundation.

\end{acknowledgments}


\appendix

\section{Event-mixing procedures and validation}
\label{sec:appendix-A}

In this analysis, \ee pairs and \eeg combinations result from combining 
positrons, electrons, and photons measured in the same event. Given the 
large multiplicity of produced particles in \auau collisions, the 
combinations include a significant background from particles of 
different physical origin, for example different \piz decays. For \ee 
pairs there are two possible combinations: signal pairs, \Ninclsg, that 
have the same source and background pairs, \Ninclbg, that have different 
sources. Both types will be combined with photons to get \eeg 
combinations.  There are three possibilities:  A correlated \ee pair is 
combined with a photon from the same source (\Ntagsg); the \ee pair is 
not correlated, but the photon is correlated to the $e^+$ or $e^-$ 
(\Ntagcorr); or the photon is uncorrelated to the \ee pair, irrespective 
whether it is \Ninclsg or \Ninclbg (\Ntaguncorr).

All backgrounds are determined using event-mixing techniques that were 
developed and validated with MC studies of high-multiplicity events, 
for which a large sample of simulated \piz events 
was generated.  These events serve as pseudodata. The \piz are generated 
according to the experimentally observed \pt spectrum, uniform in 
azimuthal angle, and with a constant rapidity density of 280 \piz, 
which corresponds to the typical \piz multiplicity in the most central 
\auau collisions at \snn{200}.

From these pseudodata, \Nincl and \Ntag are extracted using the cuts 
and event-mixing schemes developed for the analysis of real data. They 
are corrected by \ef, resulting in \Rg.  Because in the pseudodata there 
are no other hadronic decay channels contributing to \gammahadr other 
than \piz, the \Rg from this pseudodata is given by:
\begin{equation}
R_{\gamma}^{\rm pseudo} = \frac{N^{\rm incl}_{\gamma}}{N_{\gamma}^{\pi^{0},\rm tag}}\times \ef
\end{equation}\label{eq:Rg_pseudoA}
As there are no direct photons in the pseudodata, the expected result 
would be \Rg = 1, within the statistical uncertainties of the 
simulation.  The rest of this sections details each step of the \Rg 
determination from the pseudodata.  The exact same procedure is also 
applied to the real data.

\subsection{Determination of the inclusive photon yield {\Nincl}} 
\label{sec:method_Nincl}

Photon conversion candidates are created by combining $e^+$ and $e^-$ 
from the same pseudodata event by requiring a valid conversion point 
within $1<R<29$ cm. This results in a foreground, \Ninclfg, containing a 
signal, \Ninclsg, that is, conversions of \piz decay photons, and a 
background, \Ninclbg, where the $e^+$ and $e^-$ come from conversion of 
two different \piz decay photons. The background is determined by 
combining electrons and positrons from different pseudodata events, 
which are paired and subjected to the same cuts and conversion selection 
criteria. The mixed event background thus obtained, \Minclbg, is 
normalized to the foreground, \Ninclfg, in the mass region $0.16 < \mee 
< $ 0.3~\gevcc, which does not contain \ee pairs from conversions (see 
\fig{fig:massPeaks} for reference).

Figure~\ref{fig:mee1}(a) shows the background, \Minclbg, obtained from 
the mixed event technique together with the true background, \Ninclbg, 
which was obtained from the MC ancestry information.  
Figure~\ref{fig:mee1}(b) shows the results (solid curve) after 
subtracting the mixed-event background from the foreground and (open 
symbols) subtracting the true background.  Note that the two are 
practically indistinguishable, which means that \Ninclbg is equal to 
\Minclbg.

Even though the background can be subtracted accurately with the 
mixed-event technique to obtain \Nincl, the subtraction can only be done 
statistically. Thus in the next step, where conversion photons from \piz 
decays are tagged, the background pairs also need to be matched with 
EMCal showers. This substantially increases the background in the \meeg 
distribution. To reduce this background, additional cuts are applied in 
the conversion-photon selection.

The magnetic field deflects electrons and positrons in a plane 
perpendicular to the beam direction ($z$). Thus, \ee pairs from a 
conversion can be constrained by requiring a match in the beam direction 
using the PC1 information. A cut of $|\Delta z|<4$~cm is applied. 
Because the conversion reconstruction algorithm uses the projection of 
the tracks in the plane perpendicular to the beam axis, the additional 
match reduces the number of possible random-track combinations 
significantly. The $z$ cut effectively truncates the mass distribution 
as the \ee pairs are required to have the possible conversion point at 
radii below 29\,cm and only the pairs with an opening angle in the beam 
direction will create larger masses. The background rejection is clearly 
visible in \fig{fig:mee2}. The background normalization for the mixed 
events is given by the less-restrictive cuts shown in \fig{fig:mee1}, 
and applied here. For the lowest \pt and the highest-multiplicity bin, 
the background rejection is approximately a factor of eight with a 
signal efficiency of more than 85\%. The background to foreground ratio, 
$\Ninclbg/\Ninclfg$, is 12.1\%.  As \pt increases the multiplicity 
decreases and the $\Ninclbg/\Ninclfg$ ratio decreases to 0.3\% at the 
$p_{T}$ above 7~\gevc.

The analysis is repeated for the entire accessible \pt range and \Nincl 
is calculated in the mass range from 0.04 to 0.12~\gevcc by subtracting 
the background obtained from the mixed-event technique, \Minclbg, from 
the foreground, \Ninclfg. The result is compared to the true number of 
photon conversions determined from the MC-ancestry information in 
\fig{fig:incl_compare}. Panel (b) shows that the difference is less than 
1\% for all \pt.

\subsection{The tagged photon yield {\Ntag}} \label{sec:method_Ntag}

Next, the subset \Ntag of \ee pairs in the \Nincl sample that can be 
tagged as photons from a \piz decay is determined. For a given 
pseudodata event, each \ee conversion candidate is paired with all 
reconstructed showers in the EMCal, excluding the showers matched to the 
\ee pair itself.  For each combination the invariant mass \meeg is 
calculated. This constitutes the foreground, \Ntagfg, for which an 
example is given in panel (a) of \fig{fig:meeg}. Despite the large 
background the signal \Ntag is clearly visible as peak around the \piz 
mass. The background has two components: (i) combinations of \ee pairs 
with an EMCal shower from another unrelated \piz decay denoted as 
\Ntaguncorr, and (ii) a correlated background, \Ntagcorr, where the 
shower in the EMCal and the electron or positron are from the same \piz 
decay, but the \ee pair itself is a combination of an $e^+$ and $e^-$ 
from different \piz decay photons.

The uncorrelated background can be determined with a similar event 
mixing technique as used for the extraction of \Nincl; an \ee pair from 
one event is mixed with the EMCal showers from a different event 
resulting in mixed combinations, \Mtaguncorr. These are normalized to 
the foreground, \Ntagfg, in the mass region from 0.25 to 0.45~\gevcc, 
where no signal is expected. Figure~\ref{fig:meeg}(a) shows the 
corresponding distribution. There is almost no visible difference 
between the mixed-event background, \Mtaguncorr, and the true 
background, \Ntaguncorr, which is obtained using the MC-ancestry 
information. Figure~\ref{fig:meeg}(b) shows the signal and remaining 
correlated background after the uncorrelated mixed-event background is 
subtracted (\Ntagfg-\Mtaguncorr), as well as after subtracting the true 
uncorrelated background (\Ntagfg-\Ntaguncorr). Again they are 
indistinguishable.

The correlated background, \Ntagcorr, is determined with a second 
event-mixing scheme. An $e^+$ from a given event is combined with an $e^-$ 
from a different event, and the resulting \ee pair is then combined with 
the showers in the EMCal from both events; again excluding the showers 
from the $e^+$ and $e^-$. The \eeg combinations contain the correlated 
background, \Mtagcorr, plus the random background in which the $e^+$, 
$e^-$, and $\gamma$ are from three different \piz decays, \Mtagcomb. The 
normalization is per generated \ee pair, multiplied by \Ntagfg, i.e. the 
number of background pairs in the \ee pair foreground.

The random background, \Mtagcomb, can easily be determined in a third 
event-mixing step, where $e^+$, $e^-$, and $\gamma$ are from three 
different events. The \Mtagcomb is normalized to (\Mtagcorr$+$\Mtagcomb) 
in the mass range from 0.65 to 1.0~\gevcc and subtracted. 
Figure~\ref{fig:meeg_norm}(a) shows the the result, \Mtagcorr, together 
with the foreground and the other background components.
 
Last but not least, to account for any possible mismatch between the 
true background and the one obtained from our multistep event-mixing 
procedure, the ratio $(\Ntagfg-\Mtagcorr-\Mtaguncorr)/\Mtaguncorr$ is 
fit with a second-order polynomial, $f_{ee\gamma}$, excluding the \piz 
peak regions. The fit result is shown as a line on 
\fig{fig:meeg_norm}(b).  This fit is used to correct \Mtaguncorr before 
subtraction.  The final distribution for \Ntag is thus:
\begin{equation} \Ntag = \Ntagfg - \Mtagcorr - (1+f_{ee\gamma}) \ \Mtaguncorr 
\label{eq:Ntag1}
\end{equation}

For each \pt bin \Ntag is extracted by counting the number of entries in 
a window around the \piz peak ($0.09 < \meeg < 0.19$)~\gevcc. 
Figure~\ref{fig:tag_compare} shows \Ntag as function of \pt using the 
true MC-ancestry information and the event-mixing technique. Overall the 
agreement is very good, however, the result from the event-mixing 
technique is on average lower. This mismatch is accounted for in the 
systematic uncertainties on $R_{\gamma}$, which is discussed in more 
detail in the next section.

\subsection{Completing the validation by determining {\Rg}} 
\label{sec:method_ef}

With \Nincl and \Ntag established from the pseudodata, the conditional 
probability \ef remains to be determined to calculate \Rg and fully 
validate the background-subtraction procedure.  In the same way as for 
the data, a single \piz simulation is embedded into pseudodata events. 
The \ee pairs and \eeg combinations are reconstructed and counted as 
discussed in \sect{sec:analysis_ef}. The extracted \ef is shown in 
\fig{fig:280pi0ef} as a function of the conversion photon $p_T$.

With \Nincl/\Ntag from the pseudodata and \ef from the embedded single 
$\pi^0$ simulation in hand, \Rg is calculated using Eq. A1. The result 
is shown in \fig{fig:pseudo_Rg}, all points are close to unity 
indicating that the analysis procedure is self consistent. There may be 
a 1.5\% enhancement above unity, which is consistent with the slightly 
lower-than-expected value found for \Ntag. This difference is taken into 
account in the estimate of the systematic uncertainty.
 
 
\section{Uncertainty propagation with a MC sampling method}
\label{sec:appendix-B}

The uncertainties on \gammadir and any other quantity derived from 
\gammadir, such as \Teff or $\alpha$, are determined using a MC-sampling 
method, which allows taking into account the \pt and centrality 
dependent correlations of individual sources of systematic 
uncertainties, as well as the fact that the region \Rg~$<1$ is 
unphysical.

\subsection{Systematic uncertainties}

In the MC-sampling method, for each source of uncertainty, $i$, a 
variation $\delta_{i}$ of \Rg or \gammahadr is sampled from a Gaussian 
distribution centered at zero with a width corresponding to the 
associated uncertainty, $\sigma_{i}$. The size of $\delta_i$ depends not 
only on $\sigma_{i}$, but also on whether the adjacent bins in \pt and 
centrality have uncorrelated (Type~A) or correlated (Type~B/C) 
uncertainties due to the source $i$. The values of $\sigma_i$ and 
classification of each source is summarized in \tab{tab:syserr}.

If source $i$ is classified as uncorrelated, $\delta_i$ is calculated 
independently for neighboring bins from Gaussian distributions of width 
$\sigma_i$. For correlated uncertainties of Type~C in \pt or centrality, 
$\delta_i$ is calculated with one common fraction $w$ so that 
$\delta_i=w\sigma_i$ for all points. The fraction $w$ is determined 
randomly from a Gaussian distribution of width~1. And finally, for 
Type~B uncertainties, $\delta_i$ is determined separately for the 
minimum and maximum of the \pt or centrality range using the same 
procedure as Type C. All intermediate points are varied proportionally 
to create a smooth transition from the minimum to the maximum of the 
range. Uncertainties on the input \piz \pt distribution are a special 
case of Type~B uncertainties, as it is known that the systematic 
uncertainties move simultaneously either up or down. In this case, 
$\delta_i$ at the minimum and maximum of the range are chosen to have 
the same sign.

After applying all variations $\delta_i$ to recalculate \Rg and 
\gammahadr, new values of \gammadir, \Teff, and $\alpha$ are determined. 
This process is repeated multiple times, taking into account the 
different sources of uncertainties, to obtain a distributions of 
\gammadir, \Teff, and $\alpha$. The width of these distribution is 
quoted as the systematic uncertainty.  For individual \gammadir points, 
it is possible that $\langle\gammadir\rangle-\sigma$ is less than 0, 
that is, unphysical. In such cases, an upper limit of 90\% confidence 
level (CL) is quoted based on the part of the probability distribution 
in the physical region $\int_{0}^{\rm upper}/\int_{0}^{+\infty}=90\%$.

\subsection{Statistical uncertainties}

The statistical uncertainties on \Rg are assumed to have a Gaussian 
probability distribution and for most cases the statistical uncertainty 
on \gammadir can be calculated with the usual error propagation. 
However, there are two cases that need to be treated separately:

\begin{itemize}

\item $\Rg<1$: In this case \gammadir is unphysical, and hence an upper 
limit at 90\%~CL is quoted, based on the physical part of the 
probability distribution $\int_{0}^{\rm upper}/\int_{0}^{+\infty} = 
90\%$.

\item $\Rg-\sigma_{\rm stat}<1$: In this case \gammadir is in the 
physical region, but consistent with zero within less than one standard 
deviation. For these situations the central value is shown, but the 
uncertainty is given as 90\%~CL, calculated as above.

\end{itemize}


 
%

\end{document}